\documentclass[pdflatex,sn-mathphys,Numbered]{sn-jnl}
\pdfoutput=1

\usepackage{manyfoot}

\usepackage{amsmath}
\usepackage{amssymb}
\usepackage{amsthm}
\usepackage{thmtools}

\usepackage{algorithmic}
\usepackage[linesnumbered,titlenumbered,ruled,vlined,resetcount,algosection]{algorithm2e}

\usepackage{pdflscape}

\usepackage[usenames,dvipsnames]{xcolor}
\usepackage{graphicx}
\usepackage{multirow}
\usepackage{float}
\usepackage{booktabs}

\usepackage{xparse}

\usepackage[labelformat=simple]{caption}
\usepackage[labelformat=simple]{subcaption}

\usepackage[normalem]{ulem}

\usepackage{etoolbox}

\usepackage[capitalize,noabbrev]{cleveref}
\crefformat{equation}{Equation #2#1#3}
\creflabelformat{equation}{#2\textup{#1}#3}

\usepackage{todonotes}

\newtheoremstyle{theoremstyle}
	{15pt} % Space above
	{} % Space below
	{\itshape} % Body font
	{} % Indent amount
	{\bfseries} % Theorem head font
	{.} % Punctuation after theorem head
	{.5em} % Space after theorem head
	{} % Theorem head spec (can be left empty, meaning `normal')

\theoremstyle{theoremstyle}

\newtheorem{theorem}{Theorem}[section]
\newtheorem{lemma}[theorem]{Lemma}
\newtheorem{proposition}[lemma]{Proposition}

\newcommand \red[1]		{\textcolor{red}{#1}}

\newcommand \rla {\mathrm{rla}}
\newcommand \la {\mathrm{la}}
\newcommand \rsa {\mathrm{rsa}}

\newcommand \expe[2]	{\mathbb{E}_{#1}{\left[#2\right]}}
\newcommand \gexpe[1]	{\expe{*}{#1}}
\newcommand \lexpe[1]	{\expe{\rla}{#1}}

\newcommand \expet[2]	{\expe{#1}{\gamma_{#2}}}
\newcommand \gexpet[1]	{\expet{*}{#1}}

\newcommand \expetw[1]	{\expet{#1}{\omega}}
\newcommand \gexpetw	{\expetw{*}}
\newcommand \lexpetw	{\expetw{\rla}}

\newcommand \var[2]		{\mathbb{V}_{#1}{\left[#2\right]}}
\newcommand \gvar[1]	{\var{*}{#1}}
\newcommand \lvar[1]	{\var{\rla}{#1}}

\newcommand \probdelta[1]	{\delta_{#1}}
\newcommand \gprobdelta		{\probdelta{*}}
\newcommand \lprobdelta		{\probdelta{\rla}}
\newcommand \sprobdelta		{\probdelta{\rsa}}

\newcommand \probalphast[2]		{p_{#1,\:#2}}
\newcommand \gprobalphast[1]	{\probalphast{*}{#1}}
\newcommand \lprobalphast[1]	{\probalphast{\rla}{#1}}
\newcommand \sprobalphast[1]	{\probalphast{\rsa}{#1}}
\newcommand \probalphastw[1]	{\probalphast{#1}{\omega}}
\newcommand \gprobalphastw		{\probalphastw{*}}
\newcommand \lprobalphastw		{\probalphastw{\rla}}

\newcommand \reals {\mathbb{R}}

\NewDocumentCommand \crossG { O{G} }{C_{#1}}

\newcommand \arr[1] {\pi_{#1}}
\newcommand \garr {\arr{*}}

\newcommand \tree {T}
\newcommand \forest {F}

\newcommand \maxdeg {k_{max}}

\NewDocumentCommand \cycle { O{n} }{ \mathcal{C}_{#1}}
\NewDocumentCommand \lintree { O{n} }{ \mathcal{L}_{#1}}
\NewDocumentCommand \startree { O{n} }{ \mathcal{S}_{#1}}
\NewDocumentCommand \complete { O{n} }{ \mathcal{K}_{#1}}
\NewDocumentCommand \quasistar { O{n} }{ \mathcal{Q}_{#1}}
\NewDocumentCommand \compbip { O{n_1} O{n_2} }{ \mathcal{K}_{#1,#2}}

\NewDocumentCommand \zeroreg { O{n} }{ \textbf{0}_{#1}}
\NewDocumentCommand \onereg { O{n} }{ \textbf{1}_{#1}}
\NewDocumentCommand \tworeg { O{n} }{ \textbf{2}_{#1}}
\NewDocumentCommand \maxdegreg { O{n} }{ \textbf{n-1}_{#1}}
\NewDocumentCommand \kreg { O{n} }{ \textbf{k}_{#1}}

\NewDocumentCommand \npathsfour { O{G} }{n_{#1}(\lintree[4])}
\NewDocumentCommand \npathsfive { O{G} }{n_{#1}(\lintree[5])}
\NewDocumentCommand \nsquares { O{G} } { n_{#1}(\cycle[4]) }
\NewDocumentCommand \ntris { O{G} } { n_{#1}(\cycle[3]) }

\newcommand \graphpaw {Z}
\newcommand \CoL {Y}
\newcommand \CoLlong {\cycle[3] \oplus \lintree[2]}

\newcommand \ngraphpaw {n_G(\graphpaw)}
\newcommand \nCoL {n_G(\CoL)}

\newcommand \mmtdeg[1] { \langle k^{#1} \rangle }

\newcommand \bigO[1] {O{\left(#1\right)}}
\newcommand \smallo[1] {o{\left(#1\right)}}
\newcommand \bigOmega[1] {\Omega{\left(#1\right)}}

\NewDocumentCommand \probnp { m O{n} O{p} } {\mathbb{P}_{#2,#3}{\left(#1\right)}}
\newcommand \condprobnp[2] {\probnp{#1\;|\;#2}}
\NewDocumentCommand \expectednp {m O{n} O{p} } {\mathbb{E}_{#2,#3}{\left[#1\right]}}

\NewDocumentCommand \condexpectednp { m m O{n} O{p} } {\expectednp{#1\;|\;#2}[#3][#4]}

\NewDocumentCommand \randgraphp { O{n} O{p} }{\mathcal{G}_{#1,#2}}

\newcommand \pair[2] {\left\{ #1, #2 \right\}}

\mathchardef\mhyphen="2D
\NewDocumentCommand \gcrn { m O{*} } { \mathrm{cr}_{#2}(#1) }
\NewDocumentCommand \crn { m } { \gcrn{#1}[] }

\NewDocumentCommand \gmaxcr { m O{*} } { \mathrm{max\mhyphen cr}_{#2}(#1) }
\NewDocumentCommand \maxcr { m } { \gmaxcr{#1}[] }

\NewDocumentCommand \grectcr { m O{*} } { \overline{\mathrm{cr}}_{#2}(#1) }

\NewDocumentCommand \gmaxrectcr { m O{*} } { \mathrm{max\mhyphen\overline{\mathrm{cr}}}_{#2}(#1) }

\NewDocumentCommand \gmaxconvcr { m O{*} } { \mathrm{max\mhyphen\overline{\mathrm{cr}}}_{#2}^{\circ}(#1) }

\NewDocumentCommand \gconvcr { m O{*} } { \overline{\mathrm{cr}}_{#2}^{\circ}(#1) }

\begin{document}
\allowdisplaybreaks

\title[Calculation of the variance of edge crossings]{Fast calculation of the variance of edge crossings in random arrangements}
\author*[1]{\fnm{Llu\'is} \sur{Alemany-Puig}}\email{lalemany@cs.upc.edu}
\author[1]{\fnm{Ramon} \sur{Ferrer-i-Cancho}}\email{rferrericancho@cs.upc.edu}

\affil[1]{
	\orgdiv{Departament de Ci\`encies de la Computaci\'o},
	\orgname{Universitat Polit\`ecnica de Catalunya},
	\orgaddress{Jordi Girona 3},
	\city{Barcelona},
	\postcode{08034},
	\state{Catalonia},
	\country{Spain}
}

\date{Received: date / Accepted: date}

\abstract{The crossing number of a graph $G$, $\crn{G}$, is the minimum number of edge crossings arising when drawing a graph on a certain surface. Determining $\crn{G}$ is a problem of great importance in Graph Theory. Its maximum variant, i.e. the maximum crossing number, $\maxcr{G}$, is receiving growing attention. Instead of an optimization problem on the number of crossings, here we consider the variance of the number of edge crossings, when embedding the vertices of an arbitrary graph uniformly at random in some space. In his pioneering research, Moon derived this variance on random linear arrangements of complete unipartite and bipartite graphs. Given the need of efficient algorithms to support this sort of research and given also the growing interest of the number of edge crossings in spatial networks, networks where vertices are embedded in some space, here we derive an algorithm to calculate the variance in arbitrary graphs in time $\smallo{nm^2}$ that we transform into one that runs in time $\bigO{nm}$ by reusing computations. We also derive one for forests that runs in time $\bigO{n}$. These algorithms work on a wide range of random layouts (not only on Moon's) and are based on novel arithmetic expressions for the calculation of the variance that we develop from previous theoretical work. This paves the way for many applications that rely on a fast but exact calculation of the variance.}


\maketitle

%~ \tableofcontents

%------------------------------------------%
% automatic inline of '1-introduction.tex' %
%------------------------------------------%
\section{Introduction}
\label{sec:introduction}

When drawing a graph on a plane or other spaces, edges may cross \cite{Schaefer2017a}. Subfields of Graph Theory are concerned about the range of variation of the number of edge crossings, denoted as $C$, among all {\em good} drawings of $G$ in $\reals^2$. A good drawing of a graph $G$ is defined by the following three properties: (1) no edge crosses itself; (2) an independent pair of edges\footnote{Two edges are independent if they do not have common vertices.} can only cross once, and (3) a pair of non-independent edges cannot cross \cite[p. 46]{Buchheim2013a}. It is easy to see that there exist a potentially infinite number of ways to draw an edge between two fixed points. One of the most popular problems on the range of variation of $C$ is the so-called `crossing number' of a graph $G$, $\crn{G}$, defined as the minimum value of $C$ among all good drawings of $G$ \cite{Schaefer2017a}. Recently, the maximum variant of $\crn{G}$, $\maxcr{G}$, has been receiving increasing attention \cite{Chimani2018a,Fallon2018a,Bennet2019a}. It is known that the calculation of $\crn{G}$ or $\maxcr{G}$ is NP-hard \cite{Garey1983a,Bald2016a}.

In his pioneering research, Moon \cite{Moon1965a} investigated other aspects of the distribution of $C$. For instance, the expectation and the variance of $C$ when the graph's vertices are embedded on the surface of a sphere and the edges are uniquely determined by the shortest geodesic on said surface \cite{Moon1965a,Alemany2018b}. In a broader scenario, one can consider two problems on the distribution of $C$ in some layout $*$ of the vertices of an arbitrary graph $G$: the expectation $\gexpe{C}$, a trivial problem, and the variance $\gvar{C}$, which is a far more complex problem \cite{Alemany2018a}. These layouts, among which we find Moon's embedding on the surface of a sphere as particular case \cite{Moon1965a}, have to satisfy the three properties of good drawings and, additionally, (4) that $e$ edges crossing at the same point incur in $\binom{e}{2}$ crossings. In this paper, we condense the notation for layouts, $*$, to denote both the physical space in which the graph's vertices are embedded and the distribution of the vertex positions. Notice that the range of variation of $C$ (i.e. $\crn{G}$ and $\maxcr{G}$) and $\gexpe{C}$ and $\gvar{C}$ are interrelated by definition, e.g., \cite{Bathia2000a},
\begin{equation}
\gvar{C} \leq (\gmaxcr{G} - \gexpe{C})(\gexpe{C} - \gcrn{G}).
\end{equation}

\begin{figure}
	\centering
	\includegraphics[scale=0.8]{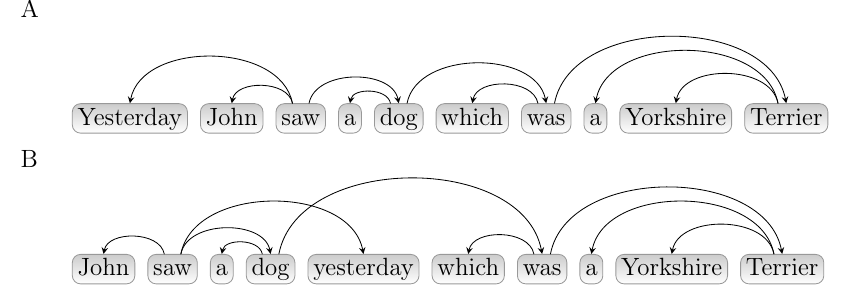}
	\caption{A) A syntactic dependency tree without edge crossings ($C = 0$). B) A reordering of the syntactic dependency tree with one edge crossing ($C = 1$). These are examples adapted from \cite[Figure 2]{McDonald05b}. }
	\label{fig:introduction:syntactic_dependency_trees}
\end{figure}

\paragraph{Motivation} In the classic setting of Graph Theory, one aims to find an optimal embedding of a given graph, i.e. an embedding that yields $\crn{G}$ or $\maxcr{G}$. In this context, research on $\gexpe{C}$ and $\gvar{C}$ complements our understanding of the distribution of $C$. Besides, $\gexpe{C}$ and $\gvar{C}$ are of paramount importance in the context of the Theory of Spatial Networks, networks whose vertices are embedded in some space \cite{Barthelemy2018a}. In this sort of networks, the layout is given by some random distribution as in Moon's classic work \cite{Moon1965a,Alemany2018b} or given by the physical position of every vertex of some real network \cite{Barthelemy2018a}. Prototypical examples of the latter case are streets, roads and transportation networks (e.g., subway and train networks); these are spatial networks on a space that is usually assumed to be two-dimensional. The study of spatial networks is driven by development of that theory in non-Euclidean geometries \cite{Barthelemy2018a}. However, the interest in networks whose vertices are embedded in a one-dimensional Euclidean space cannot be neglected. Remarkable examples are syntactic dependency networks and RNA structures. As for the former, the syntactic dependency network of a sentence can be defined as a network (usually a tree\footnote{Trees are undirected, connected, acyclic graphs.}) where vertices are words, edges indicate syntactic dependencies, and their layout is a one-dimensional space where vertices are allocated in integer positions in $[1,n]$ and edges are drawn as semicircles above the sentence (\cref{fig:introduction:syntactic_dependency_trees}). Syntactic dependency structures have become the {\em de facto} standard to represent the syntactic structure of sentences in Computational Linguistics \cite{kubler09book} and the fuel for many quantitative studies \cite{Liu2017a}. In that setup, edges may cross when drawn above the sentence (\cref{fig:introduction:syntactic_dependency_trees}). The one-dimensional layout of a syntactic dependency structure defines a linear arrangement of the vertices of the network, henceforth referred to simply as `linear arrangement'. Crossings may also occur in linear arrangements of RNA secondary structures, where vertices are nucleotides {\tt A}, {\tt G}, {\tt U}, and {\tt C}, and edges are Watson-Crick ({\tt A}-{\tt U}, {\tt G}-{\tt C}, {\tt U}-{\tt G}) base pairs \cite{Chen2009a}. A linear arrangement of a graph without crossings is called a one-page (book) embedding \cite{Chung1987a,Hochberg2003a} or a planar embedding \cite{Iordanskii1987a}.

The needs of computing the exact value of $\gexpe{C}$ and $\gvar{C}$ efficiently are many. First, as a computational backup for mathematical research on the distribution of crossings in random layouts \cite{Moon1965a}. The algorithms presented here were crucial to find and correct some inaccuracies in Moon's pioneering research \cite{Alemany2018b}. Second, these two properties allow one to standardize real values of $C$ using a $z$-score, defined as
\begin{equation}
\label{eq:introduction:z-score}
z = \frac{C - \gexpe{C}}{\sqrt{\gvar{C}}}.
\end{equation}
$z$-scores have been used to detect scale invariance in empirical curves \cite{Cocho2015a,Morales2016a} or motifs in complex networks \cite{Milo2002a} (see \cite{Stone2019a} for a historical overview). Thus, $z$-scores of $C$ would allow one to discover new statistical patterns involving $C$ in syntactic dependency trees, to name one example. Moreover, $z$-scores of $C$ can help aggregate or compare values of $C$ from graphs with different structural properties (number of vertices, degree distribution,...), as it happens with syntactic dependency trees, when calculating the average number of crossings in collections of sentences of a given language \cite{Ferrer2017a}. Third, the crossing number is trivial when considering linear arrangements (`la') of trees: since trees are outerplanar graphs they admit a one-page book embedding \cite{Chung1987a}, therefore, the minimum number of edge crossings that can be achieved in a linear arrangement of a tree $T$ is $\gcrn{T}[\la] = 0$. \cref{fig:introduction:syntactic_dependency_trees} illustrates two linear arrangements of the same tree.

However, other aspects of the distribution of $C$ are also very important, because random linear arrangements (henceforth denoted as `rla') are used as baselines to answer many research questions. For instance, to check if the real number of $C$ in syntactic dependency structures is actually lower than expected by chance \cite{Ferrer2017a}. The large and growing collections of syntactic dependency structures that are available \cite{ud26_APS,sud} calls for efficient algorithms to calculate $\lexpe{C}$ and $\lvar{C}$. This can be applied to the improvement of tests for the significance of the real value of $C$ with respect to random linear arrangements which are based on a Monte Carlo estimation of the $p$-value \cite{Ferrer2017a}. Using fast algorithms, an upper bound of the $p$-value could be obtained immediately using Chebyshev-like inequalities, which require knowledge of $\lexpe{C}$ and $\lvar{C}$. If the $p$-value was below the significance level, the null hypothesis could be rejected quickly, skipping the time-consuming estimation of an accurate enough $p$-value.

\paragraph{Overview of previous work} On the positive side, the calculation of $\gexpe{C}$ of a given graph $G$ is straightforward. It has been shown that 
\begin{equation*}
\gexpe{C} = q\gprobdelta
\end{equation*}
where $\gprobdelta$ is the probability that two independent edges cross in the given layout and $q$ is the size of the set of independent pairs of edges of the graph, denoted as $Q$, namely, the number of pairs of edges that do not share vertices \cite{Alemany2018a}. $q$ can be computed in constant time given $n$, the number of vertices of the networks, $m$, its number of edges, $\mmtdeg{2}$, the second moment of degree about zero (i.e. the mean of squared vertex degrees) because
\begin{equation}
\label{eq:introduction:overview:size_Q}
q = |Q| = \frac{1}{2} \left( m(m+1)- n \mmtdeg{2} \right).
\end{equation}

\begin{figure}
	\centering
	\includegraphics[scale=1.4]{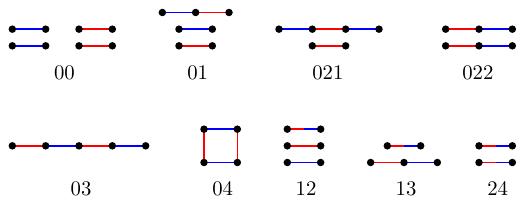}
	\caption{The subgraphs of each type $\omega\in\Omega$ (\cref{eq:introduction:overview:Omega}); below each graph is indicated the code that identifies it. Each type is an element of $Q\times Q$. Equally-colored edges belong to the same element of $Q$, and bi-colored edges (as in types $12$, $13$ and $24$) denote the same edge of the two elements of $Q$ of the type.}
	\label{fig:introduction:overview:types_of_subgraphs}
\end{figure}

On the negative side, the calculation of $\gvar{C}$ is more complex \cite{Alemany2018a}. Fortunately, its computation is a particular instance of a subgraph counting problem\footnote{ See \cite[Chapter 4]{Estrada2011a} for details on subgraph counting.} \cite{Alemany2018a}. The subgraphs to count in order to calculate $\gvar{C}$ are shown in \cref{fig:introduction:overview:types_of_subgraphs}; a similar characterization can be found in \cite[Figure 2]{Arizmendi2019a} for trees whose vertices are in convex position\footnote{The vertices of a graph are in convex position when they are arranged in the boundary of a convex, closed, simple curve, assumed to be, without loss of generality, a circle.}. Then, $\gvar{C}$ in an arbitrary graph $G$ becomes \cite{Alemany2018a}
\begin{align}
\label{eq:introduction:overview:var_C:general:freq__x__exp}
\gvar{C}
	&= \sum_{\omega \in \Omega} a_\omega n_G(F_\omega) \gexpetw \\
	&= \sum_{\omega \in \Omega} f_\omega \gexpetw. \nonumber
\end{align}
$\gexpetw$ is a layout-dependent term associated to type $\omega$, defined later in \cref{sec:background} (\cref{eq:background:expected_gamma_omega}). $\Omega$ is a set of subgraphs, where each is indexed by a code of two or three digits,
\begin{equation}
\label{eq:introduction:overview:Omega}
\Omega = \{00, 01, 021, 022, 03, 04, 12, 13, 24\}.
\end{equation}
These codes were devised in \cite{Alemany2018a}. In \cref{eq:introduction:overview:Omega}, $a_\omega$ is an integer constant for each $\omega\in\Omega$, and $n_G(F_\omega)$ is the number of times the subgraph $F_\omega$ appears in the graph. Each $F_\omega$ is depicted in \cref{fig:introduction:overview:types_of_subgraphs}; each of these graphs can be a connected graph or disjoint union of connected graphs. The values of $a_\omega$, $F_\omega$ and $\lexpetw$ are summarized in \cref{table:introduction:overview:types_as_subgraph_counting}. In sum, $\gvar{C}$ is expressed as a function of the number of subgraphs of each kind that appear in \cref{table:introduction:overview:types_as_subgraph_counting}. Notice that $n_G(F_\omega)$ is the only graph-dependent term, and $\gexpetw$ is the only layout-dependent term. The constant $a_\omega$ depends only on the type. The meaning of the codes listed in $\Omega$ (\cref{eq:introduction:overview:Omega}), which identify the graphs in \cref{fig:introduction:overview:types_of_subgraphs}, is explained in \cref{sec:background}.

\begin{table*}
	\centering
	\caption{The values of $a_\omega$, $F_\omega$ and $\lexpetw$ as function of $\omega$. $\lintree$ and $\cycle$ denote, respectively, linear trees (or path graphs) and cycle graphs of $n$ vertices, and the operator $\oplus$ indicates the disjoint union of graphs.}
	\label{table:introduction:overview:types_as_subgraph_counting}
	\begin{tabular}{@{}lllll}
		\hline
		$\omega\in\Omega$ & $a_\omega$ & $F_\omega$ & $\lexpetw$ \\
		\hline
		$00$	&	$6$ & $\lintree[2]\oplus\lintree[2]\oplus\lintree[2]\oplus\lintree[2]$ & 0 \\
		$24$	&	$1$ & $\lintree[2]\oplus\lintree[2]$ & 2/9\\
		$13$	&	$2$ & $\lintree[3]\oplus\lintree[2]$ & 1/18 \\
		$12$	&	$6$ & $\lintree[2]\oplus\lintree[2]\oplus\lintree[2]$ & 1/45 \\
		$04$	&	$2$ & $\cycle[4]$ & -1/9 \\
		$03$	&	$2$ & $\lintree[5]$ & -1/36 \\
		$021$	&	$2$ & $\lintree[4]\oplus\lintree[2]$ & -1/90\\
		$022$	&	$4$ & $\lintree[3]\oplus\lintree[3]$ & 1/180 \\
		$01$	&	$4$ & $\lintree[3]\oplus\lintree[2]\oplus\lintree[2]$ & 0 \\
		\hline
	\end{tabular}
\end{table*}

\cref{eq:introduction:overview:var_C:general:freq__x__exp} can be seen as a particular case of the general equation to define a molecular property $P$ of a graph $G$ as a summation over all subgraphs $F$, i.e. \cite{Smolenskii1964a,Klein1986a}
\begin{equation*}
P(G) = \sum_{F} \eta_F n_G(F)
\end{equation*}
where $\eta_F$ is the contribution of graph $F$ to the molecular property. In our case, we have that $\eta_F = a_\omega \gexpetw$ if $F$ is in \cref{fig:introduction:overview:types_of_subgraphs} and $\eta_F = 0$ otherwise (in our case the summation is not restricted to connected subgraphs).

In this article, we aim to develop fast algorithms to calculate the exact value of $\gvar{C}$ in arbitrary graphs as well as {\em ad hoc} algorithms for forests. Forests are a straightforward generalization of trees which are the kind of graphs that are typically found in syntactic dependency structures. Moreover, notice that RNA secondary structures are graphs such that the maximum vertex degree is 1, thus a particular case of forests\footnote{It is worth mentioning that a formula of $\gvar{C}$ for 2-regular graphs was derived in \cite{Alemany2018a}.}. In addition, the syntactic structures in recent experiments with deep agents are forests \cite{Chaabouni2019a}. By providing algorithms for forests and, also, more general graphs, we are accommodating all the possible exceptions and variants that have been discussed to define the syntactic structure of sentences, e.g., allowing for cycles (see for instance, \cite[{\em Section 4.9 Graph-theoretic properties}]{deMarneffe2008a}).

% (hide) The present article is a piece of a broader research program on the statistical properties of network measures on linear arrangements. Recently, the distribution of $D$, the sum of edge distances in random linear arrangements, has been investigated \cite{Ferrer2018a}, and so has been the distribution of $C$ \cite{Alemany2018a}. While the calculation of $\lexpe{D}$ and $\lvar{D}$ is straightforward, the efficient calculation of $\lvar{C}$, and that of $\gvar{C}$, requires further investigation.

\paragraph{Organization of the article} The remainder of the article has two parts. The first part develops gradually a formula for the variance by involving in the calculation the amount of certain subgraphs. \cref{sec:background} provides formal definitions, reviews the theoretical arguments in \cite{Alemany2018a} which allows one to express $\gvar{C}$ as in \cref{eq:introduction:overview:var_C:general:freq__x__exp}, and outlines a naive $\bigO{m^4}$-time algorithm to calculate the exact value of $\gvar{C}$ that is based on subgraph counting. \cref{sec:arithmetic_expressions} builds on \cite{Alemany2018a}: we first provide expressions to identify other, smaller subgraphs (\cref{sec:arithmetic_expressions:general_results}) in the expansion of the expressions for all $f_\omega$ (\cref{sec:arithmetic_expressions:frequencies}). The expressions in \cref{sec:arithmetic_expressions:frequencies} are obtained, essentially, by expanding the initial formal expressions in \cite{Alemany2018a} and rearranging and regrouping the results. In \cref{sec:arithmetic_expressions:general_results} we also give expressions to easily count these new subgraphs in the algorithms devised in subsequent sections. We finish with \cref{sec:arithmetic_expressions:var_C} where we obtain a general arithmetic expression for $\gvar{C}$ via \cref{eq:introduction:overview:var_C:general:freq__x__exp} applying the expressions for all $f_\omega$ obtained in \cref{sec:arithmetic_expressions:frequencies}.

The second part is devoted to solving the counting problem algorithmically and is novel. In \cref{sec:algorithms}, we use the expressions in \cref{sec:arithmetic_expressions:var_C} to devise faster algorithms (\cref{table:background:summary_algorithms}). We present two algorithms that calculate $\gvar{C}$ for arbitrary graphs. \cref{algo:algorithms:graphs:no_reuse} runs in time $\smallo{nm^2}$ and \cref{algo:algorithms:graphs:reuse} runs in time $\bigO{nm}$. We will show that the former is faster for sparse graphs and the latter is faster for dense graphs. Their time complexities are also given as a function of important graph structural parameters, such as the maximum vertex degree $\maxdeg$; the second moment of degree about zero, which is an instantiation of the more general $p$-th moment of degree about zero
\begin{equation}
\label{eq:introduction:mmt_deg}
\mmtdeg{p} = \frac{1}{n} \sum_{s=1}^n k_s^{p},
\end{equation}
where $k_s$ denotes the degree of vertex $s$; the sum of the product of degrees
\begin{equation}
\label{eq:introduction:psi}
\psi = \sum_{st\in E} k_sk_t;
\end{equation}
the transitivity index \cite[Chapter 4.5.1]{Estrada2011a}
\begin{equation}
\label{eq:introduction:transitivity_index}
T = \frac{3\ntris}{n_G(\lintree[3])}
\end{equation}
where $n_G(F)$ is the number of subgraphs isomorphic to $F$ in $G$; and the degree of connectivity between pairs of vertices $\pair{u}{v}$ via a third vertex, denoted as an indicator function
\begin{equation}
\label{eq:introduction:delta_connectivity}
\delta_{uv} =
\begin{cases}
1, \text{ exists } w \text{ such that } a_{uw}=a_{vw}=1 \\
0, \text{ otherwise.}
\end{cases}
\end{equation}
We also present an algorithm for forests that runs in time $\bigO{n}$ (\cref{algo:algorithms:forests}).

Finally, \cref{sec:discussion} discusses our findings and their implications, and suggests future work.

Readers whose primary interest are the algorithms can jump directly to \cref{sec:algorithms} after reading \cref{sec:background} (optional) and reading \cref{sec:arithmetic_expressions:general_results,sec:arithmetic_expressions:var_C} (mandatory). All the algorithms presented in this article were tested thoroughly (\cref{sec:testing_protocol}) and are publicly available in the Linear Arrangement Library\footnote{Available online at: \url{https://github.com/LAL-project/linear-arrangement-library}} \cite{Alemany2021c}.

% It is important to notice that the first part has been already been published at [1] in larger parts by the authors. The expressions for the f_* values that are needed for the algorithm are (in my understanding) a reformulation of the expressions of [1], often by expanding the old expression and regrouping the results. The level of novelty  in this part is therefore at most medium.

\begin{table*}
	\centering
	\caption{Summary of the time and space complexity algorithms devised for computing $\gvar{C}$ for general graphs and forests.}
    \label{table:background:summary_algorithms}
	\begin{tabular}{lll}
		\toprule
		Algorithm								& Time complexity								& Space complexity \\
		\midrule
		Naive algorithm							& $\bigO{m^4}$									& $\bigO{1}$	\\
												&												& \\
		\cref{algo:algorithms:graphs:no_reuse}	& $\bigO{\psi + m - n\mmtdeg{2}}$				& $\bigO{n}$	\\
		(general graphs)						& $\bigO{\maxdeg n\mmtdeg{2}}$					& \\
												& $\smallo{nm^2}$								& \\
												&												& \\
		\cref{algo:algorithms:graphs:reuse}		& $\bigO{n + \maxdeg (n\mmtdeg{2} - \ntris)}$	& $\bigO{n\mmtdeg{2} - \ntris}$	\\
		(general graphs)						& $\bigO{nm - \sum_{st\notin E} \delta_{st}}$	& $\bigO{Tm + (1 - T)n\mmtdeg{2}}$ \\
												&												& \\
		\cref{algo:algorithms:forests}			& $\bigO{n}$									& $\bigO{n}$ \\
		(forests)								&												& \\
		\bottomrule
    \end{tabular}
\end{table*}

%----------------------------------------------------%
% automatic inline of '2-theoretical-background.tex' %
%----------------------------------------------------%
\section{Theoretical background}
\label{sec:background}

Consider a graph $G=(V,E)$ of $n=|V|$ vertices and $m=|E|$ edges whose vertices are arranged with a given function $\garr$ that indicates the position of every vertex in a given, fixed, layout. The number of crossings in $\garr$, $C(\garr)$, can be defined by making explicit the dependence on a particular arrangement, $\garr$ as
\begin{equation*}
C(\garr) = \sum_{ \{e_1,e_2\}\in Q } \alpha_{\garr}(e_1,e_2),
\end{equation*}
where, recall, $Q$ denotes the set of pairs of independent edges (i.e. the set of pairs of edges that may potentially cross \cite{Alemany2018a}, \cref{eq:introduction:overview:size_Q}), and where $\alpha_{\garr}(e_1,e_2)$ is an indicator function equal to $1$ if, and only if the edges $e_1$ and $e_2$ cross in the given arrangement $\garr$ embedded in the layout under consideration. Throughout this article we use $q=|Q|$. We omit $\garr$ from the notation when we denote a random variable. For example, hereafter we use $\alpha(e_1,e_2)$ to refer to $\alpha_{\garr}(e_1,e_2)$ for simplicity.

Next we review the steps devised in \cite{Alemany2018a} towards \cref{eq:introduction:overview:var_C:general:freq__x__exp}. By definition of a random variable's variance,
\begin{equation*}
\gvar{C} = \gexpe{(C - \gexpe{C})^2}.
\end{equation*}
Notice that
\begin{align*}
C - \gexpe{C}
	&= \sum_{\{e_1,e_2\} \in Q} \alpha(e_1,e_2) - \gexpe{\sum_{\{e_1,e_2\} \in Q} \alpha(e_1,e_2)} \\
	&= \sum_{\{e_1,e_2\} \in Q} (\alpha(e_1,e_2) - \gexpe{\alpha(e_1,e_2)}).
\end{align*}
Let $\gprobdelta = \gexpe{\alpha(e_1,e_2)}$, which is the probability that any two independent edges cross in a given layout since $\alpha$ is an indicator random variable, and, further, let $\beta(e_1,e_2) = \alpha(e_1,e_2) - \gprobdelta$ and rewrite
\begin{equation*}
C - \gexpe{C} = \sum_{\{e_1,e_2\} \in Q} \beta(e_1,e_2).
\end{equation*}
Now we can express $\gvar{C}$ as
\begin{equation*}
\gvar{C} = \gexpe{\left(\sum_{\{e_1,e_2\} \in Q} \beta(e_1,e_2)\right)^2}.
\end{equation*}
Expanding the square in the previous expression, $\gvar{C}$ can be decomposed into a sum of $|Q \times Q| = |Q|^2$ summands of the form $\gexpe{\beta(e_1,e_2)\beta(e_3,e_4)}$, i.e.
\begin{align}
\gvar{C}
	&= \gexpe{\sum_{\{e_1,e_2\} \in Q} \sum_{\{e_3,e_4\} \in Q} \beta(e_1,e_2)\beta(e_3,e_4)} \nonumber\\
	&= \sum_{\{e_1,e_2\} \in Q} \sum_{\{e_3,e_4\} \in Q} \gexpe{\beta(e_1,e_2)\beta(e_3,e_4)} \label{eq:background:origins_of_classification}.
\end{align}
The two elements of $Q$ involved in the product $\beta(e_1,e_2)\beta(e_3,e_4)$ can be classified into a type, which we denoted as $\omega$, based on the relationship between the edges $e_1=st$, $e_2=uv$, $e_3=wx$ and $e_4=yz$. This classification system eventually leads to $\Omega$ (\cref{eq:introduction:overview:Omega}). This relationship was defined in \cite{Alemany2018a} with two parameters:
\begin{itemize}
\item $\tau$, the number of common edges among $st$, $uv$, $wx$ and $yz$, and
\item $\phi$, the number of common vertices among $s$, $t$, $u$, $v$, $w$, $x$, $y$, and $z$.
\end{itemize}
For example, consider the two elements of $Q$, $\{e_1,e_2\}=\{\{1,2\}, \{3,4\}\}$ and $\{e_3,e_4\}=\{\{1,3\}, \{4,5\}\}$. In this example, there are no shared edges, $\tau=0$; but there are three common vertices, $\phi=3$. Therefore, these two pairs are of type $\omega=03$. It is depicted as a path graph of five vertices in \cref{fig:introduction:overview:types_of_subgraphs}.

It is easy to see that there are only 9 types \cite{Alemany2018a} of pairs of elements of $Q$. These are shown in \cref{fig:introduction:overview:types_of_subgraphs}, where most are labeled using codes of two digits (the parameters $\tau$ and $\phi$) and only two types using three digits (the parameters $\tau$ and $\phi$ plus a disambiguation digit). The only two types that require a third digit are $021$ and $022$ since there are two possible graphs that can be made with $\{e_1,e_2\},\{e_3,e_4\}\in Q$ such that $\tau=0$ and $\phi=2$. We denote the set of all the types as $\Omega$ (\cref{eq:introduction:overview:Omega}). The features of each type of product are summarized in \cref{table:background:types:parameters}, which also includes the total number of distinct vertices in each type, denoted as $\upsilon$.

Notice that, given any two $\{e_1,e_2\}, \{e_3,e_4\}\in Q$, it is easy to see that
\begin{equation*}
\gexpe{ \beta(e_1,e_2)\beta(e_3,e_4) } = \gexpe{ \alpha(e_1,e_2)\alpha(e_3,e_4) } - \gprobdelta^2.
\end{equation*}
For the sake of brevity, let $\gamma_\omega = \beta(e_1,e_2)\beta(e_3,e_4)$ and $\gprobalphastw = \gexpe{\alpha(e_1,e_2)\alpha(e_3,e_4)}$ when the pairs of independent edges $\{e_1,e_2\},\{e_3,e_4\}$ can be classified into $\omega$, i.e., when their type is one of the types in $\Omega$. Therefore,
\begin{equation}
\label{eq:background:expected_gamma_omega}
\gexpetw = \gexpe{ \beta(e_1,e_2)\beta(e_3,e_4) } = \gprobalphastw - \gprobdelta^2.
\end{equation}
For two fixed $\{e_1,e_2\}, \{e_3,e_4\}\in Q$ the value $\gprobalphastw$ denotes the probability that $\alpha(e_1,e_2)\alpha(e_3,e_4)=1$ in a uniformly random embedding (in the layout) of all the vertices of the edges $e_1$, $e_2$, $e_3$, $e_4$. Now, as a result of the classification described above, the expected values $\gexpetw$ can be grouped into the different types in $\Omega$ and thus $\gvar{C}$ can be expressed compactly as
\begin{equation}
\label{eq:background:var_C:freq_times_exp}
\gvar{C} = \sum_{\omega\in\Omega} f_\omega \gexpetw,
\end{equation}
where, for a fixed $\omega$, $f_\omega$ is the number of occurrences of $\gexpetw$ in the summation, defined as
\begin{equation}
\label{eq:background:frequency_template}
f_\omega =
	\sum_{\{e_1,e_2\} \in Q}
	\sum_{\substack{\{e_3,e_4\} \in Q\; : \\ \text{type of }\{e_1,e_2\}, \{e_3,e_4\} \text{ is } \omega}}
	1.
\end{equation}
\cref{eq:background:frequency_template} is the formal definition of $f_\omega$.

Notice, therefore, that computing $\gvar{C}$ can be done via countings of subgraphs. \cref{fig:introduction:overview:types_of_subgraphs} relates each type of product with the subgraph that has to be counted within the graph so as to obtain the exact value of $\gvar{C}$. These graphs may be elementary graphs, a simple linear tree of a fixed number of vertices, e.g., $\lintree[4]$ or $\lintree[5]$, a cycle graph, e.g., $\cycle[4]$, or a combination with the operator $\oplus$, the disjoint union of graphs, all listed in \cref{table:introduction:overview:types_as_subgraph_counting}.

In random linear arrangements (`rla'), $\lprobdelta = 1/3$ \cite{Alemany2018a}, and in uniformly random spherical arrangements (`rsa'), $\sprobdelta = 1/8$ \cite{Moon1965a}. In the latter layout, the vertices of the graph are placed on the surface of a sphere uniformly at random, and each edge of the graph becomes the geodesic between each of its corresponding vertices. It was shown that \cite{Alemany2018a}
\begin{align*}
\gprobalphast{00} &= \gprobalphast{01} = \gprobdelta^2, \\
\gprobalphast{24} &= \gprobdelta, \\
\lprobalphast{04} &= \sprobalphast{04} = 0.
\end{align*}
From these identities we can deduce that $\gexpet{00}=\gexpet{01}=0$.

\cref{table:background:types:parameters} gives all the values of $\lprobalphastw$ and $\lexpetw$. These probabilities were calculated by brute force enumeration of all possible linear arrangements of the vertices of each type of graph.

\begin{table*}
	\centering
	\caption{The classification of the elements of $Q \times Q$ into types of products abstracting away from the order of the elements of the pair. Each $\omega\in\Omega$ is the code that identifies the product type. Source \cite[Table 2, p. 18]{Alemany2018a}.}
	\label{table:background:types:parameters}
	\begin{tabular}{@{}llllllll}
		\hline
		$\omega\in\Omega$ & $(\{e_1,e_2\},\{e_3,e_4\})$ & $|\upsilon|$ & $\tau$ & $\phi$ & $\lprobalphastw$ & $\lexpetw$ \\
		\hline
		00 & $(\{st,uv\},\{wx,yz\})$ & 8 & 0 & 0 & 1/9  & 0 \\
		24 & $(\{st,uv\},\{st,uv\})$ & 4 & 2 & 4 & 1/3  & 2/9 \\
		13 & $(\{st,uv\},\{st,uw\})$ & 5 & 1 & 3 & 1/6  & 1/18 \\
		12 & $(\{st,uv\},\{st,wx\})$ & 6 & 1 & 2 & 2/15 & 1/45 \\
		04 & $(\{st,uv\},\{su,tv\})$ & 4 & 0 & 4 & 0    & -1/9 \\
		03 & $(\{st,uv\},\{su,vw\})$ & 5 & 0 & 3 & 1/12 & -1/36 \\
		021 & $(\{st,uv\},\{su,wx\})$ & 6 & 0 & 2 & 1/10 & -1/90 \\
		022 & $(\{st,uv\},\{sw,ux\})$ & 6 & 0 & 2 & 7/60 & 1/180 \\
		01 & $(\{st,uv\},\{sw,xy\})$ & 7 & 0 & 1 & 1/9  & 0 \\
		\hline
	\end{tabular}
\end{table*}

Interestingly, $f_\omega$ is proportional to the amount of subgraphs of type $\omega$, i.e.
\begin{equation}
\label{eq:background:type_prod_subgraph}
f_\omega = a_\omega n_G(F_\omega),
\end{equation}
where $a_\omega$ is an integer constant, and $n_G(F_\omega)$ is the number of occurrences of the subgraph $F_\omega$ in the graph under consideration (\cref{table:introduction:overview:types_as_subgraph_counting}).

\cref{table:introduction:overview:types_as_subgraph_counting} allows one to obtain $f_\omega$ via \cref{eq:background:type_prod_subgraph}. In \cite{Alemany2018a}, these expressions were used to obtain arithmetic expressions for $f_\omega$ in certain types of graphs, which were in turn used to obtain an arithmetic expression for $\lvar{C}$ in those classes of graphs, e.g., complete graphs, complete bipartite graphs, cycle graphs, etc. Such arithmetic expressions depend only on the number of vertices of the graph. In arbitrary graphs, we can still use \cref{table:introduction:overview:types_as_subgraph_counting} to derive a $\bigO{m^4}$-time algorithm to calculate the exact value of $\gvar{C}$. The naive algorithm is based on brute force enumeration of all elements of $Q\times Q$ to count each $F_\omega$. To do this, one can simply list all elements in $Q\times Q$ and identify their respective types $\omega$ and count all the instances per type, namely, calculate $f_\omega$. By doing this for all $\omega\in\Omega$ we can obtain all needed countings. This is formalized in \cref{algo:background:brute_force}. Here we develop algorithms with lower asymptotic complexity for general graphs and forests. \cref{table:background:summary_algorithms} summarizes the cost of the first approximations discussed here and that of the algorithms that are presented in subsequent sections.

\begin{algorithm}
	\caption{Calculate $\gvar{C}$ by brute force in an arbitrary graph.}
	\label{algo:background:brute_force}
	\DontPrintSemicolon
	
	\KwIn{$G=(V,E)$ a graph.}
	\KwOut{$\gvar{C}$, the variance of the number of crossings.}
	
	\SetKwProg{Fn}{Function}{ is}{end}
	\Fn{\textsc{VarianceC}$(G)$} {
		
		$f_\omega \gets 0$ for all $\omega\in \Omega$ \;
		\For {$\{\{e_1, e_2\}, \{e_3, e_4\}\} \in Q\times Q$} {
			$\omega \gets $ type of $\{\{e_1, e_2\}, \{e_3, e_4\}\}$ according to \cref{table:introduction:overview:types_as_subgraph_counting,table:background:types:parameters} \;
			$f_\omega \gets f_\omega + 1$ \;
		}
		
		$V\gets 0$\;
		\For {$\omega\in\Omega$} {
			$V \gets V + f_\omega\gexpetw$ \;
		}
		\Return $V$
	}
\end{algorithm}
%---------------------------------------------------%
% automatic inline of '3-0-expression-variance.tex' %
%---------------------------------------------------%
\section{Arithmetic expressions for $\gvar{C}$}
\label{sec:arithmetic_expressions}

Here we further develop the algebraic expressions of the $f_\omega$'s for $\omega \in \Omega$ (\cref{eq:introduction:overview:Omega}) but we focus on the $\omega$'s with non-null contribution to $\gvar{C}$ (\cref{table:background:types:parameters}). We use the formalization of the $f_\omega$'s in \cite{Alemany2018a} as a starting point to obtain an arithmetic expression for $\gvar{C}$ in general graphs. These formalizations were originally used in \cite{Alemany2018a} to obtain the expressions of the form of \cref{eq:background:type_prod_subgraph}. 

The purpose of these algebraic expressions is to bridge the gap between the algorithms to calculate $\gvar{C}$ in \cref{sec:algorithms} (\cref{algo:algorithms:graphs:no_reuse,algo:algorithms:forests}) and the countings over the set of subgraphs described in \cref{sec:background} (\cref{eq:background:var_C:freq_times_exp,eq:background:frequency_template}).

In \cref{sec:arithmetic_expressions}, we first describe the notation that we use in later sections to obtain compact expressions. In \cref{sec:arithmetic_expressions:general_results}, we present general results that relate summations over the set $Q$ with subgraph countings. These are used in \cref{sec:arithmetic_expressions:frequencies}, to obtain arithmetic expressions for $f_\omega$. Finally, we derive an expression for $\gvar{C}$ in \cref{sec:arithmetic_expressions:var_C} using the $f_\omega$.

%-------------------------------------------%
% automatic inline of '3-1-definitions.tex' %
%-------------------------------------------%
\subsection{Preliminaries}
\label{sec:arithmetic_expressions:preliminaries}

Throughout this article, we use letters $s,t,...,z$ to indicate distinct vertices. We define $A^{p} = \{ a_{st}^{(p)} \}$ as the $p$-th power of the adjacency matrix $A = \{a_{st}\}$ of the graph, where $a_{st}=1$ when $st\in E$, and $a_{st}=0$ if $st\notin E$. Further, we define $m_p$ as the sum of one of half of the values of $A^p$ excluding the diagonal, i.e.
\begin{equation*}
m_p = \sum_{s < t} a_{st}^{(p)},
\end{equation*}
thus $m_1 = m$. We denote the sum of the degrees of the neighbors of a vertex $s$ as
\begin{equation}
\label{eq:arithmetic_expressions:preliminaries:sum_degrees_neighbors}
\xi(s) = \sum_{t\in V : t\in\Gamma(s)} k_t = \sum_{st\in E} k_t,
\end{equation}
The neighborhood intersection of two vertices $s$ and $t$ is denoted as
\begin{equation}
\label{eq:arithmetic_expressions:preliminaries:common_neighs_vers}
c(s,t) = \Gamma(s) \cap \Gamma(t),
\end{equation}
and the sum of the degrees of the vertices in $c(s,t)$ as
\begin{equation}
\label{eq:arithmetic_expressions:preliminaries:sum_degs_common}
S_{s,t} = \sum_{u\in c(s,t)} k_u.
\end{equation}
Notice that if $s=t$ then $S_{s,t}=\xi(s)=\xi(t)$. We also use $\psi$ (\cref{eq:introduction:psi}) which is the sum of the product of degrees\footnote{Notice that this product is involved in the calculation of degree correlations \cite{Serrano2006a}.} at both ends of an edge. The values $\xi(s)$, $c(s,t)$, $S_{s,t}$ and $\psi$ prove useful in making the expressions of the $f_\omega$'s compact as well as in deriving the algorithms to compute the exact value of $\gvar{C}$.

For any undirected simple graph $G=(V,E)$, let $G_{-s}$ be the induced graph resulting from removing vertex $s$ from $G$. More generally, we define $G_{-L}$ as the induced graph resulting from the removal of the vertices in $L \subseteq V$. We use $Q=Q(G)$ to refer to the set of pairs of independent edges of $G$.

\begin{figure}
	\centering
	\begin{subfigure}[t]{0.31\textwidth}
		\centering
		\includegraphics[scale=0.65]{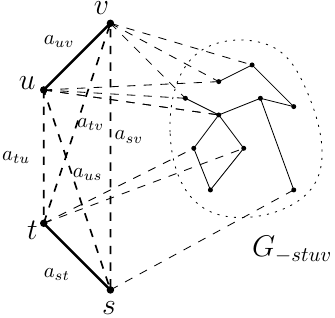}
		\caption{The result of removing vertices $s,t,u,v$ such that $\{st,uv\} \in Q$ from a graph $G$ to produce the graph $G_{-stuv}$. Source \cite[Figure 7, p. 24]{Alemany2018a}.}
		\label{fig:arithmetic_expressions:preliminaries:graph_without_stuv}
	\end{subfigure}
	\;
	\begin{subfigure}[t]{0.31\textwidth}
		\centering
		\includegraphics[scale=0.65]{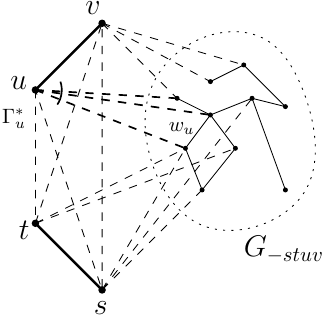}
		\caption{Illustration of the first inner summation in \cref{eq:arithmetic_expressions:frequencies:13:general}. In the figure, $\{st,uv\}\in Q$, and $w_u\in\Gamma_u^* = \Gamma(u,-stuv)$. Source \cite[Figure 8, p. 27]{Alemany2018a}.}
		\label{fig:arithmetic_expressions:preliminaries:general_13}
	\end{subfigure}
	\;
	\begin{subfigure}[t]{0.31\textwidth}
		\centering
		\includegraphics[scale=0.65]{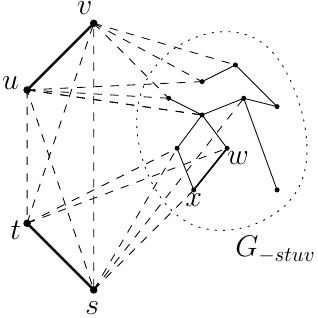}
		\caption{Illustration of the summations in \cref{eq:arithmetic_expressions:frequencies:12:general}, the first of which represents the elements in the Cartesian product of edge $st$ and the edges in $G_{-stuv}$. Source \cite[Figure 9, p. 29]{Alemany2018a}.}
		\label{fig:arithmetic_expressions:preliminaries:general_12}
	\end{subfigure}
	
	\begin{subfigure}[t]{0.31\textwidth}
		\centering
		\includegraphics[scale=0.65]{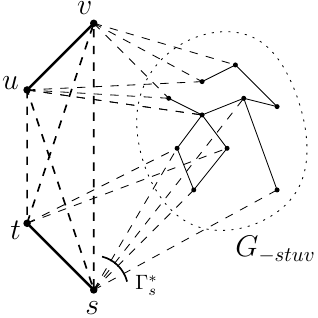}
		\caption{Illustration of $\varphi_{tus}$ (\cref{eq:arithmetic_expressions:frequencies:03:general:varphis}). It is exactly $|\Gamma_s^*| = |\Gamma(s,-stuv)|$ (the amount of neighbors of $s$ in $G_{-stuv}$), provided that the edge $tu$ exists (i.e. $a_{tu} = 1$). Source \cite[Figure 10, p. 30]{Alemany2018a}.}
		\label{fig:arithmetic_expressions:preliminaries:general_03}
	\end{subfigure}
	\;
	\begin{subfigure}[t]{0.31\textwidth}
		\centering
		\includegraphics[scale=0.65]{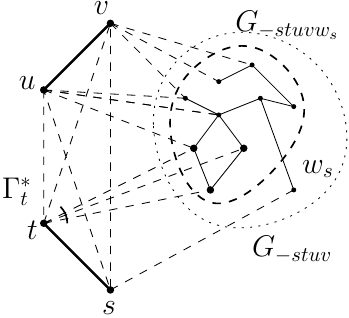}
		\caption{Illustration of $\varphi_{st}$ (\cref{eq:arithmetic_expressions:frequencies:021:varphis}). $w_s$ represents the only neighbor of $s$ different from $t,u,v$. Therefore, in this case, $\varphi_{st}$ is exactly the amount of vertices in $G_{-stuvw_s}$ neighbors of $t$, indicated with $\Gamma_t^* = \Gamma(t, -stuvw_s)$. Source \cite[Figure 12, p. 33]{Alemany2018a}.}
		\label{fig:arithmetic_expressions:preliminaries:general_021_phi}
	\end{subfigure}
	\;
	\begin{subfigure}[t]{0.31\textwidth}
		\centering
		\includegraphics[scale=0.65]{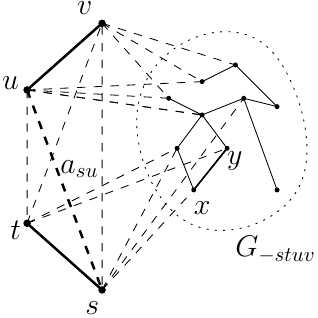}
		\caption{Illustration of $\varepsilon_{su}$ (\cref{eq:arithmetic_expressions:frequencies:021:epsilons}), which requires the existence of an edge between $s$ and $u$, indicated with $a_{su}$, and is equal to the amount of edges in $G_{-stuv}$. Source \cite[Figure 12, p. 33]{Alemany2018a}.}
		\label{fig:arithmetic_expressions:preliminaries:general_021_epsilon}
	\end{subfigure}
	
	\caption{Solid thick lines indicate the edges formed by these vertices in $G$, solid thin lines indicate edges in $G_{-stuv}$ and dashed lines indicate potential edges, namely that may not exist, between vertices $s,t,u,v$ and vertices in $G_{-stuv}$.}
\end{figure}

We denote $G_{-\{s,t,u,v\}}$ simply as $G_{-stuv}$, illustrated in \cref{fig:arithmetic_expressions:preliminaries:graph_without_stuv}. The number of edges in a graph $G_{-stuv}$ is easy to calculate as a function of the number of edges in $G$. If $s,t,u,v$ are four distinct vertices, then
\begin{equation}
\label{eq:arithmetic_expressions:preliminaries:amount_edges_no_stuv}
|E(G_{-stuv})| = m - (k_s + k_t + k_u + k_v) + a_{st} + a_{su} + a_{sv} + a_{tu} + a_{tv} + a_{uv}.
\end{equation}
Unless stated otherwise, network features refer to $G$. Therefore, $m$, $k_u$, $a_{uv}$, ... in \cref{eq:arithmetic_expressions:preliminaries:amount_edges_no_stuv} refer to $G$, and not to $G_{-stuv}$. We use $\Gamma(s,-L) = \Gamma(s)\setminus L$ to denote the set of neighbors of vertices $s \in L \subseteq V$ in $V(G_{-L})$. Its size is:
\begin{equation}
\label{eq:arithmetic_expressions:preliminaries:amount_neighs_in_G_minus}
|\Gamma(s,-L)| = k_s - |\{ w \in L \;|\; ws \in E \}|
= k_s - \sum_{w \in L} a_{sw}.
\end{equation}
We denote $\Gamma(k,-\{s,t,u,v\})$ simply as $\Gamma(k,-stuv)$, with $k\in\{s,t,u,v\}$. We use the term {\em $n$-path} to refer to a sequence of $n$ pairwise distinct vertices $v_0v_1\cdots v_{n-1}$ such that $v_iv_{i+1}\in E$. We consider $v_0v_1\cdots v_{n-1}$ and $v_{n-1}\cdots v_1v_0$ to be two different paths. Lastly, we use $n_G(F)$ to denote the amount of subgraphs isomorphic to $F$ in $G$.

The calculations of the $f_\omega$'s in subsequent sections require a clear notation that states the vertices shared between each pair of elements of $Q$ for an arbitrary graph $G$. Throughout this article we use summations of the form
\begin{equation*}
\sum_{
	\substack{ s,t,u,v \in V \;: \\ \{st,uv\} \in Q(G) }
}
\sum_{
	\substack{ w,x,y,z \in V \;: \\ \{wx,yz\} \in Q(G_{-\{s,t,u,v\}}) }
} \circ,
\end{equation*}
where below each summation operand there is a scope on top of a condition. The ``$\circ$'' represents any term. For the sake of brevity, we contract them as
\begin{equation*}
\sum_{ \{st,uv\} \in Q } \sum_{ \{wx,yz\} \in Q(G_{-stuv}) } \circ.
\end{equation*}
Notice that the scope is omitted in the new notation. This detail is crucial for the countings performed with the help of these compact summations. Likewise, if we want to denote when two elements of $Q$ from each of the summations share one or more vertices, we use:
\begin{equation*}
\sum_{ \{st,uv\} \in Q } \sum_{ \{sx,tz\} \in Q(G_{-uv}) } \circ =
\sum_{
	\substack{ s,t,u,v \in V \;: \\ \{st,uv\} \in Q(G) }
}
\sum_{
	\substack{ x,z \in V \;: \\ \{sx,tz\} \in Q(G_{-\{u,v\}})}
} \circ.
\end{equation*}
This expression denotes the summation over the pairs of elements of $Q$ in which the second one shares two vertices with the first one. Again, the expression to the left is a shorthand for the one to the right. For the sake of brevity, we also use two more compact summations:
\begin{equation*}
\sum_{ \{st,uv\} \in Q } \sum_{ \{st,yz\} \in Q(G_{-uv}) } \circ =
\sum_{
	\substack{ s,t,u,v \in V \;: \\ \{st,uv\} \in Q(G) }
}
\sum_{
	\substack{ y,z \in V \;: \\ \{st, yz\} \in Q(G_{-\{u,v\} }) }
} \circ,
\end{equation*}
\begin{equation*}
\sum_{ \{st,uv\} \in Q } \sum_{ \{sv,yz\} \in Q(G_{-tu}) } \circ =
\sum_{
	\substack{ s,t,u,v \in V \;: \\ \{st,uv\} \in Q(G) }
}
\sum_{
	\substack{ y,z \in V \;: \\ \{sv,yz\} \in Q(G_{-tu}) }
} \circ.
\end{equation*}
%-----------------------------------------------%
% automatic inline of '3-2-general-results.tex' %
%-----------------------------------------------%
\subsection{General results}
\label{sec:arithmetic_expressions:general_results}

Here we introduce some results on summations over $Q$ that we use in \cref{sec:arithmetic_expressions:frequencies} to simplify expressions on the $f_\omega$'s; recall these were defined as summations over pairs of elements of $Q$ (\cref{eq:background:frequency_template}). These expressions pave the way towards an efficient computation of $\gvar{C}$, and hence of $\lvar{C}$. The proof of each proposition can be found in \cref{sec:proofs}.

\begin{figure}
	\centering
	\includegraphics{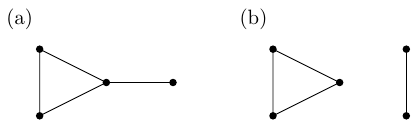}
	\caption{Two simple graphs. (a) the paw graph $\graphpaw$, (b) the $\CoL = \CoLlong = \overline{\compbip[2][3]}$ graph. }
	\label{fig:arithmetic_expressions:general_results:graphs_paw_and_C3_L2}
\end{figure}

The results introduced here serve several purposes. One is to identify new subgraphs, which are $\lintree[4]$, $\lintree[5]$, $\cycle[4]$, the paw graph (denoted as $\graphpaw$, \cref{fig:arithmetic_expressions:general_results:graphs_paw_and_C3_L2}(a)), and $\CoL=\CoLlong$ (\cref{fig:arithmetic_expressions:general_results:graphs_paw_and_C3_L2}(b)), involved in the counting of the types of graphs listed in \cref{table:background:types:parameters} (shown in \cref{fig:introduction:overview:types_of_subgraphs}). This first purpose is achieved by relating summations over elements of $Q$ to counting these new subgraphs. The other purpose is to aid in the design of the algorithms in \cref{sec:algorithms} to compute $\gvar{C}$. These results are presented here to make the article more streamlined. We also present results involving summations of degrees of vertices of elements of $Q$. All of the results below are given for general graphs; some of them have been simplified for trees, since they can be generalized to forests quite easily.

%%%%%%%%%%%%%%%%%%%%%%%%%%%%%%%%%%%%%%%%%%%%%%%%%%%%%%%%%%%%%%%%%%%%%%%%
%
The first three results involve $\lintree[4]$.
\begin{proposition}
\label{prop:arithmetic_expressions:general_results:L_4:elements_Q}
The number of subgraphs isomorphic to $\lintree[4]$, namely half the amount of $4$-paths, in a graph $G$ is
\begin{align}
\npathsfour
   &= \sum_{\{st,uv\} \in Q} (a_{su} + a_{sv} + a_{tu} + a_{tv}) \label{eq:arithmetic_expressions:general_results:L_4:elements_Q:1st} \\
   &= \frac{1}{2} \sum_{s=1}^n \sum_{ \substack{t=1 \\ t \neq s} }^n (a_{st}^{(3)} - a_{st}(2k_t - 1)) \label{eq:arithmetic_expressions:general_results:L_4:elements_Q:2nd} \\
   &= m_3 + m - n\mmtdeg{2} \label{eq:arithmetic_expressions:general_results:L_4:elements_Q:3rd}.
\end{align}
\end{proposition}

\begin{proposition}
\label{prop:arithmetic_expressions:general_results:L_4:traversal}
In any graph $G$,
\begin{equation}
\label{eq:arithmetic_expressions:general_results:L_4:traversal}
\npathsfour =
	m - n\mmtdeg{2} + \psi - \mu,
\end{equation}
where
\begin{equation}
\label{eq:arithmetic_expressions:general_results:L_4:traversal:mu}
\mu = \sum_{st\in E} |c(s,t)|,
\end{equation}
and $\psi$ and $c(s,t)$ are defined in \cref{eq:introduction:psi,eq:arithmetic_expressions:preliminaries:common_neighs_vers}.
\end{proposition}

\begin{proposition}
\label{prop:arithmetic_expressions:general_results:L_4:traversal:trees}
In any tree $\tree$,
\begin{equation}
\label{eq:arithmetic_expressions:general_results:L_4:traversal:trees}
\npathsfour[T] = \sum_{st\in E} (k_s - 1)(k_t - 1).
\end{equation}
\end{proposition}
%
%%%%%%%%%%%%%%%%%%%%%%%%%%%%%%%%%%%%%%%%%%%%%%%%%%%%%%%%%%%%%%%%%%%%%%%%

%%%%%%%%%%%%%%%%%%%%%%%%%%%%%%%%%%%%%%%%%%%%%%%%%%%%%%%%%%%%%%%%%%%%%%%%
%
The next three results involve $\lintree[5]$.
\begin{proposition}
\label{prop:arithmetic_expressions:general_results:L_5:elements_Q}
The number of subgraphs isomorphic to $\lintree[5]$, namely half the amount of $5$-paths, in a graph $G$ is
\begin{equation}
\label{eq:arithmetic_expressions:general_results:L_5:elements_Q}
\npathsfive
= \sum_{\{st,uv\} \in Q}
\left(
	\sum_{w_s \in \Gamma(s, -stuv)} (a_{uw_s} + a_{vw_s})
	+
	\sum_{w_t \in \Gamma(t, -stuv)} (a_{uw_t} + a_{vw_t})
\right).
\end{equation}
\end{proposition}

\begin{proposition}
\label{prop:arithmetic_expressions:general_results:L_5:traversal}
In any graph $G$,
\begin{equation}
\label{eq:arithmetic_expressions:general_results:L_5:traversal}
\npathsfive =
	\frac{1}{2}
	\sum_{st\in E} (g_1(s,t) + g_1(t,s)),
\end{equation}
where
\begin{equation*}
g_1(s,t) = \sum_{u\in \Gamma(s)\setminus\{t\}}
		((k_t - 1 - a_{ut})(k_u - 1 - a_{ut}) + 1 - |c(t,u)|)
\end{equation*}
and $c(s,t)$ is defined in \cref{eq:arithmetic_expressions:preliminaries:common_neighs_vers}.
\end{proposition}

\begin{proposition}
\label{prop:arithmetic_expressions:general_results:L_5:traversal:trees}
In any tree $\tree$,
\begin{equation}
\label{eq:arithmetic_expressions:general_results:L_5:traversal:trees}
\npathsfive[T] = \frac{1}{2} \sum_{st\in E} (g_2(s,t) + g_2(t,s))
\end{equation}
where $g_2(s,t) = (k_t - 1)(\xi(s) - k_t - k_s + 1)$ and $\xi(s)$ is defined in \cref{eq:arithmetic_expressions:preliminaries:sum_degrees_neighbors}.
\end{proposition}
%
%%%%%%%%%%%%%%%%%%%%%%%%%%%%%%%%%%%%%%%%%%%%%%%%%%%%%%%%%%%%%%%%%%%%%%%%

%%%%%%%%%%%%%%%%%%%%%%%%%%%%%%%%%%%%%%%%%%%%%%%%%%%%%%%%%%%%%%%%%%%%%%%%
%

The next two propositions involve the paw graph and $\CoLlong$ (\cref{fig:arithmetic_expressions:general_results:graphs_paw_and_C3_L2}).
\begin{proposition}
\label{prop:arithmetic_expressions:general_results:graphpaw:elements_Q}
The number of subgraphs isomorphic to the paw graph (\cref{fig:arithmetic_expressions:general_results:graphs_paw_and_C3_L2}(a)), denoted as $\graphpaw$, in a graph $G$ is
\begin{equation}
\label{eq:arithmetic_expressions:general_results:graphpaw:elements_Q}
n_G(\graphpaw) = \sum_{\{st,uv\}\in Q} (a_{su} + a_{tv})(a_{sv} + a_{tu}).
\end{equation}
\end{proposition}

\begin{proposition}
\label{prop:arithmetic_expressions:general_results:graphpaw:traversal}
Let $\graphpaw$ be the paw graph, depicted in \cref{fig:arithmetic_expressions:general_results:graphs_paw_and_C3_L2}(a). In any graph $G$,
\begin{equation}
\label{eq:arithmetic_expressions:general_results:graphpaw:traversal}
\ngraphpaw
	= \sum_{st\in E} \sum_{u \in c(s,t)} (k_u - 2),
\end{equation}
where $c(s,t)$ is defined in \cref{eq:arithmetic_expressions:preliminaries:common_neighs_vers}.
\end{proposition}
%
%%%%%%%%%%%%%%%%%%%%%%%%%%%%%%%%%%%%%%%%%%%%%%%%%%%%%%%%%%%%%%%%%%%%%%%%

%%%%%%%%%%%%%%%%%%%%%%%%%%%%%%%%%%%%%%%%%%%%%%%%%%%%%%%%%%%%%%%%%%%%%%%%
%
\begin{proposition}
\label{prop:arithmetic_expressions:general_results:C3_L2:elements_Q}
The number of subgraphs isomorphic to $\CoL=\CoLlong$ (\cref{fig:arithmetic_expressions:general_results:graphs_paw_and_C3_L2}(b)) in a graph $G$ is
\begin{equation}
\label{eq:arithmetic_expressions:general_results:C3_L2:elements_Q}
\nCoL =
	\frac{1}{3}
	\sum_{\{st,uv\}\in Q}
	\left(
		\sum_{w_s \in \Gamma(s, -stuv)} a_{tw_s} +
		\sum_{w_u \in \Gamma(u, -stuv)} a_{vw_u}
	\right).
\end{equation}
\end{proposition}

\begin{proposition}
\label{prop:general_reuslts:C3_L2:traversal}
Let $\CoL=\CoLlong=\overline{\compbip[2][3]}$, depicted in \cref{fig:arithmetic_expressions:general_results:graphs_paw_and_C3_L2}(b). In any graph $G$,
\begin{equation}
\label{eq:general_reuslts:C3_L2:traversal}
\nCoL =
	\frac{1}{3}
	\sum_{st\in E}
	\sum_{u \in c(s,t)} (m - k_s - k_t - k_u + 3),
\end{equation}
where $c(s,t)$ is defined in \cref{eq:arithmetic_expressions:preliminaries:common_neighs_vers}.
\end{proposition}
%
%%%%%%%%%%%%%%%%%%%%%%%%%%%%%%%%%%%%%%%%%%%%%%%%%%%%%%%%%%%%%%%%%%%%%%%%

The next two propositions involve $\cycle[4]$.

%%%%%%%%%%%%%%%%%%%%%%%%%%%%%%%%%%%%%%%%%%%%%%%%%%%%%%%%%%%%%%%%%%%%%%%%
%
\begin{proposition}
\label{prop:arithmetic_expressions:general_results:cycles_4:elements_Q}
The number of cycles of $4$ vertices, namely $\cycle[4]$, in a graph $G$ is
\begin{align}
\nsquares
	&= \frac{1}{2} \sum_{\{st,uv\} \in Q} (a_{sv}a_{tu} + a_{su}a_{tv}) \nonumber\\
	&= \frac{1}{8}\left[tr(A^4)-4n_G(\lintree[3])- 2n_G(\lintree[2])\right] \label{eq:arithmetic_expressions:general_results:num_4cyles} \\
	&= \frac{1}{8}\left[tr(A^4) + 4q - 2m^2 \right]. \label{eq:arithmetic_expressions:general_results:num_4cyles_alon}
\end{align}
\end{proposition}

\begin{proposition}
\label{prop:arithmetic_expressions:general_results:cycles_4:traversal}
In any graph $G$,
\begin{equation}
\label{eq:algorithms:graphs:cycle_4}
\nsquares
	=
	\frac{1}{4}
	\sum_{st\in E}
	\sum_{u\in\Gamma(t)\setminus\{s\}} (|c(s,u)| - 1),
\end{equation}
where $c(s,t)$ is defined in \cref{eq:arithmetic_expressions:preliminaries:common_neighs_vers}.
\end{proposition}
%
%%%%%%%%%%%%%%%%%%%%%%%%%%%%%%%%%%%%%%%%%%%%%%%%%%%%%%%%%%%%%%%%%%%%%%%%

There are other useful results regarding the sum of the degrees of all vertices involved in the elements in $Q$.

%%%%%%%%%%%%%%%%%%%%%%%%%%%%%%%%%%%%%%%%%%%%%%%%%%%%%%%%%%%%%%%%%%%%%%%%
%
\begin{proposition}
\label{prop:arithmetic_expressions:general_results:sum_degrees}
In any graph $G$,
\begin{equation}
\label{eq:arithmetic_expressions:general_results:sum_degrees}
K   = \sum_{\{st,uv\} \in Q} (k_s + k_t + k_u + k_v)
    = n[(m + 1)\mmtdeg{2} - \mmtdeg{3}] - 2\psi
\end{equation}
where $\mmtdeg{p}$ and $\psi$ are defined in \cref{eq:introduction:mmt_deg,eq:introduction:psi}.
\end{proposition}
%
%%%%%%%%%%%%%%%%%%%%%%%%%%%%%%%%%%%%%%%%%%%%%%%%%%%%%%%%%%%%%%%%%%%%%%%%

%%%%%%%%%%%%%%%%%%%%%%%%%%%%%%%%%%%%%%%%%%%%%%%%%%%%%%%%%%%%%%%%%%%%%%%%
%
Furthermore, we identify products of degrees of vertices that appear in the following derivations. These are useful to obtain the formula for $\gvar{C}$ and to compute them in an algorithm.
\begin{proposition}
\label{prop:arithmetic_expressions:general_results:Phi_1}
In any graph $G$,
\begin{equation}
\label{eq:arithmetic_expressions:general_results:Phi_1}
\Phi_1
	= \sum_{\{st,uv\}\in Q} (k_sk_t + k_uk_v)
	= (m + 1)\psi - \sum_{st \in E} k_sk_t(k_s + k_t)
\end{equation}
where $\psi$ is defined in \cref{eq:introduction:psi}.
\end{proposition}

\begin{proposition}
\label{prop:arithmetic_expressions:general_results:Phi_2}
In any graph $G$,
\begin{align}
\label{eq:arithmetic_expressions:general_results:Phi_2}
\Phi_2
	&=	\sum_{\{st,uv\}\in Q} (k_s + k_t)(k_u + k_v)  \nonumber\\
	&=	\frac{1}{2}\sum_{st\in E}
		\left[
		(k_s + k_t)
		\left(
			n\mmtdeg{2}
			- (\xi(s) + \xi(t))
			- k_s(k_s - 1) - k_t(k_t - 1)
		\right)
		\right],
\end{align}
where $\mmtdeg{2}$ and $\xi(s)$ are defined in \cref{eq:introduction:mmt_deg,eq:arithmetic_expressions:preliminaries:sum_degrees_neighbors}.
\end{proposition}
%
%%%%%%%%%%%%%%%%%%%%%%%%%%%%%%%%%%%%%%%%%%%%%%%%%%%%%%%%%%%%%%%%%%%%%%%%

%%%%%%%%%%%%%%%%%%%%%%%%%%%%%%%%%%%%%%%%%%%%%%%%%%%%%%%%%%%%%%%%%%%%%%%%
%
Finally, we define two new values which are found throughout coming derivations. These two are needed in \cref{sec:arithmetic_expressions:frequencies:03,sec:arithmetic_expressions:frequencies:021} to obtain a formula for $\gvar{C}$. First, let
\begin{equation}
\label{eq:arithmetic_expressions:general_results:Lambda_1:def}
\Lambda_1 = \sum_{\{st,uv\} \in Q} \left( a_{su}(k_t + k_v) + a_{sv}(k_t + k_u) + a_{tu}(k_s + k_v) + a_{tv}(k_s + k_u) \right)
\end{equation}
and let
\begin{equation}
\label{eq:arithmetic_expressions:general_results:Lambda_2:def}
\Lambda_2 = \sum_{\{st,uv\}\in Q} (a_{su} + a_{sv} + a_{tu} + a_{tv})(k_s + k_t + k_u + k_v).
\end{equation}
%
%%%%%%%%%%%%%%%%%%%%%%%%%%%%%%%%%%%%%%%%%%%%%%%%%%%%%%%%%%%%%%%%%%%%%%%%

The following results ease the computation of $\Lambda_1$ and $\Lambda_2$.
\begin{proposition}
\label{prop:arithmetic_expressions:general_results:Lambda_1}
In any graph,
\begin{equation}
\label{eq:arithmetic_expressions:general_results:Lambda_1}
\Lambda_1 =
	\sum_{st\in E}
	( (k_t - 1)(\xi(s) - k_t) + (k_s - 1)(\xi(t) - k_s) - 2S_{s,t} ),
\end{equation}
where $\xi(s)$, $c(s,t)$, $S_{s,t}$ and $\Lambda_1$ are defined in \cref{eq:arithmetic_expressions:preliminaries:sum_degrees_neighbors,eq:arithmetic_expressions:preliminaries:common_neighs_vers,eq:arithmetic_expressions:preliminaries:sum_degs_common,eq:arithmetic_expressions:general_results:Lambda_1:def}.
\end{proposition}

\begin{proposition}
\label{prop:arithmetic_expressions:general_results:Lambda_2}
In any graph,
\begin{equation}
\label{eq:arithmetic_expressions:general_results:Lambda_2}
\Lambda_2
	= \Lambda_1 + \sum_{st\in E} (k_s + k_t)( (k_s - 1)(k_t - 1) - |c(s,t)| ),
\end{equation}
where $c(s,t)$, $\Lambda_1$ and $\Lambda_2$ are defined in \cref{eq:arithmetic_expressions:preliminaries:common_neighs_vers,eq:arithmetic_expressions:general_results:Lambda_1:def,eq:arithmetic_expressions:general_results:Lambda_2:def}.
\end{proposition}

\cref{prop:arithmetic_expressions:general_results:Lambda_1,prop:arithmetic_expressions:general_results:Lambda_2} can be further refined for the special case of trees.

\begin{proposition}
\label{prop:arithmetic_expressions:general_results:Lambda_1:trees}
In any tree,
\begin{equation}
\label{eq:arithmetic_expressions:general_results:Lambda_1:trees}
\Lambda_1
	= \sum_{st \in E}
	\left(
		(k_t - 1)(\xi(s) - k_t) +
		(k_s - 1)(\xi(t) - k_s)
	\right),
\end{equation}
where $\xi(s)$ and $\Lambda_1$ are defined in \cref{eq:arithmetic_expressions:preliminaries:sum_degrees_neighbors,eq:arithmetic_expressions:general_results:Lambda_1:def}.
\end{proposition}

\begin{proposition}
\label{prop:arithmetic_expressions:general_results:Lambda_2:trees}
In any tree,
\begin{equation}
\label{eq:arithmetic_expressions:general_results:Lambda_2:trees}
\Lambda_2 = \Lambda_1 + \sum_{ st \in E} (k_s - 1)(k_t - 1)(k_s + k_t),
\end{equation}
where $\Lambda_1$ and $\Lambda_2$ are defined in \cref{eq:arithmetic_expressions:general_results:Lambda_1:def,eq:arithmetic_expressions:general_results:Lambda_2:def}.
\end{proposition}
%----------------------------------------------------%
% automatic inline of '3-3-0-frequency-of-types.tex' %
%----------------------------------------------------%
\subsection{Formulas for the frequencies}
\label{sec:arithmetic_expressions:frequencies}

In the following subsections, we obtain general expressions for the $f_\omega$'s based on the formalization given in \cite{Alemany2018a} and \cref{prop:arithmetic_expressions:general_results:L_4:elements_Q,prop:arithmetic_expressions:general_results:L_5:elements_Q,prop:arithmetic_expressions:general_results:graphpaw:elements_Q,prop:arithmetic_expressions:general_results:C3_L2:elements_Q,prop:arithmetic_expressions:general_results:cycles_4:elements_Q,prop:arithmetic_expressions:general_results:sum_degrees,prop:arithmetic_expressions:general_results:Phi_1,prop:arithmetic_expressions:general_results:Phi_2}. These expressions are designed based on three non-exclusive principles: easing the computation of $\gvar{C}$, linking with standard Graph Theory and linking with the recently emerging subfield of crossing theory for linear arrangements (\cite[Section 2]{Alemany2018a} and also \cite{Arizmendi2019a,Gomez2016a}). In \cite{Alemany2018a}, the expressions for the $f_\omega$'s were linked with Graph Theory via (recall \cref{sec:background})
\begin{equation*}
f_\omega = a_\omega n_G(F_\omega).
\end{equation*}
In the coming subsections, we derive simpler arithmetic expressions for the $f_\omega$'s to help derive arithmetic expression for $\gvar{C}$ (\cref{sec:arithmetic_expressions:var_C}). We focus on the $f_\omega$'s that actually contribute to $\gvar{C}$, namely those $f_\omega$ such that $\omega \neq 00, 01$ because $\gexpet{00} = \gexpet{01} = 0$ \cite{Alemany2018a}. An overview of the expressions that are derived for the $f_\omega$'s is shown in \cref{table:arithmetic_expressions:theoretical_formulas:frequencies:summary}.

\begin{table*}
	\caption{Expressions for the $f_\omega$'s as a function of $q$ and/or other network features. $q=|Q|$ is the number of pairs of independent edges of a graph $G=(V,E)$, $n=|V|$, $m=|E|$, $K$ is defined in \cref{eq:arithmetic_expressions:general_results:sum_degrees}, $\Phi_1$, $\Phi_2$, $\Lambda_1$ and $\Lambda_2$ are defined in \cref{eq:arithmetic_expressions:general_results:Phi_1,eq:arithmetic_expressions:general_results:Phi_2,eq:arithmetic_expressions:general_results:Lambda_1:def,eq:arithmetic_expressions:general_results:Lambda_2:def}. $\graphpaw$ and $\CoL$ denote the graphs depicted in \cref{fig:arithmetic_expressions:general_results:graphs_paw_and_C3_L2}. $\lintree$ and $\cycle$ denote linear trees (path graphs) and cycle graphs of $n$ vertices, respectively.}
	\label{table:arithmetic_expressions:theoretical_formulas:frequencies:summary}
	\begin{tabular}{@{}llll}
		\hline
		$f_\omega$ 																													& Equation \\
		\hline
%		$f_{00} = \sum_{\{st,uv\} \in Q} |Q(G_{-stuv})|$																			& \cite{Alemany2018a} \\
		$f_{24}  				= q$																								& \ref{eq:arithmetic_expressions:frequencies:24} \\
		$f_{13}  				= K -4q - 2\npathsfour$																				& \ref{eq:arithmetic_expressions:frequencies:13:final} \\
		$f_{12}  				= 2[(m + 2)q + \npathsfour - K]$																	& \ref{eq:arithmetic_expressions:frequencies:12:final} \\
		$f_{04}  				= 2\nsquares$																						& \\
		$\phantom{f_{04}}		= \frac{1}{4}tr(A^4) - n_G(\lintree[3])- \frac{1}{2}n_G(\lintree[2])$								& \\
		$\phantom{f_{04}}		= \frac{1}{4}tr(A^4) + q - \frac{1}{2}m^2$															& \ref{eq:arithmetic_expressions:frequencies:04:final} \\
		$f_{03}  				= \Lambda_1 - 2\npathsfour - 8\nsquares - 2\ngraphpaw$												& \ref{eq:arithmetic_expressions:frequencies:03:general:final} \\
		$f_{021} 				= 2q + (m + 5)\npathsfour + 8\nsquares + 3\ngraphpaw + \Phi_1 - 3\nCoL - \Lambda_1 - \Lambda_2 - K$	& \ref{eq:arithmetic_expressions:frequencies:021:final} \\
		$f_{022} 				= 4q + 5\npathsfour + 2\ngraphpaw + 4\nsquares + \Phi_2 - \Lambda_2 - 2K - \npathsfive$				& \ref{eq:arithmetic_expressions:frequencies:022:final} \\
%		$01$	& $\sum_{\{st,uv\} \in Q} \sum_{k \in \{s,t,u,v\}} \sum_{w \in \Gamma(k, -stuv)} |E(G_{-stuvw})|$					& \cite{Alemany2018a}\\
		\hline
	\end{tabular}
\end{table*}

Our approach consists of further developing the algebraic formalizations of the $f_\omega$ given in \cite{Alemany2018a}, which are briefly summarized below in this article to make it more self-contained. In Appendix \ref{sec:testing_protocol}, we explain the tests employed to ensure the correctness of said expressions. When reading the sections to come, we suggest the reader to recall the definition of $\tau$ and $\phi$ in \cref{sec:background}.

%-----------------------------------------%
% automatic inline of '3-3-1-freq-24.tex' %
%-----------------------------------------%
\subsubsection{$\tau=2$, $\phi=4$}
\label{sec:arithmetic_expressions:frequencies:24}

It was already shown in \cite{Alemany2018a} that
\begin{equation}
\label{eq:arithmetic_expressions:frequencies:24}
f_{24} = q.
\end{equation}
%-----------------------------------------%
% automatic inline of '3-3-2-freq-13.tex' %
%-----------------------------------------%
\subsubsection{$\tau=1$, $\phi=3$}
\label{sec:arithmetic_expressions:frequencies:13}

This type denotes the pairs of edges sharing exactly one edge ($\tau=1$) and that have 3 vertices in common ($\phi = 3$). It was shown in \cite{Alemany2018a} that
\begin{align}
f_{13}
	&=
	\sum_{\{st,uv\} \in Q}
	\left(
		\sum_{\{st,uw\} \in Q(G_{-v})} 1 +
		\sum_{\{st,vw\} \in Q(G_{-u})} 1 +
		\sum_{\{uv,sw\} \in Q(G_{-t})} 1 +
		\sum_{\{uv,tw\} \in Q(G_{-s})} 1
	\right) \label{eq:arithmetic_expressions:frequencies:13:general} \\
	&=
	\sum_{\{st,uv\} \in Q} |\Gamma(u, -stuv)| +
	\sum_{\{st,uv\} \in Q} |\Gamma(v, -stuv)| \nonumber \\
	&+	\sum_{\{st,uv\} \in Q} |\Gamma(s, -stuv)| +
	\sum_{\{st,uv\} \in Q} |\Gamma(t, -stuv)|. \label{eq:arithmetic_expressions:frequencies:13:intermediate}
\end{align}
See \cref{fig:arithmetic_expressions:preliminaries:general_13} for an illustration of the first inner summation in \cref{eq:arithmetic_expressions:frequencies:13:general}. Now we simplify \cref{eq:arithmetic_expressions:frequencies:13:intermediate}. Since $\{st, uv\} \in Q$ implies $a_{st} = a_{uv} = 1$ we obtain
\begin{equation*}
f_{13} = \sum_{\{st,uv\} \in Q} (k_s + k_t + k_u + k_v)
- 4q
- 2 \sum_{\{st,uv\} \in Q} (a_{Sun} + a_{sv} + a_{tu} + a_{tv}).
\end{equation*}
Finally, \cref{prop:arithmetic_expressions:general_results:L_4:elements_Q,prop:arithmetic_expressions:general_results:sum_degrees} allow one to rewrite $f_{13}$ as
\begin{equation}
\label{eq:arithmetic_expressions:frequencies:13:final}
f_{13} = K -4q - 2\npathsfour.
\end{equation}
%-----------------------------------------%
% automatic inline of '3-3-3-freq-12.tex' %
%-----------------------------------------%
\subsubsection{$\tau=1$, $\phi=2$}
\label{sec:arithmetic_expressions:frequencies:12}

$f_{12}$ was formalized as \cite{Alemany2018a}
\begin{align}
f_{12}	&=	\sum_{\{st,uv\} \in Q}
			\left(
				\sum_{\{st,wx\} \in Q(G_{-uv})} 1 +
				\sum_{\{uv,wx\} \in Q(G_{-st})} 1
			\right) \label{eq:arithmetic_expressions:frequencies:12:general} \\
		&= 2\sum_{\{st,uv\} \in Q} |E(G_{-stuv})|. \label{eq:arithmetic_expressions:frequencies:12:intermediate}
\end{align}
\cref{fig:arithmetic_expressions:preliminaries:general_12} illustrates \cref{eq:arithmetic_expressions:frequencies:12:general}. Using \cref{eq:arithmetic_expressions:preliminaries:amount_edges_no_stuv}, we rewrite $f_{12}$ as
\begin{equation*}
f_{12} =
2\left(
	q(m + 2) +
	\sum_{\{st,uv\} \in Q} (a_{su} + a_{sv} + a_{tu} + a_{tv}) -
	\sum_{\{st,uv\} \in Q} (k_s + k_t + k_u + k_v)
\right)
\end{equation*}
which, thanks to \cref{prop:arithmetic_expressions:general_results:L_4:elements_Q,prop:arithmetic_expressions:general_results:sum_degrees}, leads to
\begin{equation}
\label{eq:arithmetic_expressions:frequencies:12:final}
f_{12} = 2\left((m + 2)q + \npathsfour - K\right).
\end{equation}
%-----------------------------------------%
% automatic inline of '3-3-4-freq-04.tex' %
%-----------------------------------------%
\subsubsection{$\tau=0$, $\phi=4$}
\label{sec:arithmetic_expressions:frequencies:04}

All pairs of elements of $Q$ classified into this type share no edges yet have 4 vertices in common. This allowed a brief formalization for $f_{04}$ in \cite{Alemany2018a}
\begin{equation}
\label{eq:arithmetic_expressions:frequencies:04:general}
f_{04}
= \sum_{\{st,uv\} \in Q}
\left(
	\sum_{\{su,tv\} \in Q} 1 + \sum_{\{sv,tu\} \in Q} 1
\right)
= \sum_{\{st,uv\} \in Q} (a_{su}a_{tv} + a_{sv}a_{tu}).
\end{equation}
Therefore, by \cref{prop:arithmetic_expressions:general_results:cycles_4:elements_Q} we rewrite \cref{eq:arithmetic_expressions:frequencies:04:general} as
\begin{align}
\label{eq:arithmetic_expressions:frequencies:04:final}
f_{04}
	&= 2 \nsquares \nonumber \\
	&= \frac{1}{4}tr(A^4) - n_G(\lintree[3])- \frac{1}{2}n_G(\lintree[2]) \nonumber \\
	&= \frac{1}{4}tr(A^4) + q - \frac{1}{2}m^2.
\end{align}
%-----------------------------------------%
% automatic inline of '3-3-5-freq-03.tex' %
%-----------------------------------------%
\subsubsection{$\tau=0$, $\phi=3$}
\label{sec:arithmetic_expressions:frequencies:03}

$f_{03}$ was formalized as \cite{Alemany2018a}
\begin{equation}
\label{eq:arithmetic_expressions:frequencies:03:general}
f_{03} =
	\sum_{\{st,uv\} \in Q}
	(
		\varphi_{sut} + \varphi_{svt} + \varphi_{tus} + \varphi_{tvs} +
		\varphi_{svu} + \varphi_{tvu} + \varphi_{tuv} + \varphi_{suv}
	),
\end{equation}
where $\varphi_{sut}$, $\varphi_{svt}$, ... are functions over $\{st,uv\}\in Q$. In particular these $\varphi_{...}$ are defined as
\begin{equation}
\label{eq:arithmetic_expressions:frequencies:03:general:varphis}
\varphi_{xyz} = a_{xy} |\Gamma(z, -stuv)|.
\end{equation}
where $x,y,z \in \{s,t,u,v\}$ are explicit distinct parameters and $\{s,t,u,v\}\setminus \{x,y,z\}$ as implicit parameter. Examining $\varphi_{sut}, \varphi_{tus}, \cdots, \varphi_{suv}$ separately, we see that, given $\{st,uv\} \in Q$, $\varphi_{sut}$ counts the amount of neighbors of $t$ in $G_{-stuv}$ if $su \in E$, $\varphi_{tus}$ counts the amount of neighbors of $s$ in $G_{-stuv}$ if $tu \in E$, and so on. \cref{fig:arithmetic_expressions:preliminaries:general_03} illustrates $\varphi_{tus}$.

Here we obtain a simpler expression for \cref{eq:arithmetic_expressions:frequencies:03:general} by simplifying first the inner sum of all the $\varphi_{...}$. For this, we apply \cref{eq:arithmetic_expressions:preliminaries:amount_neighs_in_G_minus} to a pairwise sum of these $\varphi_{...}$ so as to obtain a series of expressions that are easier to evaluate
\begin{align*}
\varphi_{tus} + \varphi_{tvs} &= (a_{tu} + a_{tv})(k_s - 1 - a_{su} - a_{sv}), \\
\varphi_{sut} + \varphi_{svt} &= (a_{su} + a_{sv})(k_t - 1 - a_{tv} - a_{tu}), \\
\varphi_{svu} + \varphi_{tvu} &= (a_{sv} + a_{tv})(k_u - 1 - a_{ut} - a_{us}), \\
\varphi_{suv} + \varphi_{tuv} &= (a_{su} + a_{tu})(k_v - 1 - a_{vt} - a_{vs}).
\end{align*}
When adding all of them together, we can simplify the expression a bit more
\begin{align*}
\ & (a_{tu} + a_{tv})(k_s - 1) - (a_{tu} + a_{tv})(a_{su} + a_{sv}) + 
    (a_{su} + a_{sv})(k_t - 1) - (a_{su} + a_{sv})(a_{tu} + a_{tv}) + \\
+ & (a_{sv} + a_{tv})(k_u - 1) - (a_{vs} + a_{vt})(a_{ut} + a_{us}) +
    (a_{su} + a_{tu})(k_v - 1) - (a_{us} + a_{ut})(a_{vt} + a_{vs}) \\
= & (a_{tu} + a_{tv})(k_s - 1) + (a_{su} + a_{sv})(k_t - 1) +
    (a_{sv} + a_{tv})(k_u - 1) + (a_{su} + a_{tu})(k_v - 1) \\
- & 2( (a_{tu} + a_{tv})(a_{su} + a_{sv}) + (a_{vs} + a_{vt})(a_{ut} + a_{us}) ).
\end{align*}

Upon expansion of the positive part of the expression, we obtain
\begin{equation*}
k_s(a_{tu} + a_{tv}) + k_t(a_{su} + a_{sv}) + k_u(a_{sv} + a_{tv}) + k_v(a_{su} + a_{tu})
  - 2( a_{tu} + a_{tv} + a_{sv} + a_{su} ),
\end{equation*}
and, upon expansion of the negative part,
\begin{align*}
  & -2( (a_{tu} + a_{tv})(a_{su} + a_{sv}) + (a_{vs} + a_{vt})(a_{ut} + a_{us}) ) \\
= & -4(a_{vs}a_{ut} + a_{vt}a_{us}) - 2(a_{tu} + a_{vs})(a_{su} + a_{tv}).
\end{align*}

Thanks to the results for each part, the expression for $f_{03}$ becomes
\begin{align*}
f_{03}
& =	\sum_{\{st,uv\} \in Q}
	\left(
		k_s(a_{tu} + a_{tv}) + k_t(a_{su} + a_{sv}) +
		k_u(a_{sv} + a_{tv}) + k_v(a_{su} + a_{tu})
	\right) \\
&\	-2 \sum_{\{st,uv\} \in Q} (a_{tu} + a_{tv} + a_{sv} + a_{su})
	-4 \sum_{\{st,uv\} \in Q} (a_{vs}a_{ut} + a_{vt}a_{us} ) \\
&\	-2 \sum_{\{st,uv\} \in Q} (a_{tu} + a_{vs})(a_{su} + a_{tv}).
\end{align*}
By noticing that, via a simple rearrangement of the terms inside the summation,
\begin{equation*}
\Lambda_1
= \sum_{\{st,uv\} \in Q} \left( k_s(a_{tu} + a_{tv}) + k_t(a_{su} + a_{sv}) + k_u(a_{sv} + a_{tv}) + k_v(a_{su} + a_{tu}) \right) \nonumber \\
\end{equation*}
where $\Lambda_1$ is defined in \cref{eq:arithmetic_expressions:general_results:Lambda_1:def}, and thanks to \cref{prop:arithmetic_expressions:general_results:L_4:elements_Q,prop:arithmetic_expressions:general_results:graphpaw:elements_Q,prop:arithmetic_expressions:general_results:cycles_4:elements_Q}, $f_{03}$ can be simplified further,
\begin{equation}
\label{eq:arithmetic_expressions:frequencies:03:general:final}
f_{03} = \Lambda_1 - 2 \npathsfour - 8 \nsquares - 2 n_G(\graphpaw).
\end{equation}
%------------------------------------------%
% automatic inline of '3-3-6-freq-021.tex' %
%------------------------------------------%
\subsubsection{$\tau=0$, $\phi=2$, Subtype 1}
\label{sec:arithmetic_expressions:frequencies:021}

\begin{figure}
	\centering
	\includegraphics{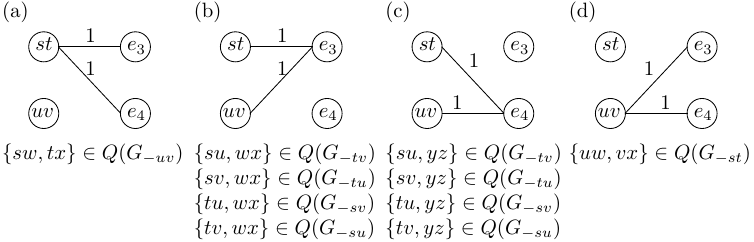}
	\caption{Elements of $Q$ such that when paired with element $\{st,uv\}\in Q$, the pair is classified as type $021$. Elements in (b) and (c) are symmetric. Source \cite[Figure 11, p. 31]{Alemany2018a}.}
	\label{fig:arithmetic_expressions:frequencies:021:combinations}
\end{figure}

$f_{021}$ (\cref{fig:arithmetic_expressions:frequencies:021:combinations}) can be formalized as
\begin{equation}
\label{eq:arithmetic_expressions:frequencies:021:general}
f_{021} = f_{021,\varphi} + f_{021,\varepsilon},
\end{equation}
where \cite{Alemany2018a}
\begin{align}
f_{021,\varphi}
	&= \sum_{\{st,uv\} \in Q} (\varphi_{st} + \varphi_{uv}), \label{eq:arithmetic_expressions:frequencies:021:f_phi}\\
f_{021,\varepsilon}
	&=
	\sum_{\{st,uv\} \in Q}
	\left(
		\varepsilon_{su} +
		\varepsilon_{sv} +
		\varepsilon_{tu} +
		\varepsilon_{tv}
	\right), \label{eq:arithmetic_expressions:frequencies:021:f_epsilon}
\end{align}
and $\varphi_{xy}$ and $\varepsilon_{xy}$ are functions over $\{st,uv\}\in Q$, being $xy$ the explicit parameters and $\{s,t,u,v\}\setminus\{x,y\}$ as implicit parameters. These functions are defined as
\begin{align}
\varphi_{xy}
	&= \sum_{w_x \in \Gamma(x, -stuv)} \sum_{w_y \in \Gamma(y, -stuvw)} 1
	&= \sum_{w_x \in \Gamma(x, -stuv)} |\Gamma(y, -stuvw)|, \label{eq:arithmetic_expressions:frequencies:021:varphis}\\
\varepsilon_{xy}
	&= a_{xy} |E(G_{-stuv})|, \label{eq:arithmetic_expressions:frequencies:021:epsilons}
\end{align}
for $x,y\in\{s,t,u,v\}$.

The functions $\varphi_{st}$ and $\varphi_{uv}$ count the elements of the form of the illustrated in Figures \ref{fig:arithmetic_expressions:preliminaries:general_021_phi}, \ref{fig:arithmetic_expressions:frequencies:021:combinations}(a) and \ref{fig:arithmetic_expressions:frequencies:021:combinations}(d). The first function counts, for each neighbor of $s$, say $w_s \neq t,u,v$, the number of neighbors of $t$, say $w_t$, such that $w_t \neq s,t,u,v,w_s$. Likewise for the second function. On the other hand, the values $\varepsilon_{su}$, $\varepsilon_{sv}$, $\varepsilon_{tu}$, $\varepsilon_{tv}$ count the edges $xy\in E$, $x,y\neq s,t,u,v$ such that when paired with $su$, $sv$, $tu$, $tv$ form an element of $Q$ whose form is that of those elements illustrated in Figures \ref{fig:arithmetic_expressions:preliminaries:general_021_epsilon}, \ref{fig:arithmetic_expressions:frequencies:021:combinations}(b) and \ref{fig:arithmetic_expressions:frequencies:021:combinations}(c). These amounts are counted only if such edges exist in the graph, hence the factor $a_{su}$ for $\varepsilon_{su}$, and likewise for the other $\varepsilon_{..}$.

On the one hand, we factor $|E(G_{-stuv})|$ out of \cref{eq:arithmetic_expressions:frequencies:021:f_epsilon} using \cref{eq:arithmetic_expressions:frequencies:021:epsilons}, giving
\begin{equation*}
f_{021,\varepsilon} =
\sum_{\{st,uv\} \in Q}
	(\varphi_{st} + \varphi_{uv} +
	(a_{su} + a_{sv} + a_{tu} + a_{tv})|E(G_{-stuv})|).
\end{equation*}
\cref{eq:arithmetic_expressions:preliminaries:amount_edges_no_stuv} produces
\begin{align*}
(a_{su} + a_{sv} + a_{tu} + a_{tv})|E(G_{-stuv})|
& = (a_{su} + a_{sv} + a_{tu} + a_{tv})m \\
&  -(a_{su} + a_{sv} + a_{tu} + a_{tv})(k_s + k_t + k_u + k_v) \\
&  +(a_{su} + a_{sv} + a_{tu} + a_{tv})^2 \\
&  + 2(a_{su} + a_{sv} + a_{tu} + a_{tv}).
\end{align*}

Recall that for $\{st,uv\}\in Q$, $a_{st} + a_{uv} = 2$. Then, thanks to \cref{prop:arithmetic_expressions:general_results:L_4:elements_Q,prop:arithmetic_expressions:general_results:Lambda_2}, we obtain
\begin{align*}
f_{021,\varepsilon} =
&\ m \sum_{\{st,uv\} \in Q} (a_{su} + a_{sv} + a_{tu} + a_{tv}) - \Lambda_2 \\
&\ + \sum_{\{st,uv\} \in Q} (a_{su} + a_{sv} + a_{tu} + a_{tv})^2
   + 2 \sum_{\{st,uv\} \in Q} (a_{su} + a_{sv} + a_{tu} + a_{tv}) \\
&= (m + 2)\npathsfour - \Lambda_2 + \sum_{\{st,uv\} \in Q} (a_{su} + a_{sv} + a_{tu} + a_{tv})^2.
\end{align*}

On the other hand, \cref{eq:arithmetic_expressions:preliminaries:amount_neighs_in_G_minus} leads to
\begin{align}
\label{eq:arithmetic_expressions:frequencies:021:phi-st}
|\Gamma(t, -stuvw_s)|
	&= k_t - \sum_{z \in \{s,t,u,v,w_s\}} a_{tz} = k_t - (a_{ts} + a_{tu} + a_{tv} + a_{tw_s}), \nonumber \\
\varphi_{st}
	&= \sum_{w_s \in \Gamma(s, -stuv)} (k_t - (a_{ts} + a_{tu} + a_{tv} + a_{tw_s})) \nonumber\\
	% &= \sum_{w_s \in \Gamma(s, -stuv)} (k_t - (a_{ts} + a_{tu} + a_{tv})) - \sum_{w_s \in \Gamma(s, -stuv)} a_{tw_s} =\\
	% &= (k_s - a_{st} - a_{su} - a_{sv})(k_t - a_{ts} - a_{tu} - a_{tv}) - \sum_{w_s \in \Gamma(s, -stuv)} a_{tw_s} =\\
	&= (k_s - a_{su} - a_{sv} - 1)(k_t - a_{tu} - a_{tv} - 1) - \sum_{w_s \in \Gamma(s, -stuv)} a_{tw_s}.
\end{align}
Likewise for $\varphi_{uv}$
\begin{equation}
\label{eq:arithmetic_expressions:frequencies:021:phi-uv}
\varphi_{uv}
	= (k_u - a_{us} - a_{ut} - 1)(k_v - a_{vs} - a_{vt} - 1) - \sum_{w_u \in \Gamma(u, -stuv)} a_{vw_u}.
\end{equation}
The negative summations in $\varphi_{st}$, \cref{eq:arithmetic_expressions:frequencies:021:phi-st} (and $\varphi_{uv}$, \cref{eq:arithmetic_expressions:frequencies:021:phi-uv}) represent the amount of vertices from $G_{-stuv}$ neighbors of $s$ (of $u$) in $G$ that are also neighbors of $t$ (of $v$), in $G$, i.e., the triangles formed by vertices $s,t,w_s$ and $u,v,w_u$ respectively. Then, \cref{eq:arithmetic_expressions:frequencies:021:f_phi} becomes
\begin{align*}
f_{021,\varphi}
&= \sum_{\{st,uv\} \in Q} (k_s -a_{su}-a_{sv}-1)(k_t -a_{tu}-a_{tv}-1) \nonumber\\
&+ \sum_{\{st,uv\} \in Q} (k_u -a_{us}-a_{ut}-1)(k_v -a_{vs}-a_{vt}-1) \nonumber\\
&- \sum_{\{st,uv\} \in Q}
	\left(
		\sum_{w_s \in \Gamma(s, -stuv)} a_{tw_s} +
		\sum_{w_u \in \Gamma(u, -stuv)} a_{vw_u}
	\right). \nonumber
\end{align*}
% We obtain another intermediate result
% \begin{align}
% \label{eq:arithmetic_expressions:frequencies:021:prev-final}
% \begin{split}
% f_{021}
% &= (m + 2)n_G(\lintree[4])\\
% &+ \sum_{\{st,uv\} \in Q} (a_{su} + a_{sv} + a_{tu} + a_{tv})^2\\
% &+ \sum_{\{st,uv\} \in Q} (k_s -a_{su}-a_{sv}-1)(k_t -a_{tu}-a_{tv}-1)\\
% &+ \sum_{\{st,uv\} \in Q} (k_u -a_{us}-a_{ut}-1)(k_v -a_{vs}-a_{vt}-1)\\
% &- \sum_{\{st,uv\} \in Q} (a_{su}+a_{sv}+a_{tu}+a_{tv})(k_s+k_t+k_u+k_v)\\
% &- \sum_{\{st,uv\} \in Q}
%	\left(
%		\sum_{w_s \in \Gamma(s, -stuv)} a_{tw_s} +
%		\sum_{w_u \in \Gamma(u, -stuv)} a_{vw_u}
%	\right). \\
% \end{split}
% \end{align}

Within $f_{021,\varepsilon}$,
\begin{align}
\label{eq:arithmetic_expressions:frequencies:021:sum_adj_squared}
               (a_{su} + a_{sv} + a_{tu} + a_{tv})^2
&=             2(a_{tu} + a_{sv})(a_{su} + a_{tv}) + 2(a_{su}a_{tv} + a_{sv}a_{tu}) \nonumber\\
&\phantom{=} + a_{su} + a_{sv} + a_{tu} + a_{tv},
\end{align}
and, within $f_{021,\varphi}$,
\begin{align*}
&\phantom{=+}   (k_s-a_{su}-a_{sv}-1)(k_t-a_{tu}-a_{tv}-1) \\
&\phantom{=}   +(k_u-a_{su}-a_{tu}-1)(k_v-a_{sv}-a_{tv}-1) \\
&=\phantom{+}   (k_sk_t + k_uk_v) - (k_s + k_t + k_u + k_v) \\
&\phantom{=}   -(k_s(a_{tu} + a_{tv}) + k_t(a_{su} + a_{sv}) + k_u(a_{sv} + a_{tv}) + k_v(a_{su} + a_{tu})) \\
&\phantom{=}   +(a_{su} + a_{tv})(a_{sv} + a_{tu}) + 2(a_{su}a_{tv} + a_{sv}a_{tu}) \\
&\phantom{=}  +2(a_{su} + a_{sv} + a_{tu} + a_{tv}) + 2.
\end{align*}
Then \cref{eq:arithmetic_expressions:general_results:Lambda_1:def}, and \cref{prop:arithmetic_expressions:general_results:graphpaw:elements_Q,prop:arithmetic_expressions:general_results:C3_L2:elements_Q,prop:arithmetic_expressions:general_results:cycles_4:elements_Q} allow one to rewrite \cref{eq:arithmetic_expressions:frequencies:021:general} as
\begin{equation}
\label{eq:arithmetic_expressions:frequencies:021:final}
f_{021}
=    (m + 5)\npathsfour + 8\nsquares + 2q - K + 3n_G(\graphpaw)
   - 3\nCoL - \Lambda_1 - \Lambda_2 + \Phi_1
\end{equation}
where $\Phi_1$ is defined in \cref{eq:arithmetic_expressions:general_results:Phi_1}.

%------------------------------------------%
% automatic inline of '3-3-7-freq-022.tex' %
%------------------------------------------%
\subsubsection{$\tau=0$, $\phi=2$, Subtype 2}
\label{sec:arithmetic_expressions:frequencies:022}

\begin{figure}
	\centering
	\includegraphics{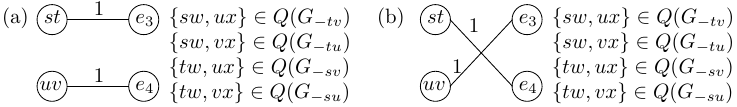}
	\caption{Elements of $Q$ such that when paired with element $\{st,uv\}\in Q$, the pair is classified as type $022$. (a) One of the bipartite graphs of type 022. (b) The other bipartite graph of type 022. The elements in (a) are symmetric to those of (b). Source \cite[Figure 13, p. 34]{Alemany2018a}.}
	\label{fig:arithmetic_expressions:frequencies:022:combinations}
\end{figure}

$f_{022}$ (\cref{fig:arithmetic_expressions:frequencies:022:combinations}) was formalized as \cite{Alemany2018a}
\begin{equation}
\label{eq:arithmetic_expressions:frequencies:022:general}
f_{022} =
	\sum_{\{st,uv\} \in Q} (\varphi_{su} +
	\varphi_{sv} +
	\varphi_{tu} +
	\varphi_{tv}),
\end{equation}
where $\varphi_{xy}$ is an auxiliary function defined as in \cref{eq:arithmetic_expressions:frequencies:021:varphis}. $\varphi_{su}$ can be understood from the case of $\varphi_{st}$ in \cref{fig:arithmetic_expressions:preliminaries:general_021_phi}.

We now derive a useful arithmetic expression for $f_{022}$. First, we expand, following \cref{eq:arithmetic_expressions:frequencies:021:phi-st,eq:arithmetic_expressions:frequencies:021:phi-uv}, the $\varphi...$ in \cref{eq:arithmetic_expressions:frequencies:022:general} as
\begin{align*}
\varphi_{su} &= (k_s - a_{su} - a_{sv} - 1)(k_u - a_{us} - a_{ut} - 1) - \sum_{w_s \in \Gamma(s, -stuv)} a_{uw_s}, \\
\varphi_{sv} &= (k_s - a_{su} - a_{sv} - 1)(k_v - a_{vs} - a_{vt} - 1) - \sum_{w_s \in \Gamma(s, -stuv)} a_{vw_s}, \\
\varphi_{tu} &= (k_t - a_{tu} - a_{tv} - 1)(k_u - a_{us} - a_{ut} - 1) - \sum_{w_t \in \Gamma(t, -stuv)} a_{uw_t}, \\
\varphi_{tv} &= (k_t - a_{tu} - a_{tv} - 1)(k_v - a_{vs} - a_{vt} - 1) - \sum_{w_t \in \Gamma(t, -stuv)} a_{vw_t}. \\
\end{align*}
Inserting these expressions into \cref{eq:arithmetic_expressions:frequencies:022:general} and taking common factors out, one obtains
\begin{align*}
f_{022}
&= \sum_{\{st,uv\} \in Q} (k_s + k_t - a_{su} - a_{sv} - a_{tu} - a_{tv} - 2)(k_u + k_v - a_{us} - a_{ut} - a_{vs} - a_{vt} - 2)\\
& -\sum_{\{st,uv\} \in Q} \left( \sum_{w_s \in \Gamma(s, -stuv)} (a_{uw_s} + a_{vw_s}) + \sum_{w_t \in \Gamma(t, -stuv)} (a_{uw_t} + a_{vw_t}) \right),
\end{align*}
which can be further developed and then simplified using \cref{prop:arithmetic_expressions:general_results:L_4:elements_Q,prop:arithmetic_expressions:general_results:L_5:elements_Q,eq:arithmetic_expressions:general_results:Lambda_2:def}, becoming
\begin{align*}
f_{022}
&= 4\npathsfour - \npathsfive - \Lambda_2\\
&+ \sum_{\{st,uv\} \in Q} (k_s + k_t - 2)(k_u + k_v - 2) + \sum_{\{st,uv\} \in Q} (a_{su} + a_{sv} + a_{tu} + a_{tv})^2.
\end{align*}
Using \cref{eq:arithmetic_expressions:general_results:Phi_2,eq:arithmetic_expressions:frequencies:021:sum_adj_squared}, \cref{prop:arithmetic_expressions:general_results:graphpaw:elements_Q,prop:arithmetic_expressions:general_results:cycles_4:elements_Q}, and by expanding $(k_s + k_t - 2)(k_u + k_v - 2)$, we finally obtain
\begin{equation}
\label{eq:arithmetic_expressions:frequencies:022:final}
f_{022} = 4q - 2K + 5\npathsfour - \npathsfive + 2n_G(\graphpaw) + 4\nsquares - \Lambda_2 + \Phi_2.
\end{equation}
%----------------------------------------%
% automatic inline of '3-4-variance.tex' %
%----------------------------------------%
\subsection{Variance of the number of crossings}
\label{sec:arithmetic_expressions:var_C}

Applying the arithmetic expressions of the $f_\omega$'s above (summarized in \cref{table:arithmetic_expressions:theoretical_formulas:frequencies:summary}) to \cref{eq:introduction:overview:var_C:general:freq__x__exp}, we obtain
\begin{align}
\label{eq:arithmetic_expressions:var_C:graphs:general}
\gvar{C}
&=\phantom{+} 	  q				(\gexpet{24} - 4\gexpet{13} + 2(m + 2)\gexpet{12} + 2\gexpet{021} + 4\gexpet{022}) \nonumber\\
&\phantom{=}	+ K				(\gexpet{13} - 2\gexpet{12} - \gexpet{021} - 2\gexpet{022}) \nonumber\\
&\phantom{=}	+ \npathsfour	(-2\gexpet{13} + 2\gexpet{12} - 2\gexpet{03} + (m + 5)\gexpet{021} + 5\gexpet{022}) \nonumber\\
&\phantom{=}	- \npathsfive	\gexpet{022} \nonumber\\
&\phantom{=}	+ \nsquares		(2\gexpet{04} - 8\gexpet{03} + 8\gexpet{021} + 4\gexpet{022}) \nonumber\\
&\phantom{=}	+ \Lambda_1		(\gexpet{03} - \gexpet{021})  \nonumber\\
&\phantom{=}	- \Lambda_2		(\gexpet{021} + \gexpet{022}) \nonumber\\
&\phantom{=}	+ \Phi_1		\gexpet{021} \nonumber\\
&\phantom{=}	+ \Phi_2		\gexpet{022} \nonumber\\
&\phantom{=}	+ \ngraphpaw	(-2\gexpet{03} + 3\gexpet{021} + 2\gexpet{022}) \nonumber\\
&\phantom{=}	- 3\nCoL		\gexpet{021},
\end{align}
where $q$ is the number of pairs of independent edges (\cref{eq:introduction:overview:size_Q}), $K$ is defined in \cref{eq:arithmetic_expressions:general_results:sum_degrees}, $\Phi_1$ and $\Phi_2$ are defined in \cref{eq:arithmetic_expressions:general_results:Phi_1,eq:arithmetic_expressions:general_results:Phi_2}, $\Lambda_1$ and $\Lambda_2$ are defined in \cref{eq:arithmetic_expressions:general_results:Lambda_1:def,eq:arithmetic_expressions:general_results:Lambda_2:def}, and $\graphpaw$ is the graph paw and $\CoL=\CoLlong$, depicted in \cref{fig:arithmetic_expressions:general_results:graphs_paw_and_C3_L2}. Since forests are acyclic, $\nsquares = \nCoL = \ngraphpaw = 0$, and then the variance of $C$ becomes
\begin{align}
\label{eq:arithmetic_expressions:var_C:general:forests}
\gvar{C}
&=\phantom{+} 	  q					(\gexpet{24} - 4\gexpet{13} + 2(m + 2)\gexpet{12} + 2\gexpet{021} + 4\gexpet{022}) \nonumber\\
&\phantom{=}	+ K					(\gexpet{13} - 2\gexpet{12} - \gexpet{021} - 2\gexpet{022}) \nonumber\\
&\phantom{=}	+ \npathsfour	(-2\gexpet{13} + 2\gexpet{12} - 2\gexpet{03} + (m + 5)\gexpet{021} + 5\gexpet{022}) \nonumber\\
&\phantom{=}	- \npathsfive	\gexpet{022} \nonumber \\
&\phantom{=}	+ \Lambda_1			(\gexpet{03} - \gexpet{021})  \nonumber\\
&\phantom{=}	- \Lambda_2			(\gexpet{021} + \gexpet{022}) \nonumber\\
&\phantom{=}	+ \Phi_1			\gexpet{021} \nonumber\\
&\phantom{=}	+ \Phi_2			\gexpet{022}.
\end{align}

For the case of uniformly random linear arrangements, the instantiation of \cref{eq:arithmetic_expressions:var_C:graphs:general} gives
\begin{equation}
\label{eq:arithmetic_expressions:var_C:graphs:rla}
\begin{split}
\lvar{C} =
	\frac{1}{180}[
	&	8(m + 2)q + 2K - (2m + 7)\npathsfour - 12\nsquares + 6\ngraphpaw \\
	& - \npathsfive + 6\nCoL - 3\Lambda_1 + \Lambda_2 - 2\Phi_1 + \Phi_2
	],
\end{split}
\end{equation}
and the result of instantiating \cref{eq:arithmetic_expressions:var_C:general:forests} is
\begin{align*}
%\label{eq:arithmetic_expressions:var_C:forests:rla}
\lvar{C} =
	\frac{1}{180}[
	8(m + 2)q + 2K - (2m + 7)\npathsfour - \npathsfive - 3\Lambda_1 + \Lambda_2 - 2\Phi_1 + \Phi_2
	].
\end{align*}

We refer the reader to \cite[Table 5]{Alemany2018a} for a summary of expressions of $\lvar{C}$ in particular classes of graphs (complete graphs, complete bipartite graphs, linear trees, cycle graphs, one-regular, star graphs and quasi-star graphs).

%---------------------------------------------------%
% automatic inline of '4-0-algorithms-variance.tex' %
%---------------------------------------------------%
\section{Algorithms to compute $\gvar{C}$}
\label{sec:algorithms}

In this section we provide algorithms that compute the exact value of $\gvar{C}$ in general graphs (based on \cref{eq:arithmetic_expressions:var_C:graphs:general}) in \cref{sec:algorithms:graphs} and in forests (based on \cref{eq:arithmetic_expressions:var_C:general:forests}) in \cref{sec:algorithms:forests}. Since the computation of $\gvar{C}$ reduces to a subgraph counting problem (as seen in previous sections), the algorithms presented below consist of solving the subgraph counting problem that we face in \cref{eq:arithmetic_expressions:var_C:graphs:general,eq:arithmetic_expressions:var_C:general:forests} and then calculating $\gvar{C}$ with knowledge of $\gexpetw$. The subgraph counting problem was presented in \cref{sec:arithmetic_expressions:general_results}.

%--------------------------------------------------------%
% automatic inline of '4-1-algorithm-general-graphs.tex' %
%--------------------------------------------------------%
\subsection{An algorithm for general graphs}
\label{sec:algorithms:graphs}

Our algorithm to calculate $\gvar{C}$ is a simple traversal of the edges of the graph with some extra work to be done for each edge, which is, basically, a traversal over the neighborhood of the endpoints. For general graphs, this yields an algorithm of time complexity $\bigO{\maxdeg n\mmtdeg{2}} = \smallo{nm^2}$.

This algorithm was derived combining three strategies. First, we have shown that the values $q$, $K$, $\Phi_1$ and $\Phi_2$ are linear-time computable in $n$ or $m$ since they have very simple arithmetic expressions (see \cref{eq:introduction:overview:size_Q,eq:arithmetic_expressions:general_results:sum_degrees,eq:arithmetic_expressions:general_results:Phi_1,eq:arithmetic_expressions:general_results:Phi_2}, respectively). Second, the subgraphs present in \cref{eq:arithmetic_expressions:var_C:graphs:general}, namely $\lintree[4]$, $\cycle[4]$, $\lintree[5]$, $\graphpaw$ and $\CoL$ (the last two are depicted in \cref{fig:arithmetic_expressions:general_results:graphs_paw_and_C3_L2}) could be counted straightforwardly by relying on previous work \cite{Alon1997a,Movarraei2014a}. For example, we could use Alon {\em et al.}'s contributions \cite{Alon1997a} on the counting of $\cycle[4]$ and $\graphpaw$ in graphs, and Movarraei's contributions \cite{Movarraei2014a} regarding formulas to count all instances of $\lintree[4]$ and $\lintree[5]$ in graphs. However, these methods require using the adjacency matrix of the graph and we have opted for a combinatorial approach that does not employ it; we discuss this in deeper detail in \cref{sec:discussion:general_subgraphs,sec:discussion:specific_subgraphs}. Third, we apply the reinterpretation of $\Lambda_1$ and $\Lambda_2$ (in \cref{eq:arithmetic_expressions:general_results:Lambda_1,eq:arithmetic_expressions:general_results:Lambda_2}) which states that their value can be computed without enumerating the elements of $Q$.

All the expressions in \cref{sec:arithmetic_expressions:general_results} are integrated in \cref{algo:algorithms:graphs:no_reuse} to compute the exact value of $\lvar{C}$ in a straightforward manner. We formalize the complexity of this algorithm in \cref{prop:algorithms:graphs:no_reuse}.

\begin{proposition}
\label{prop:algorithms:graphs:no_reuse}
Let $G=(V,E)$ be a graph with $n=|V|$ and $m=|E|$. Let the graph be implemented with sorted adjacency lists, namely, the adjacency list of each vertex contains labels sorted in increasing lexicographic order. \cref{algo:algorithms:graphs:no_reuse} computes $\gvar{C}$ in $G$ in time $\bigO{\psi + m - n\mmtdeg{2}} = \bigO{\maxdeg n\mmtdeg{2}} = \smallo{nm^2}$ and space $\bigO{n}$.
\end{proposition}
\begin{proof}
The computation of $\gvar{C}$'s value is done by putting together all the results presented in this work that involve the terms in \cref{eq:arithmetic_expressions:var_C:graphs:general}. The results' correctness has already been proved and the algorithm uses them in a straightforward manner. The space complexity is easy to calculate: we need $\bigO{n}$-space to store the values of the function $\xi(s)$ (\cref{eq:arithmetic_expressions:preliminaries:sum_degrees_neighbors}) for each vertex $s\in V$.

As for the time complexity, the cost of \textsc{Setup} (\cref{algo:algorihtms:graphs:setup}) is $\bigO{n+m}$ and then the algorithm iterates over the set of edges and performs, for each edge, three intersection operations to compute the values $|c(t,u_1)|$, $|c(s,u_2)|$ and $|c(s,t)|$. The other computations are constant-time operations as a function of the vertices of similar intersection or as a function of the vertices of the edge. Now, if we denote the cost of the intersection of two sorted lists $\Gamma(u)$ and $\Gamma(v)$ as $\Delta(u,v)$, the algorithm has cost
\begin{equation*}
\overbrace{\bigO{n + m}}^{\text{\cref{algo:algorithms:graphs:no_reuse:call_to_setup}}} + \bigO{ \sum_{st\in E}
	\left(
		1 +
		\overbrace{\Delta(s,t)}^{\text{\cref{algo:algorithms:graphs:no_reuse:for:3rd}}} +
		\overbrace{\sum_{u_1\in\Gamma(s)\setminus\{t\}} \Delta(t,u_1)}^{\text{\cref{algo:algorithms:graphs:no_reuse:for:1st}}} +
		\overbrace{\sum_{u_2\in\Gamma(t)\setminus\{s\}} \Delta(s,u_2)}^{\text{\cref{algo:algorithms:graphs:no_reuse:for:2nd}}}
	\right)
}.
\end{equation*}

It is easy to make an algorithm to calculate the intersection of two sorted lists with cost $\Delta(u,v)=\bigO{k_u + k_v}$. This leads to the following cost function
\begin{equation*}
\bigO{n + 2m + 2\sum_{st\in E} k_sk_t - \sum_{st\in E} (k_s + k_t) + \sum_{st\in E} (\xi(s) + \xi(t)) }.
\end{equation*}
Obviously, the first summation is $\psi$ (\cref{eq:introduction:psi}) and the second summation equals $n\mmtdeg{2}$. The third summation is also easy,
\begin{equation*}
\sum_{st\in E} (\xi(s) + \xi(t))
	= \sum_{u\in V} k_u\xi(u)
	= \sum_{u\in V} \sum_{v\in \Gamma(u)} k_uk_v
	= 2 \sum_{st\in E} k_sk_t
	= 2\psi.
\end{equation*}
Therefore, the algorithm has time complexity
\begin{equation*}
\bigO{4\psi + 2m - n\mmtdeg{2}} = \bigO{\psi + m - n\mmtdeg{2}}.
\end{equation*}

We can easily derive an upper bound of the cost which, although generous, gives a more understandable cost of the algorithm. Simply, notice that
\begin{equation*}
4\sum_{st\in E} k_sk_t + \sum_{u\in V} k_u - \sum_{u\in V} k_u^2
	< 4\sum_{st\in E} k_sk_t 
	%&< \maxdeg \sum_{st\in E} k_t \\
	< 4\maxdeg \sum_{st\in E} (k_s + k_t)
	= \bigO{\maxdeg n\mmtdeg{2}}.
\end{equation*}
The algorithm's cost is now expressed in terms of $\maxdeg$ and the structure of the graph, as captured by $\mmtdeg{2}$. To obtain a cost in terms of $n$ and $m$, we use $n\mmtdeg{2} \le m(m + 1)$ which follows from the fact that $q\ge 0$ (\cref{eq:introduction:overview:size_Q}); and thus the cost can be expressed as $\smallo{nm^2}$.
\end{proof}

\begin{algorithm}
	\caption{Setup.}
	\label{algo:algorihtms:graphs:setup}
	\DontPrintSemicolon

	\SetKwProg{Fn}{Function}{ is}{end}
	\Fn{\textsc{Setup}$(G)$} {
		Calculate $\mmtdeg{2}$ and $\mmtdeg{3}$ in $\bigO{n}$-time \tcp{\cref{eq:introduction:mmt_deg}}
		$\xi(s) \gets 0^n$				\tcp{\cref{eq:arithmetic_expressions:preliminaries:sum_degrees_neighbors}}
		Calculate $\xi(s)$ for all $s\in V$ in $\bigO{m}$-time. \;
		$\psi \gets 0$					\tcp{\cref{eq:introduction:psi}}
		
		$\Phi_1, \Phi_2 \gets 0$		\tcp{\cref{eq:arithmetic_expressions:general_results:Phi_1,eq:arithmetic_expressions:general_results:Phi_2}}

		$\ngraphpaw, \nCoL, \nsquares \gets 0$
										\tcp{\cref{eq:arithmetic_expressions:general_results:graphpaw:traversal,eq:general_reuslts:C3_L2:traversal,eq:algorithms:graphs:cycle_4}}

		$\mu \gets 0$					\tcp{\cref{eq:arithmetic_expressions:general_results:L_4:traversal:mu}}
		$\npathsfive \gets 0$			\tcp{\cref{eq:arithmetic_expressions:general_results:L_5:traversal}}
		$\Lambda_1, \Lambda_2 \gets 0$	\tcp{\cref{eq:arithmetic_expressions:general_results:Lambda_1:def,eq:arithmetic_expressions:general_results:Lambda_2:def}}
	}
\end{algorithm}

\begin{algorithm}
	\caption{Calculate $\gvar{C}$ in general graphs.}
	\label{algo:algorithms:graphs:no_reuse}
	\DontPrintSemicolon
	
	\KwIn{$G=(V,E)$ a graph as described in \cref{prop:algorithms:graphs:no_reuse}.}
	\KwOut{$\gvar{C}$, the variance of the number of crossings.}
	
	\SetKwProg{Fn}{Function}{ is}{end}
	\Fn{\textsc{VarianceC}$(G)$} {
		
		\textsc{Setup}($G$) \label{algo:algorithms:graphs:no_reuse:call_to_setup} \tcp{\cref{algo:algorihtms:graphs:setup}}

		\For {$\{s,t\}\in E$} {
			\For {$u\in\Gamma(s)\setminus\{t\}$} { \label{algo:algorithms:graphs:no_reuse:for:1st}
			
				\red{$|c(t,u)| \gets 0$} \tcp{\cref{eq:arithmetic_expressions:preliminaries:common_neighs_vers}}
				\lFor {$\red{w\in \Gamma(t)\cap\Gamma(u)}$} {
					$\red{|c(t,u)|} \gets \red{|c(t,u)|} + \red{1}$
				}
			
				$\npathsfive \gets \npathsfive -
					\red{|c(t,u)|}
					+ (k_t - 1 - a_{tu})(k_u - 1 - a_{tu}) + 1$ \;
			}
			\For {$u\in\Gamma(t)\setminus\{s\}$} { \label{algo:algorithms:graphs:no_reuse:for:2nd}
			
				\red{$|c(s,u)| \gets 0$} \tcp{\cref{eq:arithmetic_expressions:preliminaries:common_neighs_vers}}
				\lFor {$\red{w\in \Gamma(s)\cap\Gamma(u)}$} {
					$\red{|c(s,u)|} \gets \red{|c(s,u)|} + \red{1}$
				}
				
				$\npathsfive \gets \npathsfive -
					\red{|c(s,u)|} + (k_s - 1 - a_{su})(k_u - 1 - a_{su}) + 1$ \;
				$\nsquares \gets \nsquares + \red{|c(s,u)|} - 1$ \;
			}
			$\red{|c(s,t)|}$, $\red{S_{s,t}} \gets 0$ \tcp{\cref{eq:arithmetic_expressions:preliminaries:common_neighs_vers,eq:arithmetic_expressions:preliminaries:sum_degs_common}}
			\For {$\red{u\in \Gamma(s)\cap\Gamma(t)}$} { \label{algo:algorithms:graphs:no_reuse:for:3rd}
				$\red{|c(s,t)|} \gets \red{|c(s,t)|} + \red{1}$ \;
				$\red{S_{s,t}} \gets \red{S_{s,t}} + \red{k_u}$ \;
			}
			
			$\ngraphpaw \gets \ngraphpaw + \red{S_{s,t}} - 2\red{|c(s,t)|}$ \;
			$\nCoL \gets \nCoL + (m - k_s - k_t + 3)\red{|c(s,t)|} - \red{S_{s,t}}$ \;
			
			$\psi \gets \psi + k_sk_t$ \;
			
			$\Phi_1 \gets \Phi_1 - k_sk_t(k_s + k_t)$ \;
			$\Phi_2 \gets \Phi_2 + (k_s + k_t)(n\mmtdeg{2} - (\xi(s) + \xi(t)) - k_s(k_s - 1) - k_t(k_t - 1))$ \;
			
			$\mu \gets \mu + \red{|c(s,t)|}$\;
			
			$\Lambda_1 \gets \Lambda_1 + (k_t - 1)(\xi(s) - k_t) + (\xi(t) - k_s) - 2\red{S_{s,t}}$\;
			$\Lambda_2 \gets \Lambda_2 + (k_s + k_t)( (k_s - 1)(k_t - 1) - \red{|c(s,t)|})$\;
		}

		$q \gets \frac{1}{2}[m(m + 1) - n\mmtdeg{2}]$ \tcp{\cref{eq:introduction:overview:size_Q}}
		
		$K \gets (m + 1)n\mmtdeg{2} - n\mmtdeg{3} - 2\psi$ \tcp{\cref{eq:arithmetic_expressions:general_results:sum_degrees}}
		
		$\Phi_1 \gets \Phi_1 + (m + 1)\psi$ \tcp{\cref{eq:arithmetic_expressions:general_results:Phi_1}}
		$\Phi_2 \gets \frac{1}{2}\Phi_2$ \tcp{\cref{eq:arithmetic_expressions:general_results:Phi_2}}
		
		$\nCoL \gets \frac{1}{3}\nCoL$ \tcp{\cref{eq:general_reuslts:C3_L2:traversal}}
		$\nsquares \gets \frac{1}{4}\nsquares$ \tcp{\cref{eq:algorithms:graphs:cycle_4}}
		
		$\npathsfour \gets m - n\mmtdeg{2} + \psi - \mu$ \tcp{\cref{eq:arithmetic_expressions:general_results:L_4:traversal}}
		
		$\npathsfive \gets \frac{1}{2}\npathsfive$ \tcp{\cref{eq:arithmetic_expressions:general_results:L_5:traversal}}
		$\Lambda_2 \gets \Lambda_1 + \Lambda_2$ \tcp{\cref{eq:arithmetic_expressions:general_results:Lambda_2}}
		
		Compute $\gvar{C}$ by instating \cref{eq:arithmetic_expressions:var_C:graphs:general} appropriately \;
	}
\end{algorithm}

Notice that the assumption that the graph's adjacency list being sorted merely simplifies the algorithm. In case it was not, sorting it, prior to the algorithm's execution, has cost $\bigO{\sum_{u\in V} k_u\log{k_u}} = \smallo{n\mmtdeg{2}}$ when using a comparison-based algorithm. Algorithms that are not based on comparisons may have lower time complexity. For instance, counting sort \cite{Cormen2001a} can sort $k$ numbers in time $\bigO{k}$ and space $\bigO{k}$ if their values range within the interval $[1,k]$. In an adjacency list, the entry corresponding to vertex $u$ contains $k_u$ values which range within the interval $[1,n]$; sorting can be done in space $\bigO{n}$, and time $\bigO{\sum_{u\in V} k_u}=\bigO{m}$.

In \cref{sec:algorithms:graphs:reuse} we extend \cref{algo:algorithms:graphs:no_reuse} to reuse computations and show, using empirical results, that doing so produces an algorithm several times faster in practice. 

\subsubsection{Improving the algorithm by reusing computations}
\label{sec:algorithms:graphs:reuse}

Here we improve \cref{algo:algorithms:graphs:no_reuse} by reusing the computations that are marked in red. The new algorithm is detailed in \cref{algo:algorithms:graphs:reuse} and its complexity is analyzed in \cref{prop:algorithms:graphs:reuse}, where we show that the transitivity index of a graph (\cref{eq:introduction:transitivity_index}) influences its space complexity.

\begin{algorithm}
	\caption{Update the hash table $H$ of \cref{algo:algorithms:graphs:reuse}, if necessary.}
	\label{algo:algorithms:graphs:reuse:update_hash}
	\DontPrintSemicolon
	
	\KwIn{$H$ hash table, $G=(V,E)$, $u,v\in V$.}
	\KwOut{Returns the values $|c(u,v)|$ and $S_{u,v}$.}

	\SetKwProg{Fn}{Function}{ is}{end}
	\Fn{\textsc{ComputeAndStore}$(H, G, u,v)$} {
		$|c(u,v)| \gets 0$ \;
		$S_{u,v} \gets 0$ \tcp{\cref{eq:arithmetic_expressions:preliminaries:sum_degs_common}}
		\If{$\pair{u}{v}\notin H$} {
			\tcp{Compute values $|c(u,v)|$ and $S_{u,v}$}
			\For {$w\in \Gamma(u)\cap\Gamma(v)$} {
				$|c(u,v)| \gets |c(u,v)| + 1$ \;
				$S_{u,v} \gets S_{u,v} + k_w$ \;
			}
			\tcp{Store $|c(u,v)|$ and $S_{u,v}$ in $H$ indexed with key $\pair{u}{v}$}
			$H\gets H \cup \{ \pair{u}{v} , \{|c(u,v)|, S_{u,v} \} \}$
		}
		\Else {
			\tcp{Retrieve $|c(u,v)|$ and $S_{u,v}$ from the table}
			$|c(u,v)| \gets H(\pair{u}{v}).|c(u,v)|$ \;
			$S_{u,v} \gets H(\pair{u}{v}).S_{u,v}$ \;
		}
		\Return $\{ |c(u,v)|, S_{u,v} \}$
	}
\end{algorithm}

\cref{algo:algorithms:graphs:reuse} reuses the calculation of the number of common neighbors of two not-necessarily-adjacent vertices $u$, $v$, i.e., $|c(u,v)|$, defined in \cref{eq:arithmetic_expressions:preliminaries:common_neighs_vers}, and the sum of the degrees of the vertices that are neighbors of both vertices $S_{u,v}$, defined in \cref{eq:arithmetic_expressions:preliminaries:sum_degs_common}. These values are marked in red in \cref{algo:algorithms:graphs:no_reuse}. In order to reuse them, \cref{algo:algorithms:graphs:reuse} makes use of a hash table $H$ whose keys are unordered pairs of vertices $u$ and $v$, denoted as $\pair{u}{v}$, and the associated values are $|c(u,v)|$ and $S_{u,v}$. Keys are made up of vertices that are either adjacent ($\{u,v\}\in E$) or there exists another vertex $w$ such that $a_{uw}=a_{wv}=1$. Notice that these cases are not mutually exclusive: if two vertices are both adjacent and connected via a third vertex (i.e., when $u$ and $v$ are vertices of a $\cycle[3]$) then the values associated are computed only once and the pair $\pair{u}{v}$ is stored only once. The same applies if $u$ and $v$ are not adjacent but are connected via another vertex (and then $u$ and $v$ are the ends of some $\lintree[3]$).

Whenever $|c(u,v)|$ or $S_{u,v}$ are needed, the pair $\pair{u}{v}$ is first searched in $H$. If $H$ has such pair, its associated values are retrieved. If it does not, both $|c(u,v)|$ and $S_{u,v}$ are computed and stored in $H$. This update step is detailed in \cref{algo:algorithms:graphs:reuse:update_hash}. These ideas yield \cref{algo:algorithms:graphs:reuse}, where changes with respect to \cref{algo:algorithms:graphs:no_reuse} are marked in red.

\begin{algorithm}
	\caption{Calculate $\gvar{C}$ in general graphs reusing computations.}
	\label{algo:algorithms:graphs:reuse}
	\DontPrintSemicolon
	
	\KwIn{$G=(V,E)$ a graph as described in \cref{prop:algorithms:graphs:no_reuse}.}
	\KwOut{$\gvar{C}$, the variance of the number of crossings.}
	
	\SetKwProg{Fn}{Function}{ is}{end}
	\Fn{\textsc{VarianceC}$(G)$} {
		
		\textsc{Setup}($G$) \tcp{\cref{algo:algorihtms:graphs:setup}}
		$H\gets \emptyset$ \tcp{Empty hash table}
		\For {$\{s,t\}\in E$} {
			\label{line:algorithms:graphs:reuse:edge_loop}
			
			\For {$u\in\Gamma(s)\setminus\{t\}$} {
				\red{$|c(t,u)|, \_\_ \gets$ \textsc{ComputeAndStore}$(H,G, t,u)$}
					\label{line:algorithms:graphs:reuse:1st_call_H} \;
				$\npathsfive \gets \npathsfive -
					|c(t,u)|
					+ (k_t - 1 - a_{tu})(k_u - 1 - a_{tu}) + 1$ \;
			}
			\For {$u\in\Gamma(t)\setminus\{s\}$} {
				\red{$|c(s,u)|, \_\_ \gets$ \textsc{ComputeAndStore}$(H,G, s,u)$}
					\label{line:algorithms:graphs:reuse:2nd_call_H}\;
				$\npathsfive \gets \npathsfive -
					|c(s,u)| + (k_s - 1 - a_{su})(k_u - 1 - a_{su}) + 1$ \;
				$\nsquares \gets \nsquares + |c(s,u)| - 1$ \;
			}
			\red{$|c(s,t)|, S_{s,t} \gets$ \textsc{ComputeAndStore}$(H,G, s,t)$}
				\label{line:algorithms:graphs:reuse:3rd_call_H}\;
			
			$\ngraphpaw \gets \ngraphpaw + S_{s,t}- 2|c(s,t)|$ \;
			$\nCoL \gets \nCoL + (m - k_s - k_t + 3)|c(s,t)| - S_{s,t}$ \;
			
			$\psi \gets \psi + k_sk_t$ \;
			
			$\Phi_1 \gets \Phi_1 - k_sk_t(k_s + k_t)$ \;
			$\Phi_2 \gets \Phi_2 + (k_s + k_t)(n\mmtdeg{2} - \xi(s) - \xi(t) - k_s(k_s - 1) - k_t(k_t - 1))$ \;
			
			$\mu \gets \mu + |c(s,t)|$\;
			
			$\Lambda_1 \gets \Lambda_1 + (k_t - 1)(\xi(s) - k_t) + (\xi(t) - k_s) - 2S_{s,t}$\;
			$\Lambda_2 \gets \Lambda_2 + (k_s + k_t)( (k_s - 1)(k_t - 1) - |c(s,t)|)$\;
		}
		
		$q \gets \frac{1}{2}[m(m + 1) - n\mmtdeg{2}]$ \tcp{\cref{eq:introduction:overview:size_Q}}
		
		$K \gets (m + 1)n\mmtdeg{2} - n\mmtdeg{3} - 2\psi$ \tcp{\cref{eq:arithmetic_expressions:general_results:sum_degrees}}
		
		$\Phi_1 \gets \Phi_1 + (m + 1)\psi$ \tcp{\cref{eq:arithmetic_expressions:general_results:Phi_1}}
		$\Phi_2 \gets \frac{1}{2}\Phi_2$ \tcp{\cref{eq:arithmetic_expressions:general_results:Phi_2}}
		
		$\nCoL \gets \frac{1}{3}\nCoL$ \tcp{\cref{eq:general_reuslts:C3_L2:traversal}}
		$\nsquares \gets \frac{1}{4}\nsquares$ \tcp{\cref{eq:algorithms:graphs:cycle_4}}
		
		$\npathsfour \gets m - n\mmtdeg{2} + \psi - \mu$ \tcp{\cref{eq:arithmetic_expressions:general_results:L_4:traversal}}
		
		$\npathsfive \gets \frac{1}{2}\npathsfive$ \tcp{\cref{eq:arithmetic_expressions:general_results:L_5:traversal}}
		$\Lambda_2 \gets \Lambda_1 + \Lambda_2$ \tcp{\cref{eq:arithmetic_expressions:general_results:Lambda_2:def}}
		
		Compute $\gvar{C}$ by instating \cref{eq:arithmetic_expressions:var_C:graphs:general} appropriately \;
	}
\end{algorithm}

To better analyze the cost of the algorithm, we use $\delta_{uv}$ as defined in \cref{eq:introduction:delta_connectivity}.

\begin{proposition}
\label{prop:algorithms:graphs:reuse}
Consider a graph $G=(V,E)$ that is implemented as in \cref{prop:algorithms:graphs:no_reuse}. \cref{algo:algorithms:graphs:reuse} computes $\gvar{C}$ of $G$ in time
\begin{equation*}
\bigO{nm - \sum_{uv\notin E} \delta_{uv}}
\end{equation*}
and space complexity $\bigO{n + |H|}$, where $|H|$ is the size of the hash table $H$ at the end of the algorithm
\begin{equation}
\label{eq:algorithms:graphs:reuse:size_H}
|H| = \bigO{n\mmtdeg{2} - \ntris} = \bigO{Tm + (1 - T)n\mmtdeg{2}} = \bigO{n^2}
\end{equation}
where $T$ is the transitivity index of a graph (\cref{eq:introduction:transitivity_index}).
\end{proposition}
\begin{proof}
The space complexity of this algorithm depends on the memory used to store the values of function $\xi(s)$ (\cref{eq:arithmetic_expressions:preliminaries:sum_degrees_neighbors}) and the size of the hash table $H$ at the end of its execution. The former needs $\bigO{n}$-space. Now follows a derivation of $H$'s size.

Recall that the size of any hash table is proportional to the amount of keys plus the values associated to each of them. In our case, the keys have fixed size (a pair of vertices) and the amount of values associated to the keys is always constant (two integers), so we only need to know the amount of keys it contains at the end of the algorithm.

As explained above, pairs of vertices $\pair{u}{v}$ are added to $H$ in two not necessarily mutually exclusive cases. When (1) the vertices are an edge of the graph ($a_{uv}=1$), and when (2) the two vertices are connected via a third vertex ($\delta_{uv}=1$). Since the second event includes the first, we simplify (2) to the case when $a_{uv}=0$ and $\delta_{uv}=1$. Define $\rho_1$ as the amount of pairs in (1), and $\rho_2$ as the amount of pairs added in (2). Then, $|H| = \bigO{\rho_1 + \rho_2}$.

The first case is easy: $\rho_1=m$. Consider now case (2). In this case, $u$ and $v$ are the vertices of an open $\lintree[3]$ (if $u$ and $v$ were adjacent then the $\lintree[3]$ would be closed, also a $\cycle[3]$). The exact value of $\rho_2$ is the amount of pairs of vertices $s$, $t$ such that $a_{st}=0$ and $a_{st}^{(2)}\neq 0$,
\begin{equation*}
\rho_2
	= \sum_{uv\notin E} \delta_{uv}.
\end{equation*}
An upper bound of $\rho_2$ is the number of open $\lintree[3]$ in a graph, i.e.,
\begin{align*}
\rho_2
	& \le \text{no. open $\lintree[3]$} \\
	&= \text{no. open and closed $\lintree[3]$} - \text{no. closed $\lintree[3]$} \\
	&= n_G(\lintree[3]) - 3n_G(\cycle[3]) \\
    &= (1 - T)n_G(\lintree[3]).
\end{align*}
% Therefore, the space complexity of this algorithm is the size of $H$ plus $\bigO{n}$ space for the values $\xi(s)$.
The size of $H$ can be expressed in two different ways. First, easily enough, we can express the size of $H$ using the transitivity index $T$
\begin{equation}
\label{eq:algorithms:graphs:reuse:size_H_using_T_pre}
|H| = \bigO{m + (1 - T)n_G(\lintree[3])}.
\end{equation}
In the worst case, $|H|=\bigO{n^2}$. For example, in a complete graph $|H|={n \choose 2}$. Second, since \cite[p. 103]{Estrada2015a}
\begin{equation}
\label{eq:algorithms:graphs:reuse:num_lintree3}
n_G(\lintree[3]) = \sum_{i=1}^n {k_i \choose 2} = \frac{n}{2}\mmtdeg{2} - m,
\end{equation}
we get that
\begin{equation*}
|H| = \bigO{n\mmtdeg{2} - n_G(\cycle[3])}.
\end{equation*}
Applying \cref{eq:algorithms:graphs:reuse:num_lintree3} to \cref{eq:algorithms:graphs:reuse:size_H_using_T_pre}, we obtain via straightforward arithmetic operations the cost in \cref{eq:algorithms:graphs:reuse:size_H}.

Its time complexity is given by the amount of work to be done for each pair of vertices $u,v$ to be passed as an argument of \cref{algo:algorithms:graphs:reuse:update_hash} in \cref{algo:algorithms:graphs:reuse} (\cref{line:algorithms:graphs:reuse:1st_call_H,line:algorithms:graphs:reuse:2nd_call_H,line:algorithms:graphs:reuse:3rd_call_H}). Let $\sigma_{uv}$ be the number of times the pair $u,v$ is passed as an argument of a call to \cref{algo:algorithms:graphs:reuse:update_hash}. It is easy to see that $\sigma_{uv}=a_{uv} + |c(u,v)|$. For every such pair, the algorithm performs an intersection operation of cost $\Delta(u,v)$ once, and exactly $\sigma_{uv}-1$ operations of constant-time cost.

The total running time assuming that $\Delta(u,v)=\bigO{k_u + k_v}$ is then
\begin{equation*}
\bigO{n + m} + \bigO{ \sum_{\pair{u}{v}\in H} (k_u + k_v + \sigma_{uv} - 1) }
\end{equation*}
where $\bigO{n + m}$ is the cost of \cref{algo:algorihtms:graphs:setup}. We will bound the second term by parts. First notice that 
\begin{equation*}
\sum_{\pair{u}{v}\in H} (k_u + k_v)
	\le \sum_{\pair{u}{v}\in {V\choose 2}} (k_u + k_v)
	\le 2nm
\end{equation*}
where ${V \choose 2}$ denotes the set of all unordered pairs of {\em different} vertices of $V$. Second, notice that 
\begin{align*}
\sum_{\pair{u}{v}\in H} (\sigma_{uv} - 1)
	%&= \sum_{\pair{u}{v}\in {V\choose 2}} a_{uv}(\sigma_{uv} - 1) + \sum_{\pair{u}{v}\in {V\choose 2}} (1 - a_{uv})\delta_{uv}(\sigma_{uv} - 1) \\
	&\leq \sum_{uv\in E} (\sigma_{uv} - 1) + \sum_{uv\notin E} \delta_{uv}(\sigma_{uv} - 1) \\
	&= \sum_{uv\in E} |c(u,v)| + \sum_{uv\notin E} \delta_{uv}(|c(u,v)| - 1) \\
	%&= \sum_{\pair{u}{v}\in E} |c(u,v)| + \sum_{\pair{u}{v}\notin E} \delta_{uv}|c(u,v)| - \sum_{\pair{u}{v}\notin E} \delta_{uv}\\
	&= \sum_{uv\in E} |c(u,v)| + \sum_{uv\notin E} |c(u,v)| - \sum_{uv\notin E} \delta_{uv} \\
	&= \sum_{\pair{u}{v}\in {V\choose 2}} |c(u,v)| - \sum_{uv\notin E} \delta_{uv},
\end{align*}
where, since $|c(u,v)|\le k_u + k_v - a_{uv}$,
\begin{equation*}
\sum_{\pair{u}{v}\in {V\choose 2}} |c(u,v)|
	%\le -m + \sum_{\pair{u}{v}\in {V\choose 2}} (k_u + k_v)
	%\le -m + 2nm
	\le 2nm.
\end{equation*}
The total time complexity of reusing computations is, then,
\begin{equation*}
\bigO{n + m} + \bigO{nm- \sum_{uv\notin E} \delta_{uv}}=\bigO{nm - \sum_{uv\notin E} \delta_{uv}}.
\end{equation*}
\end{proof}

\subsubsection{Analysis in Erd\H{o}s-R\'enyi random graphs}
\label{sec:algorithms:graphs:analysis_erdos_renyi}

\begin{figure}[H]
	\centering
	\includegraphics[scale=0.7]{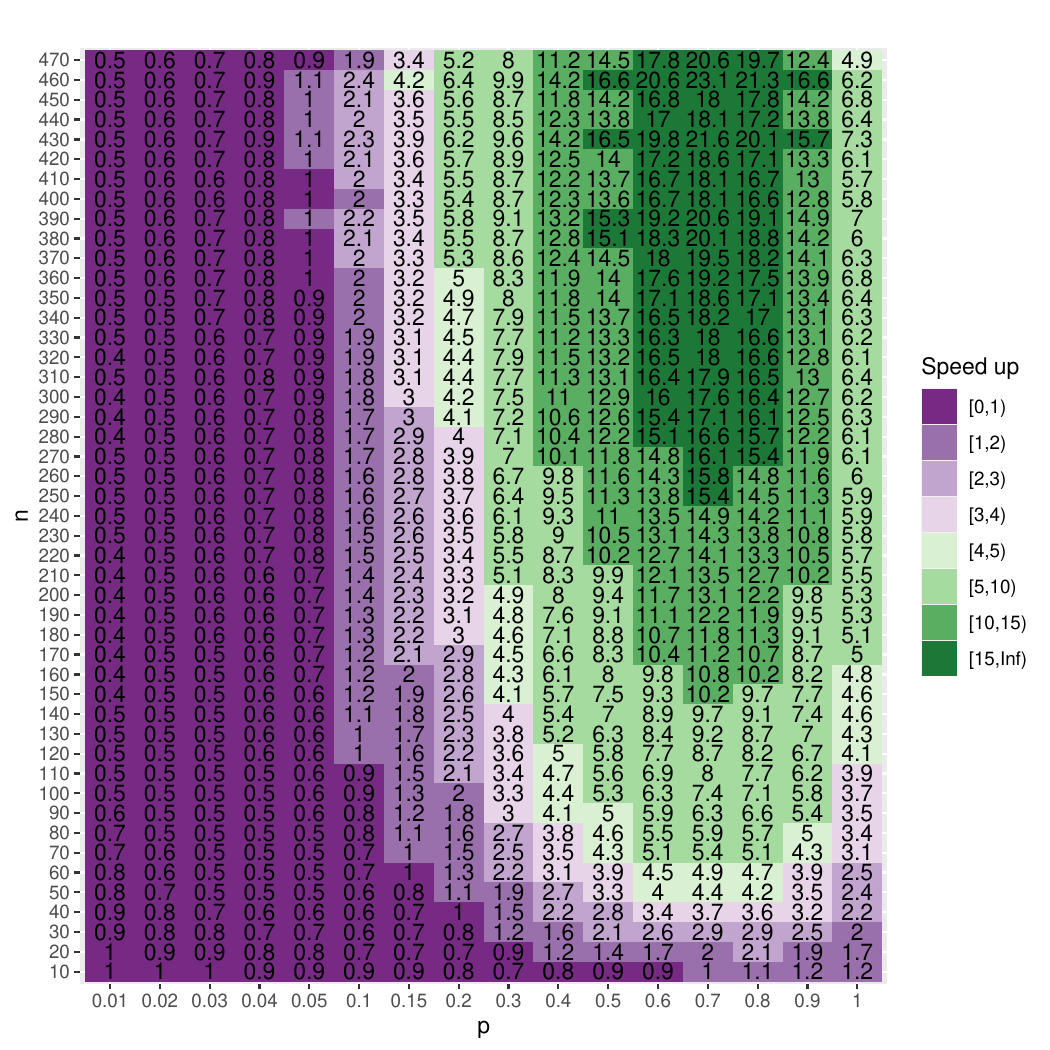}
	\caption{Speed-up values of reusing computations, that is, the execution time of not reusing computations (\cref{algo:algorithms:graphs:no_reuse}) divided by the execution of reusing them (\cref{algo:algorithms:graphs:reuse}), as a function of the parameters $n$ and $p$ of the Erd\H{o}s-R\'enyi model \cite[Section V]{Bollobas1998a}. The execution time for \cref{algo:algorithms:graphs:reuse} is the average of its execution time on 30 graphs in $\randgraphp$. The execution time for a single graph in $\randgraphp$ is the average over $r$ executions of the algorithm on a single graph in $\randgraphp$. The execution time of \cref{algo:algorithms:graphs:no_reuse} was measured likewise. The speedup is calculated as the ratio of the two final averages. We used different values of $r$ depending on the value of $p$: for $p<0.05$ we used $r=1000$, for $0.05\le p\le 0.15$ we used $r=100$, and for $p\ge 0.2$ we used $r=10$.}
	\label{fig:algorithms:graphs:analysis_erdos_renyi:speedup}
\end{figure}

Comparing the running time of \cref{algo:algorithms:graphs:no_reuse} and \cref{algo:algorithms:graphs:reuse}, it is easy to see that the latter reduces the asymptotic upper bound of the time cost (\cref{table:background:summary_algorithms}), suggesting that it should be faster in denser graphs. However, this does not imply that \cref{algo:algorithms:graphs:reuse} is faster in general: the time costs are asymptotic estimates of upper bounds. Therefore, we evaluate the potential speed-up of \cref{algo:algorithms:graphs:reuse} with the help of Erd\H{o}s-R\'enyi random graphs $\randgraphp$ \cite[Section V]{Bollobas1998a}. \cref{fig:algorithms:graphs:analysis_erdos_renyi:speedup} shows that reusing computations (\cref{algo:algorithms:graphs:reuse}) is faster for sufficiently dense $n$-vertex graphs, given a fixed $n$. Both algorithms are available in the Linear Arrangement Library \cite{Alemany2021c}\footnote{We used 2017--C++ standard. The hash table was implemented using the template class {\tt std::unordered\_map}, in which insertions and lookups have constant-time complexity. We used GNU's {\tt gcc} compiler, version {\tt 11.1}, and we used the usual optimization flags, such as {\tt -O3}. The hashing scheme we used can be found in the implementation of an R package for cluster analytics \cite{Arratia2022a}. Execution time was measured on an Intel Core i5-10600 CPU 3.30 GHz.}.

It is easy to see from \cref{fig:algorithms:graphs:analysis_erdos_renyi:speedup} that the speed up depends on both $n$ and $p$. To prove this we give the expected running time of both algorithms in $\randgraphp$. We start with \cref{algo:algorithms:graphs:no_reuse}.

\begin{proposition}
\label{prop:algorithms:graphs:analysis_erdos_renyi:expected_time:no_reuse}
The expected running time of \cref{algo:algorithms:graphs:no_reuse} in a graph $G\in\randgraphp$ is
\begin{equation*}
\bigO{ \expectednp{\psi} + \expectednp{m} - \expectednp{n\mmtdeg{2}} } = \bigO{n^4p^3}.
\end{equation*}
\end{proposition}
\begin{proof}
Recall the cost of the algorithm in \cref{prop:algorithms:graphs:no_reuse}. First, it is well known that $\expectednp{m} = p{n \choose 2}$. Second,
\begin{align*}
\expectednp{\psi}
	&= \expectednp{\sum_{st\in E} k_sk_t}
	 = \expectednp{m}\condexpectednp{k_sk_t}{a_{st}=1} \\
	&= p{n\choose 2}p(2n - 3 + p(n - 2)^2).
\end{align*}
We obtained the value $\condexpectednp{k_sk_t}{a_{st}=1}$ by applying the Total Law of Expectation
\begin{equation*}
\expectednp{k_sk_t} = \condexpectednp{k_sk_t}{a_{st}=1}\probnp{a_{st}=1} + \condexpectednp{k_sk_t}{a_{st}=0}\probnp{a_{st}=0}
\end{equation*}
and 
\begin{equation*}
\condexpectednp{k_sk_t}{a_{st}=0} = \condexpectednp{k_s}{a_{st}=0} \condexpectednp{k_t}{a_{st}=0} = p^2(n - 2)^2
\end{equation*}
as $\condexpectednp{k_s}{a_{st}=0}[n-1] = \expectednp{k_s}[n-1] = p(n-2)$. Lastly,
\begin{equation*}
\expectednp{n\mmtdeg{2}} = \sum_{u\in V}\expectednp{k_u^2} = n(n - 1)p(1 + (n - 2)p),
\end{equation*}
since $\expectednp{k_u^2}=(n - 1)p(1 + (n - 2)p)$ \cite{Ferrer2018a}.
\end{proof}

Before we derive the expected time and space complexities of \cref{algo:algorithms:graphs:reuse} in Erd\H{o}s-R\'enyi graphs, we first present an intermediate result.
\begin{lemma}
\label{lemma:algorithms:graphs:analysis_erdos_renyi:probability_duv_eq_1}
Consider a graph $G\in\randgraphp$. For any two $u,v\in V$ we have that
\begin{equation*}
\probnp{\delta_{uv}=1} = 1 - (1 - p^2)^{n - 2}.
\end{equation*}
\end{lemma}
\begin{proof}
The proof is straightforward
\begin{align*}
\probnp{\delta_{uv}=1}
	&= \probnp{\exists w \text{ such that } w\neq u,v \text{ and } a_{uw}=a_{vw}=1 } \\
	&= 1 - \probnp{\nexists w \text{ such that } w\neq u,v \text{ and } a_{uw}=a_{vw}=1 } \\
	&= 1 - \left(1 - p^2\right)^{n - 2}.
\end{align*}
\end{proof}

We continue with \cref{algo:algorithms:graphs:reuse}.

\begin{proposition}
\label{prop:algorithms:graphs:analysis_erdos_renyi:expected_time:reuse}
The expected running time of \cref{algo:algorithms:graphs:reuse} in a graph $G\in\randgraphp$ is
\begin{equation*}
\bigO{ {n\choose 2}(np - (1 - p)(1 - (1 - p^2)^{n-2}) ) } = \bigO{n^3p}.
\end{equation*}
\end{proposition}
\begin{proof}
We calculate the expected running time from its asymptotic cost in \cref{prop:algorithms:graphs:reuse}.
\begin{equation*}
\expectednp{nm - \sum_{uv\notin E} \delta_{uv}}
	= n{n \choose 2}p - {n \choose 2}(1 - p) \condexpectednp{\delta_{uv}}{a_{uv}=0}.
\end{equation*}
As $\delta_{uv}$ is an indicator variable,
\begin{equation*}
\condexpectednp{\delta_{uv}}{a_{uv}=0}
	= \condprobnp{\delta_{uv}=1}{a_{uv}=0}
	= \probnp{\delta_{uv}=1}.
\end{equation*}
The value of $\probnp{\delta_{uv}=1}$ is given in \cref{lemma:algorithms:graphs:analysis_erdos_renyi:probability_duv_eq_1}. The expected asymptotic cost now follows from straightforward calculations.
\end{proof}

We still cannot calculate the theoretical speedup of reusing computations in graphs that follow the Erd\H{o}s-R\'enyi random graph model since the expected costs obtained are approximate upper bounds of the actual cost. Their quotient, $n^4p^3/(n^3p) = np^2$ therefore, gives us an approximate lower bound of the speed up of reusing computations. The speed-up of reusing computations is then $\bigOmega{np^2}$. Put differently, the boundary of the region where reusing computations is warranted to be faster is given by $cnp^2 = 1$ for some constant $c$. Besides, the empirical analysis of the actual speed up of reusing computations in \cref{fig:algorithms:graphs:analysis_erdos_renyi:speedup} shows two regions: one where reusing computations is slower and another where reusing computations is faster. Interestingly, the boundary between regions (the points in \cref{fig:algorithms:graphs:analysis_erdos_renyi:speedup} where the speed up is one) is defined by a  curve that decreases as $p$ increases (for small $p$), consistently with the approximate theoretical analysis.

Finally, we also analyze the expected size of $H$ in Erd\H{o}s-R\'enyi. \cref{prop:algorithms:graphs:analysis_erdos_renyi:expected_size_H} shows that the probability that a pair of vertices is in $H$ grows quickly to 1 for sufficiently large $n$, indeed the effect is sharper as $n$ increases. Consequently, the expected size $H$ grows as $p$ increases but reaches ${n \choose 2}$, its maximum value, quickly for sufficiently large $n$.

\begin{proposition}
\label{prop:algorithms:graphs:analysis_erdos_renyi:expected_size_H}
Consider \cref{algo:algorithms:graphs:reuse} on a graph $G\in\randgraphp$. Let $I_{uv}$ be an indicator random variable that equals $1$ when $\pair{u}{v}$ is in $H$ and 0 otherwise. Then the probability that a $\pair{u}{v}$ is in $H$ is 
\begin{equation}
\label{eq:algorithms:graphs:analysis_erdos_renyi:probability_I_uv}
\probnp{I_{uv}=1} = 1 - (1 - p)(1 - p^2)^{n - 2}
\end{equation}
and the expected size of $|H|$ at the end of the algorithm is then
\begin{equation}
\label{eq:algorithms:graphs:analysis_erdos_renyi:expected_size_H}
\expectednp{|H|} = {n\choose 2}\probnp{I_{uv}=1}.
\end{equation}
\end{proposition}
\begin{proof}
Simply,
\begin{equation*}
\expectednp{|H|}
	= \sum_{\pair{u}{v}\in{V\choose 2}} \probnp{I_{uv}=1} % \probnp{\pair{u}{v}\in H}
	= {n \choose 2} \probnp{I_{uv}=1}.
\end{equation*}
Then
\begin{align*}
\probnp{I_{uv}=1}
	&= \probnp{a_{uv}=1 \cup \delta_{uv}=1} \\
	&= \probnp{a_{uv}=1} + \probnp{\delta_{uv}=1} - \probnp{a_{uv}=1 \cap \delta_{uv}=1}.
\end{align*}
Easily enough, $\probnp{a_{uv}=1}= p$, the value of $\probnp{\delta_{uv}=1}$ is given in \cref{lemma:algorithms:graphs:analysis_erdos_renyi:probability_duv_eq_1}, and
\begin{equation*}
\probnp{a_{uv}=1 \cap \delta_{uv}=1}
	= \probnp{a_{uv}=1}\probnp{\delta_{uv}=1}
	= p\left(1 - (1 - p^2)^{n - 2}\right),
\end{equation*}
which leads directly to $\probnp{I_{uv}=1} = 1 - (1 - p)(1 - p^2)^{n - 2}$.
\end{proof}

\begin{figure}
	\centering
	\includegraphics[scale=0.4]{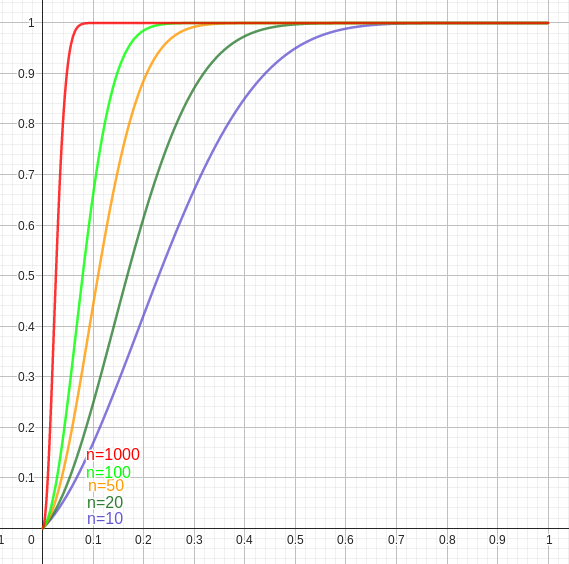}
	\caption{$\probnp{I_{uv}=1}$, the probability that any two vertices $u$, $v$ are connected via a third vertex ($y$ axis) in a graph $G\in\randgraphp$ (\cref{eq:algorithms:graphs:analysis_erdos_renyi:probability_I_uv}), for several fixed values of $n$ as a function of $p$ ($x$ axis). Made with GeoGebra \cite{Geogebra6}.}
	\label{fig:algorithms:graphs:analysis_erdos_renyi:probability_I_uv}
\end{figure}

\cref{fig:algorithms:graphs:analysis_erdos_renyi:probability_I_uv} shows the growth of the probability function in \cref{eq:algorithms:graphs:analysis_erdos_renyi:probability_I_uv}, $\probnp{I_{uv}=1}$, for several values of $n$. It can be seen that the probability of any pair of two vertices to be in $H$ at the end of the algorithm surges with $p$ and reaches values close to $1$ rapidly, even for low $n$. Therefore, in Erd\H{o}s-R\'enyi graphs, the function in \cref{eq:algorithms:graphs:analysis_erdos_renyi:expected_size_H} reaches ${n\choose 2}$ at the same rate as $\probnp{I_{uv}=1}$ reaches $1$.

It is worth studying at what values of $p$ we should expect $H$ to have size $\bigO{n}$ as this is the size cost of Algorithm \cref{algo:algorithms:graphs:no_reuse}. These are the values of $p$ for which $1 - (1 - p)(1 - p^2)^{n - 2} = c/n$, for some constant $c$. There does not seem to be an exact solution, but we can easily derive an upper bound that indicates that these values vanish as $n$ increases.

\begin{proposition}
\label{prop:algorithms:graphs:analysis_erdos_renyi:value_of_p_linear_size}
Consider a graph $G\in\randgraphp$. The value of $p$ such that the expected size of $H$ at the end of \cref{algo:algorithms:graphs:reuse} is $c n$, satisfies
\begin{equation*}
p \le \sqrt{1 - \sqrt[n-1]{1 - \frac{c}{n}}}.
\end{equation*}
\end{proposition}
\begin{proof}
We start by rearranging terms and applying logarithms
\begin{align*}
1 - (1 - p)(1 - p^2)^{n - 2} &= c/n \\
(n - 1)\ln{(1 - p)} + (n - 2)\ln{(1 + p)} &= \ln{(1 - c/n)}.
\end{align*}
Notice that
\begin{equation*}
(n - 1)\ln{(1 - p)} + (n - 1)\ln{(1 + p)} \ge \ln{(1 - c/n)}
\end{equation*}
and
\begin{equation*}
                 \ln{(1 - p^2)} \ge (1/(n-1))\ln{(1 - c/n)}
\Longleftrightarrow  1 - p^2 \ge \sqrt[n-1]{ 1 - c/n } 
\end{equation*}
from which we can obtain the desired result.
\end{proof}

\begin{figure}
	\centering
	\includegraphics[scale=0.4]{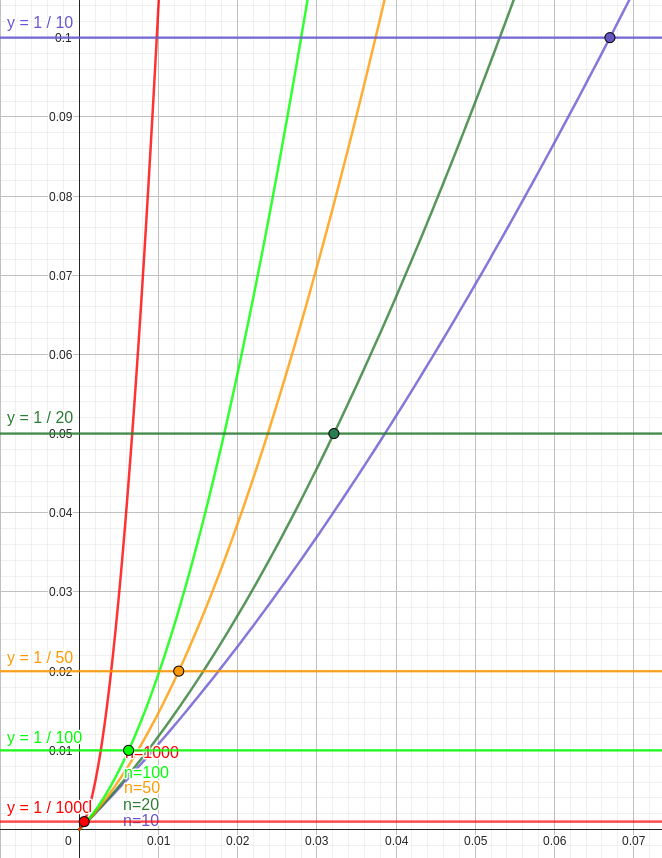}
	\caption{Zoom in of \cref{fig:algorithms:graphs:analysis_erdos_renyi:probability_I_uv}. The points indicate the intersection of a curve obtained with a fixed value of $n$ and its corresponding curve $1/n$. Made with GeoGebra \cite{Geogebra6}.}
	\label{fig:algorithms:graphs:analysis_erdos_renyi:value_of_p_linear_size}
\end{figure}

\cref{fig:algorithms:graphs:analysis_erdos_renyi:value_of_p_linear_size} shows the points $(p,1/n)$, where $p$ is the probability in $\randgraphp$ for which $\probnp{I_{uv}=1}$ (\cref{eq:algorithms:graphs:analysis_erdos_renyi:probability_I_uv}) equals $1/n$.

%-------------------------------------------------%
% automatic inline of '4-2-algorithm-forests.tex' %
%-------------------------------------------------%
\subsection{An algorithm for trees and forests}
\label{sec:algorithms:forests}

When instantiating \cref{algo:algorithms:graphs:reuse} on forests, its time complexity can be easily estimated to be $\bigO{n^2}$, and its space complexity to be $\bigO{n\mmtdeg{2}}$ (since $\ntris[\forest]=0$ for any forest $\forest$). However, $\gvar{C}$ can be computed in time and space $\bigO{n}$ in forests by exploiting their acyclicity. We first draw the attention of the reader to the simplified subgraph counting formulas for $\npathsfour[T]$, $\npathsfive[T]$ given in \cref{sec:arithmetic_expressions:general_results} (\cref{prop:arithmetic_expressions:general_results:L_4:traversal:trees,prop:arithmetic_expressions:general_results:L_5:traversal:trees}). We draw the attention of the reader to the simplification of $\Lambda_1$ and $\Lambda_2$ (\cref{eq:arithmetic_expressions:general_results:Lambda_1:def,eq:arithmetic_expressions:general_results:Lambda_2:def}) in \cref{sec:arithmetic_expressions:general_results} (\cref{prop:arithmetic_expressions:general_results:Lambda_1:trees,prop:arithmetic_expressions:general_results:Lambda_2:trees}). We prove the algorithm's correctness and complexity in \cref{prop:algorithms:forests}.

\begin{proposition}
\label{prop:algorithms:forests}
Let $\forest$ be a forest of $n$ vertices. \cref{algo:algorithms:forests} computes $\gvar{C}$ in $F$ in time and space $\bigO{n}$.
\end{proposition}
\begin{proof}
The terms $q$, $K$, $\Phi_1$ and $\Phi_2$ are defined for general graphs as sums of values that only depend on vertices that are adjacent (\cref{eq:introduction:overview:size_Q,eq:arithmetic_expressions:general_results:sum_degrees,eq:arithmetic_expressions:general_results:Phi_1,eq:arithmetic_expressions:general_results:Phi_2}). The terms $\npathsfour[F]$ and $\npathsfive[F]$, $\Lambda_1$ and $\Lambda_2$ can be computed in forests as easily as they can be computed in trees. Namely, the arguments in the proofs of \cref{prop:arithmetic_expressions:general_results:L_4:traversal:trees,prop:arithmetic_expressions:general_results:L_5:traversal:trees,prop:arithmetic_expressions:general_results:Lambda_1:trees,prop:arithmetic_expressions:general_results:Lambda_2:trees} that led to their respective equations can be extended to forests straightforwardly while keeping the same conclusions thanks to the fact that forests, as their definition states, are merely a disjoint union of trees.

Recall that $m = n-c$ in a forest with $c$ connected components. Then, it is easy to see that the time complexity of the algorithm is $\bigO{n}$ since the loop in \cref{algo:algorithms:forests:main_loop} iterates over the set of edges of $\forest$, and the computation of the $\xi(s)$ is also done in $\bigO{n}$ steps. It is also obvious that the space complexity is $\bigO{n}$ since we have to store the values of $\xi(s)$ for each $s\in V$.
\end{proof}

\begin{algorithm}
	\caption{Computing $\gvar{C}$ in forests.}
	\label{algo:algorithms:forests}
	\DontPrintSemicolon
	
	\KwIn{$\forest$ a forest.}
	\KwOut{$\gvar{C}$, the variance of the number of crossings in $\forest$.}
	
	\SetKwProg{Fn}{Function}{ is}{end}
	\Fn{\textsc{VarianceC}$(\forest)$} {
		Calculate $\mmtdeg{2}$ and $\mmtdeg{3}$ in $\bigO{n}$-time \tcp{\cref{eq:introduction:mmt_deg}}
		$\xi(s) \gets 0^n$ \tcp{\cref{eq:arithmetic_expressions:preliminaries:sum_degrees_neighbors}}
		Calculate $\xi(s)$ for each $s\in V$ in $\bigO{n}$-time \\
		$\psi \gets 0$ \tcp{\cref{eq:introduction:psi}}
		$\Phi_1, \Phi_2 \gets 0$ \tcp{\cref{eq:arithmetic_expressions:general_results:Phi_1,eq:arithmetic_expressions:general_results:Phi_2}}
		$\npathsfour[F], \npathsfive[F] \gets 0$ \tcp{\cref{eq:arithmetic_expressions:general_results:L_4:traversal:trees,eq:arithmetic_expressions:general_results:L_5:traversal:trees}}
		$\Lambda_1, \Lambda_2 \gets 0$ \tcp{\cref{eq:arithmetic_expressions:general_results:Lambda_1:trees,eq:arithmetic_expressions:general_results:Lambda_2:trees}}
		
		\For {$\{s,t\}\in E(F)$} { \label{algo:algorithms:forests:main_loop}
			$\psi \gets \psi + k_sk_t$ \;
			$\npathsfive[F] \gets \npathsfive[F] + (k_t - 1)(\xi(s) - k_t - k_s + 1) + (k_s - 1)(\xi(t) - k_t - k_s + 1)$ \;
			
			$\npathsfour[F] \gets \npathsfour[F] + (k_s - 1)(k_t - 1)$ \;
			
			$\Lambda_1 \gets \Lambda_1 + (k_t - 1)(\xi(s) - k_t) + (k_s - 1)(\xi(t) - k_s)$ \;
			$\Lambda_2 \gets \Lambda_2 + (k_s - 1)(k_t - 1)(k_s + k_t)$ \;
			$\Phi_1 \gets \Phi_1 - k_sk_t(k_s + k_t)$ \;
			$\Phi_2 \gets \Phi_2 + (k_s + k_t)(n\mmtdeg{2} - \xi(s) - \xi(t) - k_s(k_s - 1) - k_t(k_t - 1))$ \;
					%\label{algo:algorithms:forests:last-loop-trees}
		}
		
		$q \gets \frac{1}{2}[m(m + 1) - n\mmtdeg{2}]$ \tcp{\cref{eq:introduction:overview:size_Q}}
		$K \gets (m + 1)n\mmtdeg{2} - n\mmtdeg{3} - 2\psi$ \tcp{\cref{eq:arithmetic_expressions:general_results:sum_degrees}}
		
		$\Phi_1 \gets \Phi_1 + (m + 1)\psi$ \tcp{\cref{eq:arithmetic_expressions:general_results:Phi_1}}
		$\Phi_2 \gets \frac{1}{2}\Phi_2$ \tcp{\cref{eq:arithmetic_expressions:general_results:Phi_2}}
		
		$\npathsfive[F] \gets \frac{1}{2}\npathsfive[F]$ \tcp{\cref{eq:arithmetic_expressions:general_results:Lambda_1:trees}}
		
		$\Lambda_2 \gets \Lambda_2 + \Lambda_1$ \tcp{\cref{eq:arithmetic_expressions:general_results:Lambda_2:trees}}
		
		Compute $\gvar{C}$ by instating \cref{eq:arithmetic_expressions:var_C:general:forests} appropriately \;
	}
\end{algorithm}
%----------------------------------------%
% automatic inline of '5-discussion.tex' %
%----------------------------------------%
\section{Discussion and future work}
\label{sec:discussion}

We start discussing the primary goal of this article, namely the problem of computing $\gvar{C}$ efficiently (\cref{sec:discussion:computation_of_variance}). Then, since the calculation of $\gvar{C}$ reduces to a problem of counting the number of subgraphs of a certain type (\cref{eq:introduction:overview:var_C:general:freq__x__exp} and \cref{table:introduction:overview:types_as_subgraph_counting}), we proceed to discuss our contributions with respect to two research problems that are by-products of our primary goal: the counting of all subgraphs (\cref{sec:discussion:general_subgraphs}) and the counting of specific subgraphs (\cref{sec:discussion:specific_subgraphs}). Indeed, the problem of the computation of $\gvar{C}$ can be seen as an instance of a third research problem: counting on a prescribed set of subgraphs. We conclude suggesting future research (\cref{sec:discussion:applications}).

\subsection{Computation of variance}
\label{sec:discussion:computation_of_variance}

We have developed efficient algorithms to calculate $\gvar{C}$ in general graphs and also in forests. Part of our work consists of reducing the complexity of the computation of arithmetic expressions of the $f_\omega$'s for all types (except types $00$ and $01$, since $\gexpet{00}=\gexpet{01}=0$). Using \cref{eq:arithmetic_expressions:var_C:graphs:general}, we derived algorithms tailored to layouts meeting the requirements characterized in \cref{sec:introduction}.

We have alleviated the complexity of the computation of $\gvar{C}$ for such layouts by several orders of magnitude (\cref{algo:algorithms:graphs:no_reuse}) with respect to the naive $\bigO{m^4}$-time subgraph counting algorithm introduced in \cref{sec:background}. Moreover, we have also improved \cref{algo:algorithms:graphs:no_reuse} to reuse computations (\cref{algo:algorithms:graphs:reuse}) and tailored a solution for forests that is even faster than a direct instantiation of \cref{algo:algorithms:graphs:reuse}.

Furthermore, we have demonstrated the advantage of reusing computations on Erd\H{o}s-R\'enyi random graphs theoretically and measured empirically the speed up of reusing computations in Erd\H{o}s-R\'enyi random graphs. \cref{fig:algorithms:graphs:analysis_erdos_renyi:speedup} shows significant speed up values in dense graphs. However, reusing computations introduces a non-negligible cost in very sparse graphs: \cref{algo:algorithms:graphs:no_reuse} seems to be faster than \cref{algo:algorithms:graphs:reuse} for low of $p$ ($0\le p\le 0.05$) and $n\le300$. Nevertheless, it is important to notice that the speed up measurements in sparse graphs seem to have an increasing (though slow) tendency, which suggests that reusing computations in sparse graphs is favorable when $n$ is large ($n\ge500$).

\subsection{Counting subgraphs in general}
\label{sec:discussion:general_subgraphs}

Since the calculation of $\gvar{C}$ reduces to a problem of counting the number of subgraphs of a certain type (\cref{eq:introduction:overview:var_C:general:freq__x__exp} and \cref{table:introduction:overview:types_as_subgraph_counting}), one could also apply algorithms for counting graphlets or graphettes \cite{Przulj2007a,Hasan2017a}. Graphlets are connected subgraphs while graphettes are a generalization of graphlets to potentially disconnected subgraphs. The subgraphs for calculating the number of products of each types are indeed graphettes; only types 03 and 04 the subgraphs are graphlets (\cref{fig:introduction:overview:types_of_subgraphs}).  As the subgraphs we are interested in have between 4 and 6 vertices\footnote{Recall that types $00$ and $01$ do not matter.} (\cref{fig:introduction:overview:types_of_subgraphs}), we could generate all subsets of 4, 5 and 6 vertices and use the look-up table provided in \cite{Hasan2017a} to classify the corresponding subgraphs in constant time for each subset. However, that would produce and algorithm that runs in $\Theta(n^6)$ time while ours runs in $\smallo{n^5}$ if we do not reuse computations and in $\bigO{n^3}$ if we reuse them (\cref{table:background:summary_algorithms}).

\subsection{Counting specific subgraphs}
\label{sec:discussion:specific_subgraphs}

Our algorithms are based on formulas for the $f_\omega$'s, namely, the number of products of each type (\cref{table:arithmetic_expressions:theoretical_formulas:frequencies:summary}). As a side-effect of such characterization, we have contributed with expressions for the number of paths of 4 and 5 vertices of a graph (\cref{prop:arithmetic_expressions:general_results:L_4:elements_Q,prop:arithmetic_expressions:general_results:L_5:elements_Q}) that are more compact than others obtained in previous work \cite{Movarraei2014a}.

The subgraphs present in \cref{eq:arithmetic_expressions:var_C:graphs:general}, namely $\lintree[4]$, $\cycle[4]$, $\lintree[5]$, $\graphpaw$ and $\CoL$ (the last two are depicted in \cref{fig:arithmetic_expressions:general_results:graphs_paw_and_C3_L2}) could be counted easily by relying on previous work by Alon {\em et al.} \cite{Alon1997a} and Movarraei \cite{Movarraei2014a} based on powers of the adjacency matrix. For example, Alon {\em et al.} \cite{Alon1997a}, among many other contributions, showed that
\begin{align*}
n_G(\cycle[3]) &= \frac{1}{6}tr(A^3), \\
\nsquares &= \frac{1}{8}\left[ \text{tr}(A^4) - 4\sum_{u\in V} \binom{k_u}{2} - 2m \right], \\
\ngraphpaw &= \frac{1}{2}\left[ \sum_{i=1}^n a_{ii}^{(3)}(d_i - 2) \right].
\end{align*}
Moreover, Movarraei \cite{Movarraei2014a} showed that
\begin{align*}
\npathsfour
	&= \frac{1}{2} \sum_{i \neq j} (a_{ij}^{(3)} - (2k_j - 1)a_{ij}), \\
\npathsfive
	&= \frac{1}{2}
	\left[
		  \sum_{i \neq j} (a_{ij}^{(4)} - 2a_{ij}^{(2)}(k_j - a_{ij}))
		- \sum_{i = 1}^n
		  \left(
			(2k_i - 1)a_{ii}^{(3)} + 6\binom{k_i}{3}
		  \right)
	\right].
\end{align*}
It easy to see that the expressions by Alon {\em et al.} and those of Movarraei can all be computed in time $\bigO{n^\omega}$, where $\omega$ is the exponent of the fastest algorithm for matrix-matrix multiplication, as already pointed out by Alon {\em et al.} \cite[Theorem 6.3]{Alon1997a}. The value of $\omega$ depends on the algorithm used (obviously, $\omega \geq 2$). The starting point is Strassen's, with $\omega = 2.807$ \cite{Strassen1969a}.
% Finally, besides the fact that matrix-matrix multiplication algorithms are not sensitive to the graph's structure (in the shape of an adjacency matrix), 
This cost can only be lowered by applying improvements of increasing complexity over already-complex algorithms (such as Strassen's \cite{Strassen1986a}, Coppersmith-Winograd's \cite{Coppersmith1981a,Coppersmith1990a}) which may introduce large constants that multiply execution time \cite{Iliopoulos1989a}. Moreover, some algorithms may need to enlarge the matrices to operate properly\footnote{For example, Strassen's algorithm requires the number of rows (and of columns) to be a power of $2$ in the matrices being multiplied \cite{Strassen1969a}. This can be achieved by padding the matrix with $0$'s, but it may become infeasible quite quickly.}.

Now, in this paper we followed an alternative approach, combinatorial in nature, to reduce the time and space complexities with respect to the naive algorithm (\cref{algo:background:brute_force}), in which we developed and applied supporting expressions to count many distinct subgraphs. It is clear that the complexity of this approach is determined by the complexity of the more complex subgraphs to count, namely, $\lintree[5]$ and $\cycle[4]$. Our main goal was to devise an algorithm whose time complexity adapted to the graph's structure. 

The first algorithm we devised (\cref{algo:algorithms:graphs:no_reuse}) has time complexity $\smallo{nm^2}$ and space complexity $\bigO{n}$ (\cref{prop:algorithms:graphs:no_reuse}). Although there is some compensation as far as space complexity is concerned, its complexity is obviously much higher than that of a naive matrix-matrix multiplication algorithm. We developed a second algorithm (\cref{algo:algorithms:graphs:reuse}) to reduce time complexity at the expense of space complexity; \cref{algo:algorithms:graphs:reuse} has time complexity $\bigO{nm}$ and space complexity $\bigO{n^2}$ (\cref{prop:algorithms:graphs:reuse}). While the space complexity of this algorithm is asymptotically the same as using adjacency matrices, it is, in the worst case, as fast as that of a naive matrix-matrix multiplication algorithm. We can conclude, therefore, that, when compared to an algorithm that uses adjacency matrices, \cref{algo:algorithms:graphs:reuse} has the same drawbacks as far as space complexity is concerned, but it has advantages concerning time complexity since it is adaptive to the graph's structure.

\subsection{Future work}
\label{sec:discussion:applications}

Concerning the algorithms and their underlying theory, future work should investigate simpler formulas for the number of products of the types, specially types that are relevant for the calculation of variance but for whom a simple arithmetic formula is not forthcoming yet, i.e. types 03, 021 and 022 (\cref{table:arithmetic_expressions:theoretical_formulas:frequencies:summary}). In this and previous articles \cite{Ferrer2018a}, we have provided a satisfactory solution to the computation of the first and the second moments of $C$ about zero, i.e. $\gexpe{C}$ and $\gexpe{C^2} = \gvar{C} + \gexpe{C}^2$. Similar methods could be applied to derive algorithms for higher moments, e.g., $\gexpe{C^3}$.

Concerning applications, the algorithms above pave the way for many applications. A crucial one is offering computational support for theoretical research on the distribution of crossings in random layouts of vertices \cite{Moon1965a,Alemany2018b}. In the domain of statistical research on crossings in syntactic dependency trees \cite{Ferrer2017a}, $C$ has been shown to be significantly low with respect to random linear arrangements with the help of Monte Carlo statistical tests in syntactic dependency structures \cite{Ferrer2017a}. Faster algorithms for such a test could be developed with the help of Chebyshev-like inequalities that allow one to calculate upper bounds of the real $p$-values. These inequalities typically imply the calculation of $\lexpe{C}$, which is straightforward (simply, $\lexpe{C} = q/3$ \cite{Alemany2018a}) and the calculation of $\lvar{C}$, that thanks to the present article, has become simpler to compute. Similar tests could be applied to determine if the number of crossings in RNA structures \cite{Chen2009a} is lower or greater than that expected by chance. For the same reasons, our algorithms allow one to calculate $z$-scores of $C$ (\cref{eq:introduction:z-score}) efficiently. $z$-scores have been used to detect scale invariance in empirical curves \cite{Cocho2015a,Morales2016a} or motifs in complex networks \cite{Milo2002a} (see \cite{Stone2019a} for a historical overview). Thus, $z$-scores of $C$ could lead to the discovery of new statistical patterns involving $C$. Moreover, $z$-scores of $C$ can help aggregate or compare values of $C$ from graphs with different structural properties (number of vertices, degree distribution,...). In addition, $z$-scores of $C$ can help to aggregate or compare values of $C$ from heterogeneous sources as it happens in the context syntactic dependency trees, where one has to aggregate or compare values of $C$ of syntactic dependency trees from the same language but that differ in parameters such as size $n$ or internal structure (e.g., different value of $\mmtdeg{2}$) \cite{Ferrer2017a}. We hope that our algorithms stimulate further theoretical research on the distribution of crossing as well as research on crossings in spatial network, specially in the domain of linguistic and biological sequences.

\bmhead{Acknowledgments}

We are grateful to J. L. Esteban for helpful comments. LAP is supported by Secretaria d’Universitats i Recerca de la Generalitat de Catalunya and the Social European Fund. This research was supported by the grant TIN2017-89244-R from MINECO (Ministerio de Econom{\'i}a y Competitividad) and the recognition 2017SGR-856 (MACDA) from AGAUR (Generalitat de Catalunya). The authors are currently supported by a recognition 2021SGR-Cat (01266 LQMC) from AGAUR (Generalitat de Catalunya).

\section*{Declarations}

\paragraph{Conflict of interest} The authors declare that they have no conflict of interest.

\appendix

%------------------------------------%
% automatic inline of '6-proofs.tex' %
%------------------------------------%
\section{Proofs}
\label{sec:proofs}

Here we give the proofs of many of the propositions given throughout this work. In particular, we give the proofs of those non-trivial propositions that are relevant enough regarding the goal of this paper.

\subsection{Proof of \cref{prop:arithmetic_expressions:general_results:L_4:elements_Q}}
\label{sec:proofs:sum_adjs_equals_4_paths}

\begin{figure}
	\centering
	\includegraphics[scale=0.95]{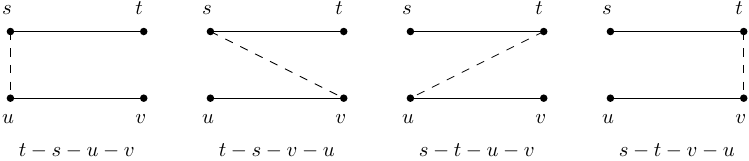}
	\caption{The four $\lintree[4]$ we can make with $\{st,uv\} \in Q$. }
	\label{fig:paths-4-in-element-Q}
\end{figure}

Let $\{st,uv\}$ be a pair of edges from $Q$. It is easy to see that the expression $a_{su} + a_{sv} + a_{tu} + a_{tv}$ counts how many $\lintree[4]$ we can make with these two edges. By definition of $Q$, $a_{st} = a_{uv} = 1$. To these two edges, we only have to add one of the four edges in the expression (i.e., any of the edges that connect a vertex of $st$ with another vertex of $uv$) to make a $\lintree[4]$. Each of the edges in the expression produces a distinct $\lintree[4]$ (\cref{fig:paths-4-in-element-Q}).

Let $\lintree[4](q_1)$ be the set of $\lintree[4]$ into which $q_1 \in Q$ is mapped. The $\lintree[4]$ of $\lintree[4](q_1)$ follow a concrete pattern: the edges of $q$ are at each end of the $\lintree[4]$ (\cref{fig:paths-4-in-element-Q}). \cref{eq:arithmetic_expressions:general_results:L_4:elements_Q:1st} could be false for two reasons. Firstly, some $\lintree[4]$ is not counted. Suppose that there exists a $\lintree[4]$ with vertices $s-t-u-v$ in $G$ that is not counted in the summation. If this was true then the pair of independent edges we can make using its vertices ($\{st,uv\}$) would not be in $Q$. But this cannot happen by definition of $Q$. Secondly, some $\lintree[4]$ is counted more than once. It can only happen when there exists a $q_2 \in Q, q_1 \neq q_2$ such that $\lintree[4](q_1) \cap \lintree[4](q_2) \neq \emptyset$. For this to happen, $q_2$ must place the same edges as $q_1$ at the end of the $\lintree[4]$, which is a contradiction because $q_2 \neq q_1$. Therefore, such a $q_2$ does not exist.

\cref{eq:arithmetic_expressions:general_results:L_4:elements_Q:2nd} follows from \cite{Movarraei2014a} and can be expressed as
\begin{equation*}
\frac{1}{2} \sum_{s \neq t} (a^{(3)}_{st} - (2k_t - 1)a_{st})
	= m_3 + m_1 - \sum_{s \neq t} a_{st} k_t \\
	= m_3 + m - n\mmtdeg{2},
\end{equation*}
hence \cref{eq:arithmetic_expressions:general_results:L_4:elements_Q:3rd}.

%%%%%%%%%%%%%%%%%%%%%%%%%%%%%%%%%%%%%%%%%%%%%%%%%%%%%%%%%%%%%%%%%%%%%%%%

\subsection{Proof of \cref{prop:arithmetic_expressions:general_results:L_4:traversal}}
\label{sec:proofs:graphs:paths_4}

Consider three vertices $s$,$t$,$u$ inducing a path of $3$ vertices in $G$: $(s,t,u)$. We can count all induced subgraphs $\lintree[4]$ that start with vertices $s,t,u$ and finish at $v\in\Gamma(u)$ by counting how many $v$ are different from $s$ and $t$ in the neighborhood of $u$, $\Gamma(u)$. Then,
\begin{equation*}
\npathsfour =
	\frac{1}{2}
	\sum_{s\in V}
	\sum_{t\in \Gamma(s)}
	\sum_{u\in \Gamma(t)\setminus\{s\}}
	\sum_{v\in \Gamma(u)\setminus\{s,t\}} 1.
\end{equation*}
We replace the inner-most summation with the expression $k_u - 1 - a_{su}$. This expression, when summed over the vertices $u\in\Gamma(t)\setminus\{s\}$, can be simplified further leading to
\begin{align*}
\npathsfour
	& =
	\frac{1}{2}
	\sum_{s\in V}
	\sum_{t\in \Gamma(s)} (\xi(t) - (k_s + k_t) + 1 - |c(s,t)|) \\
	&=
	\frac{1}{2}
	\sum_{st\in E} (\xi(s) + \xi(t) - 2(k_s + k_t) + 2 - 2|c(s,t)|).
\end{align*}
It is easy to see that
\begin{equation*}
\sum_{st\in E} (\xi(s) + \xi(t)) = \sum_{s\in V} k_s\xi(s) = \sum_{s\in V} \sum_{t\in \Gamma(s)} k_sk_t = 2\sum_{st\in E} k_sk_t = 2\psi.
\end{equation*}
Obtaining the expression in \cref{eq:arithmetic_expressions:general_results:L_4:traversal} is now straightforward.

%%%%%%%%%%%%%%%%%%%%%%%%%%%%%%%%%%%%%%%%%%%%%%%%%%%%%%%%%%%%%%%%%%%%%%%%

\subsection{Proof of \cref{prop:arithmetic_expressions:general_results:L_5:elements_Q}}
\label{sec:proofs:sum_next_adjs_equals_5_paths}

\begin{figure}
	\centering
	\includegraphics{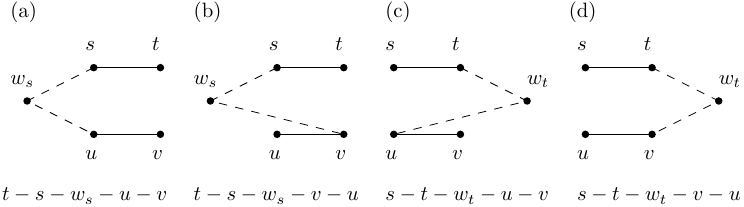}
	\caption{The four different families of $\lintree[5]$ we can make with two edges $st$ and $uv$ such that $q=\{st,uv\} \in Q$, namely $\lintree[5](q)$. (a) $w_s$ is a neighbor of $s$ and $u$, (b) $w_s$ is a neighbor of $s$ and $v$, (c) $w_t$ is a neighbor of $t$ and $u$, (d) $w_t$ is a neighbor of $t$ and $v$. }
	\label{fig:paths-5-in-element-Q}
\end{figure}

The proof is similar to the proof of \cref{prop:arithmetic_expressions:general_results:L_4:elements_Q}. For a given $q=\{st,uv\}\in Q$, the inner summations of \cref{eq:arithmetic_expressions:general_results:L_5:elements_Q}, i.e.
\begin{equation*}
	\sum_{w_s \in \Gamma(s, -stuv)} (a_{uw_s} + a_{vw_s})
	+
	\sum_{w_t \in \Gamma(t, -stuv)} (a_{uw_t} + a_{vw_t}),
\end{equation*}
count the number of $\lintree[5]$ we can make with edges $st$ and $uv$ of $q$. This set of $\lintree[5]$ is denoted as $\lintree[5](q)$ and contains the $\lintree[5]$ that follow a concrete pattern: each edge of $q$ is at one end of the $\lintree[5]$ and the edges of $q$ are linked via a fifth vertex $w$, such that $w\neq s,t,u,v$ (\cref{fig:paths-5-in-element-Q}). For example, $\lintree[5](q)$ may contain $t-s-w-u-v$ if $a_{sw}=a_{wu}=1$. Therefore, the graphs $\lintree[5](q)$, for any $q\in Q$, have $4$ different forms determined by the choice of the vertex of each edge of $q$ that will be at one of the ends of the $\lintree[5]$.

Similarly, \cref{eq:arithmetic_expressions:general_results:L_5:elements_Q} could be wrong for two reasons: some path may be counted more than once or not counted at all.

All $\lintree[5]$ are counted: by contradiction, given $\{st,uv\}\in Q$, suppose that there is a $\lintree[5]$, $s-t-w-u-v$, not counted in the inner summation. By definition of $\lintree[5]$, $a_{st}=a_{tw}=a_{wu}=a_{uv}=1$, the vertices are distinct and we have $\{st,uv\} \in Q$. Therefore, if such $\lintree[5]$ is not counted then $\{st,uv\}$ would not be in $Q$.

Some $\lintree[5]$ may be counted more than once: if this was true then for some $q_1 = \{st, uv\} \in Q$ there would exist a $q_2 \in Q$, $q_1 \neq q_2$, such that $\lintree[5](q_1) \cap \lintree[5](q_2) \neq \emptyset$. For this to happen, $q_2$ must place the same edges as $q_1$ at the end of the $\lintree[5]$, which is a contradiction because $q_2 \neq q_1$. Therefore, such a $q_2$ does not exist.

As a conclusion, all paths are counted, and no path is counted more than once. Therefore, \cref{eq:arithmetic_expressions:general_results:L_5:elements_Q} evaluates to exactly all $\lintree[5]$ in $G$.

%%%%%%%%%%%%%%%%%%%%%%%%%%%%%%%%%%%%%%%%%%%%%%%%%%%%%%%%%%%%%%%%%%%%%%%%

\subsection{Proof of \cref{prop:arithmetic_expressions:general_results:L_5:traversal}}
\label{sec:proofs:graphs:paths_5}

The proof is similar to the proof of \cref{prop:arithmetic_expressions:general_results:L_4:traversal}.
\begin{align*}
\npathsfive
&=
	\frac{1}{2}
	\sum_{s\in V}
	\sum_{t\in \Gamma(s)}
	\sum_{u\in \Gamma(s)\setminus\{t\}}
	\sum_{v\in \Gamma(t)\setminus\{s,u\}}
	\sum_{w\in \Gamma(u)\setminus\{s,t,v\}} 1 \\
&=
	\frac{1}{2}
	\sum_{s\in V}
	\sum_{t\in \Gamma(s)}
	\sum_{u\in \Gamma(s)\setminus\{t\}}
	\sum_{v\in \Gamma(t)\setminus\{s,u\}}
		(k_u - 1 - a_{ut} - a_{uv}) \\
&=
	\frac{1}{2}
	\sum_{s\in V}
	\sum_{t\in \Gamma(s)}
	\sum_{u\in \Gamma(s)\setminus\{t\}}
		((k_t - 1 - a_{ut})(k_u - 1 - a_{ut}) + 1 - |c(t,u)|) \\
&=
	\frac{1}{2}
	\sum_{s\in V}
	\sum_{t\in \Gamma(s)}
	g_1(s,t) \\
&=
	\frac{1}{2}
	\sum_{st\in E}
	\left( g_1(s,t) + g_1(t,s) \right).
\end{align*}

%%%%%%%%%%%%%%%%%%%%%%%%%%%%%%%%%%%%%%%%%%%%%%%%%%%%%%%%%%%%%%%%%%%%%%%%

\subsection{Proof of \cref{prop:arithmetic_expressions:general_results:L_5:traversal:trees}}
\label{sec:proofs:trees:paths_5}

Any $\lintree[5]$ has only one centroidal vertex $s\in V$. For any pair of different neighbors of $s$, $t\in\Gamma(s)$ and $u\in\Gamma(s)\setminus\{t\}$, the product $(k_t - 1)(k_u - 1)$ gives the amount of $\lintree[5]$ with centroidal vertex $s$ and through vertices $t$ and $u$. Therefore
\begin{equation}
\label{eq:variance-C:trees:paths-5:notice}
n_\tree(\lintree[5])
	=
	\sum_{s \in V}
	\left(
	\frac{1}{2}
	\sum_{t\in\Gamma(s)}
	\sum_{u\in\Gamma(s)\setminus\{t\}}
	(k_t - 1)(k_u - 1)
	\right).
\end{equation}
Notice that the two inner summations of \cref{eq:variance-C:trees:paths-5:notice} count such paths, twice. As
\begin{equation*}
  \sum_{u\in\Gamma(s)\setminus\{t\}} (k_u - 1)
= -k_s + 1 + \sum_{u\in\Gamma(s)\setminus\{t\}} k_u
= \xi(s) - k_t - k_s + 1
\end{equation*}
we finally obtain
\begin{equation*}
n_\tree(\lintree[5])
	=
	\frac{1}{2}
	\sum_{s \in V}
	\sum_{t\in\Gamma(s)}
	g_2(s,t)
	=
	\frac{1}{2}
	\sum_{st \in E}
	(g_2(s,t) + g_2(t,s))
\end{equation*}
with $g_2(s,t) = (k_t - 1)(\xi(s) - k_t - k_s - 1)$.

%%%%%%%%%%%%%%%%%%%%%%%%%%%%%%%%%%%%%%%%%%%%%%%%%%%%%%%%%%%%%%%%%%%%%%%%

\subsection{Proof of \cref{prop:arithmetic_expressions:general_results:graphpaw:elements_Q}}
\label{sec:proofs:num_paw_graphs}

\begin{figure}
	\centering
	\includegraphics{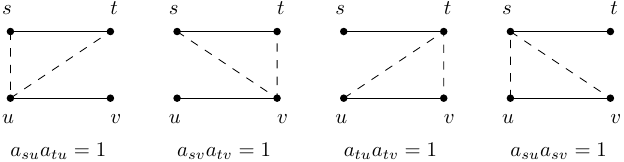}
	\caption{All possible paw graphs (\cref{fig:arithmetic_expressions:general_results:graphs_paw_and_C3_L2}(a)), that can be made with $\{st,uv\}\in Q$ given the adjacencies at the bottom of each graph.}
	\label{fig:variance:graphpaw:F-in-stuv}
\end{figure}

The proof is similar to that of \cref{prop:arithmetic_expressions:general_results:L_4:elements_Q}. We first show that the term in the summation
\begin{equation*}
(a_{tu} + a_{sv})(a_{tv} + a_{su}) =
a_{tu}a_{tv} + a_{tu}a_{su} + a_{sv}a_{tv} + a_{sv}a_{su}
\end{equation*}
counts the amount of subgraphs isomorphic to $\graphpaw$ in $G$ that we can form given $\{st,uv\}\in Q$. This can be easily seen in \cref{fig:variance:graphpaw:F-in-stuv} where, for a given $q=\{st,uv\}\in Q$, are shown all possible labeled graphs, $\graphpaw(q)$, isomorphic to $\graphpaw$, that can be made with $st$ and $uv$ assuming the existence of the pairs of edges in the summation $a_{tu}a_{tv} + a_{tu}a_{su} + a_{sv}a_{tv} + a_{sv}a_{su}$. This means that when counting how many of these pairs of adjacencies exist we are actually counting how many subgraphs isomorphic to $\graphpaw$ exist that have these four vertices.

Now we prove the claim in this proposition by contradiction. Since we know that the term inside the summation counts all labeled $\graphpaw$ that can be made with every element of $Q$, the claim can only be false for two reasons: in the whole summation of \cref{eq:arithmetic_expressions:general_results:graphpaw:elements_Q} some $\graphpaw$ is not counted, and/or some of these $\graphpaw$ is counted more than once.

Firstly, it is clear that all $\graphpaw$ are counted at least once. If one was not counted then the element of $Q$ we can make with its vertices would not be in $Q$ which cannot happen by definition of $Q$.

Secondly, none is counted more than once. Let $\graphpaw(q_1)$ be the set of labeled graphs $q_1=\{e_1,e_2\}\in Q$ is mapped to. A paw is a triangle with a link attached to it. The paws in $\graphpaw(\{e_1, e_2\})$ follow a concrete pattern: $e_1$ is linked to a triangle containing $e_2$, or the other way around, $e_2$ is linked to a triangle containing $e_1$ (\cref{fig:variance:graphpaw:F-in-stuv}). If any $\graphpaw\in \graphpaw(q_1)$ is counted twice then there exists a different $q_2\in Q$ such that $\graphpaw(q_1) \cap \graphpaw(q_2) \neq \emptyset$. For this to happen, $q_2$ must place the same edges as $q_1$ in and outside the triangle of the paw, which is a contradiction because $q_2 \neq q_1$. Therefore, such a $q_2$ does not exist.

%%%%%%%%%%%%%%%%%%%%%%%%%%%%%%%%%%%%%%%%%%%%%%%%%%%%%%%%%%%%%%%%%%%%%%%%

\subsection{Proof of \cref{prop:arithmetic_expressions:general_results:C3_L2:elements_Q}}
\label{sec:proofs:num_tri_edge}

\begin{figure}
	\centering
	\includegraphics{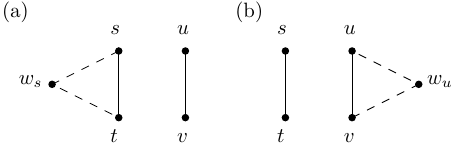}
	\caption{All the labeled graphs isomorphic to $\CoLlong$ that can be formed using a fixed element $\{st,uv\}\in Q$. In this figure, $w_s$ ($w_u$) is one of the common neighbors of $s$ and $t$ ($u$ and $v$) different from $s,t,u,v$. The graphs in (a) correspond to the first summation of \cref{eq:tri-edge:inner} and the graphs in (b) correspond to the second.}
	\label{fig:tri-edge}
\end{figure}

We use $\CoL=\CoLlong$ for brevity. Similarly as in previous proofs we first show that the inner summation of the right hand side of \cref{eq:arithmetic_expressions:general_results:C3_L2:elements_Q}, i.e.,
\begin{equation}
\label{eq:tri-edge:inner}
\sum_{w_s \in \Gamma(s, -stuv)} a_{tw_s} +
\sum_{w_u \in \Gamma(u, -stuv)} a_{vw_u}
\end{equation}
counts the amount of graphs isomorphic to $\CoL$ that can be made using the edges of $q=\{st,uv\}\in Q$, which we denote as $\CoL(q)$. Graphs in $\CoL(q)$ follow a concrete pattern: if edge $st$ is part of $\cycle[3]$ then $uv$ is the $\lintree[2]$, and vice versa. The $\cycle[3]$ is completed with another vertex $w$ neighbor to the other two vertices of the $\cycle[3]$. None of the graphs in $\CoL(q)$ are repeated and are illustrated in \cref{fig:tri-edge}.

Let $V(\CoL)=\{c_1,c_2,c_3,l_1,l_2\}$ be the set of vertices of a $\CoL$ where the $c_i\in V(\cycle[3])$ and $l_i\in V(\lintree[2])$. Then we have that $\{c_1c_2, l_1l_2\}, \{c_1c_3, l_1l_2\}, \{c_2c_3, l_1l_2\} \in Q$. If a $\CoL$ was not counted in \cref{eq:arithmetic_expressions:general_results:C3_L2:elements_Q} it would mean that none of these elements would be in $Q$. This cannot happen by definition of $Q$. Finally, since we can make three elements of $Q$ from $V(\CoL)$, every $\CoL$ is present in the set $\CoL(q_i)$ of three different $q_i$. Hence the $1/3$ factor in \cref{eq:arithmetic_expressions:general_results:C3_L2:elements_Q}.

%%%%%%%%%%%%%%%%%%%%%%%%%%%%%%%%%%%%%%%%%%%%%%%%%%%%%%%%%%%%%%%%%%%%%%%%

\subsection{Proof of \cref{prop:arithmetic_expressions:general_results:cycles_4:elements_Q}}
\label{sec:proofs:prod_adjs_twice_cycles_4}

The first equality was proven in \cite[section 4.4.5]{Alemany2018a}. \cref{eq:arithmetic_expressions:general_results:num_4cyles,eq:arithmetic_expressions:general_results:num_4cyles_alon} follow from previous results in \cite{Alon1997a,Harary1971a}. Suppose that $n_G({\cal C}_4)$ is the number of different cycles of length 4 that are contained in $G$ \cite{Alon1997a}. More technically, $\nsquares$ is the number of subgraphs of $G$ that are isomorphic to $\cycle[4]$.

We have that \cite{Alon1997a}
\begin{equation}
\nsquares = \frac{1}{8}\left[tr(A^4)-4n_G(H_2)- 2n_G(H_1)\right]
\label{eq:Alon_et_al}
\end{equation}
with
\begin{equation*}
n_G(H_1) = m,\qquad n_G(H_2) = \sum_{s=1}^n \binom{k_s}{2}.
\end{equation*}
A similar formula for $\nsquares$ was derived in the pioneering research by Harary and Manvel \cite{Harary1971a}. The fact that $n_G(H_1) = n_G(\lintree[2])$ and $n_G(H_2) = n_G(\lintree[3])$ \cite{Alon1997a}, transforms \cref{eq:Alon_et_al} into \cref{eq:arithmetic_expressions:general_results:num_4cyles}. Recalling the definition of $q$ in \cite{Piazza1991a}
\begin{equation*}
q = {m \choose 2} -  \sum_{s = 1}^n {k_s \choose 2},
\end{equation*}
we may write $\nsquares$ equivalently as
\begin{equation*}
\nsquares = \frac{1}{8}\left[tr(A^4) + 4q - 2m^2\right].
\end{equation*}

%%%%%%%%%%%%%%%%%%%%%%%%%%%%%%%%%%%%%%%%%%%%%%%%%%%%%%%%%%%%%%%%%%%%%%%%

\subsection{Proof of \cref{prop:arithmetic_expressions:general_results:sum_degrees}}
\label{sec:proofs:sum_degrees_from_verts_Q}

The amount of edges of the graph induced from the removal of vertices $s$ and $t$
\begin{equation}
\label{eq:independent_edges}
|E(G_{-st})| = \sum_{uv\in E(G_{-st})} 1 = m + 1 - k_s - k_t.
\end{equation}

The proof is straightforward. First,
\begin{align*}
K	&= \sum_{\{st,uv\} \in Q} (k_s + k_t + k_u + k_v) \\
    &= \sum_{\{st,uv\} \in Q} (k_s + k_t) + \sum_{\{st,uv\} \in Q} (k_u + k_v) \\
	&= \frac{1}{2} \left[ \sum_{st \in E} \sum_{uv\in E(G_{-st})} (k_s + k_t)
		+ \sum_{uv \in E} \sum_{st\in E(G_{-uv})} (k_u + k_v) \right] \\
	&= \sum_{st \in E} (k_s + k_t) \sum_{uv\in E(G_{-st})} 1.
\end{align*}
Thanks to \cref{eq:independent_edges} one obtains
\begin{align*}
K	&= \sum_{st \in E} (k_s + k_t) (m + 1 - k_s - k_t) \\
	&= n[(m + 1)\mmtdeg{2} - \mmtdeg{3}] - 2\psi,
\end{align*}
where $\psi$ is defined in \cref{eq:introduction:psi}.

%%%%%%%%%%%%%%%%%%%%%%%%%%%%%%%%%%%%%%%%%%%%%%%%%%%%%%%%%%%%%%%%%%%%%%%%

\subsection{Proof of \cref{prop:arithmetic_expressions:general_results:Phi_1}}
\label{sec:proofs:sum_prod_degs}

The proof is simple. Notice that
\begin{align*}
\Phi_1
	&= \sum_{\{st,uv\}\in Q} k_sk_t + \sum_{\{st,uv\}\in Q} k_uk_v \\
	&=	\frac{1}{2}
		\left[
			\sum_{st \in E} \sum_{uv \in E(G_{-st})} k_sk_t +
			\sum_{uv \in E} \sum_{st \in E(G_{-uv})} k_uk_v
		\right] \\
	&=	\sum_{st \in E} \sum_{uv \in E(G_{-st})} k_sk_t
	 =	\sum_{st \in E} k_sk_t \sum_{uv \in E(G_{-st})} 1.
\end{align*}
Thanks to \cref{eq:independent_edges}, \cref{eq:arithmetic_expressions:general_results:Phi_1} follows immediately.

%%%%%%%%%%%%%%%%%%%%%%%%%%%%%%%%%%%%%%%%%%%%%%%%%%%%%%%%%%%%%%%%%%%%%%%%

\subsection{Proof of \cref{prop:arithmetic_expressions:general_results:Phi_2}}
\label{sec:proofs:prod_sum_degs}

The proof is also straightforward. We start by noticing that
\begin{equation*}
\Phi_2 = \frac{1}{2} \sum_{st\in E} (k_s + k_t) \sum_{uv\in E(G_{-st})} (k_u + k_v).
\end{equation*}
For a fixed edge $\{s,t\}$, the inner summation above sums over the set
\begin{equation*}
E(G_{-st}) = E \setminus
	\left(\quad
	\{ us \in E \;:\; u\in V\setminus\{t\}\} \cup
	\{ ut \in E \;:\; u\in V\setminus\{s\}\} \cup
	\{ st \}
	\quad\right).
\end{equation*}
Then, for a fixed edge $\{s,t\}$, this leads to
\begin{align*}
\sum_{uv\in E(G_{-st})} (k_u + k_v)
&=	\sum_{uv\in E} (k_u + k_v)
	- \sum_{\substack{us\in E \\ u \neq t}} (k_u + k_s)
	- \sum_{\substack{ut\in E \\ u \neq s}} (k_u + k_t)
	- (k_s + k_t) \\
&=	\sum_{u\in V} k_u^2
	- \sum_{\substack{us\in E \\ u \neq t}} k_u - k_s(k_s - 1)
	- \sum_{\substack{ut\in E \\ u \neq s}} k_u - k_t(k_t - 1)
	- (k_s + k_t) \\
&=	n\mmtdeg{2} - (\xi(s) + \xi(t)) - k_s(k_s - 1) - k_t(k_t - 1).
\end{align*}

%%%%%%%%%%%%%%%%%%%%%%%%%%%%%%%%%%%%%%%%%%%%%%%%%%%%%%%%%%%%%%%%%%%%%%%%

\subsection{Proof of \cref{prop:arithmetic_expressions:general_results:Lambda_1}}
\label{sec:proofs:graphs:Lambda_1}

First, take notice that whenever one of the adjacencies $a_{su}$, $a_{sv}$, $a_{tu}$ or $a_{tv}$ equals $1$, the summation in \cref{eq:arithmetic_expressions:general_results:Lambda_1:def} adds the degree of the first and last vertices of the $\lintree[4]$ induced by the edges $st$,$uv$ and the adjacencies that equal $1$. Therefore, it is easy to see that
\begin{align*}
\Lambda_1
	&=
	\sum_{st\in E}
	\sum_{u\in \Gamma(s)\setminus\{t\}}
	\sum_{v\in \Gamma(t)\setminus\{s\}} (k_u + k_v) \\
	&=
	\sum_{st\in E}
	\left(
	(k_t - 1)(\xi(s) - k_t) + (k_s - 1)(\xi(t) - k_s) - 2S_{s,t}
	\right),
\end{align*}
where $S_{s,t}$ is defined in \cref{eq:arithmetic_expressions:preliminaries:sum_degs_common}.

%%%%%%%%%%%%%%%%%%%%%%%%%%%%%%%%%%%%%%%%%%%%%%%%%%%%%%%%%%%%%%%%%%%%%%%%

\subsection{Proof of \cref{prop:arithmetic_expressions:general_results:Lambda_2}}
\label{sec:proofs:graphs:Lambda_2}

Similarly as in \cref{prop:arithmetic_expressions:general_results:Lambda_1}, we can see that the summation in \cref{eq:arithmetic_expressions:general_results:Lambda_2:def} adds the degrees of the vertices of each $\lintree[4]$ in $G$. Therefore, we can express \cref{eq:arithmetic_expressions:general_results:Lambda_2:def} equivalently as
\begin{align*}
\Lambda_2
	&=
	\sum_{st\in E}
	\sum_{u\in\Gamma(s)\setminus\{t\}}
	\sum_{v\in\Gamma(t)\setminus\{s\}} (k_s + k_t + k_u + k_v) \\
	&=
	\Lambda_1 +
	\sum_{st\in E}
	\sum_{u\in\Gamma(s)\setminus\{t\}}
	\sum_{v\in\Gamma(t)\setminus\{s\}} (k_s + k_t) \\
	&=
	\Lambda_1 +
	\sum_{st\in E} (k_s + k_t)( (k_s - 1)(k_t - 1) - |c(s,t)|).
\end{align*}
%----------------------------------------------%
% automatic inline of '7-testing-protocol.tex' %
%----------------------------------------------%
\section{Testing protocol}
\label{sec:testing_protocol}

The derivations of the $f_\omega$'s in \cref{sec:arithmetic_expressions:frequencies} and the algorithms to compute $\gvar{C}$ presented in \cref{sec:algorithms} have been tested thoroughly via automated tests. For these algorithms we only consider the case of uniformly random linear arrangements, i.e., $\lvar{C}$. Here we detail how we assessed the correctness of the work presented above.

The calculation of the $f_\omega$'s have been tested comparing three different but equivalent procedures whose results must coincide. Firstly, the $f_\omega$'s are computed by classifying all elements of $Q\times Q$ into their corresponding $\omega$ (\cref{table:background:types:parameters} for the classification criteria). Secondly, the $f_\omega$'s are computed via \cref{eq:background:type_prod_subgraph} after counting by brute force the amount of subgraphs corresponding to each $\omega$ (\cref{fig:introduction:overview:types_of_subgraphs}). Finally, the $f_\omega$'s are calculated using the expressions summarized in \cref{table:arithmetic_expressions:theoretical_formulas:frequencies:summary} with a direct implementation of the corresponding arithmetic expression. Such a three-way test was performed on Erd\H{o}s-R\'enyi random graphs $\randgraphp$ \cite[Section V]{Bollobas1998a} of several sizes ($1\le n \le 50$) and three different probabilities of edge creation $p=0.1, 0.2, 0.5$. The test was also formed on particular types of graphs for which formulas for the $f_\omega$'s that depend only on $n$ are known: cycle graphs, linear trees, complete graphs, complete bipartite graphs, star trees and quasi star trees \cite{Alemany2018a}, for values of $n\le 100$.

The algorithms in \cref{sec:algorithms} have been tested in three ensembles of graphs: general graphs, forest and trees. The values of $\lvar{C}$ are always represented as an exact rational value (the GMP library (see \url{https://gmplib.org/}) provides implementations of these numbers). In each ensemble, $\lvar{C}$ is computed in a certain number of different ways. The test consists of checking that all the ways give the same result. The first way, $\lvar{C}^{(1)}$, consists of computing $\lvar{C}$ by brute force, i.e., by classifying all elements in $Q\times Q$ to compute the values of the $f_\omega$'s that are in turn used to obtain $\lvar{C}$ via \cref{eq:introduction:overview:var_C:general:freq__x__exp}. The second way, $\lvar{C}^{(2)}$, is obtained computing $\lvar{C}$ with a direct implementation of the derivations of the $f_\omega$'s (summarized in \cref{table:arithmetic_expressions:theoretical_formulas:frequencies:summary}). $\lvar{C}^{(3)}$, is obtained computing $\lvar{C}$ via \cref{algo:algorithms:graphs:no_reuse}. Finally, we also computed $\lvar{C}$ in forests using \cref{algo:algorithms:forests}, denoted as $\lvar{C}^{(4)}$, and in trees, denoted as $\lvar{C}^{(5)}$. Within in each ensemble of graphs, the details of the test are as follows:
\begin{enumerate}
\item General graphs. In this group we computed $\lvar{C}^{(i)}$ for $i=1,2,3$ for all the following graphs: Erd\H{o}s-R\'enyi graphs (for $10\le n\le 50$ and $p=0.1, .., 1.0$), complete graphs ($n\le 20$), complete bipartite graphs (all pairs of sizes with $2\le n_1,n_2\le 9$), linear trees ($2\le n\le 100$), one-regular graphs (all even values of $2\le n\le 100$), quasi-star trees ($2\le n\le 100$), star trees ($4\le n\le 100$) and cycle graphs ($2\le n\le 100$). Since the computation of $\lvar{C}^{(1)}$ is extremely time-consuming in dense Erd\H{o}s-R\'enyi graphs with a high number of vertices ($n\ge 40$, $p\ge0.6$), we computed it once for all these graphs and stored it on disk.

\item Forests. In this group we computed $\lvar{C}^{(i)}$ for $i=1,2,3,4$ in all the trees listed above and also in forests of random trees. We generated these forests by joining several trees of potentially different sizes generated uniformly at random. The total size of the forest was always kept under $n\le 270$.

\item Trees. In this group we computed $\lvar{C}^{(i)}$ for $i=1,2,3,4,5$, in all free unlabeled trees of size $n\le 17$.
\end{enumerate}

%\bibliographystyle{spmpsci}
%-----------------------------------------%
% automatic inline of 'main.bbl'
\newcommand{\beeksort}[1]{}

\end{document}